\documentclass[english,prd,superscriptaddress,nofootinbib,preprintnumbers]{revtex4-1}
\usepackage[T1]{fontenc}
\usepackage[latin1]{inputenc}
\usepackage{amsmath}
\usepackage{babel}
\usepackage{url}

\providecommand{\tabularnewline}{\\}

\usepackage{ifpdf}
\ifpdf 
  \usepackage[pdftex]{graphicx}
  \usepackage[hyperfootnotes=false,pdfstartview=,colorlinks=true,linkcolor=purple,citecolor=blue,urlcolor=blue,bookmarks=false]{hyperref}      
  \DeclareGraphicsExtensions{.pdf,.png}
  \usepackage{pgf}

\else 
  \usepackage{graphicx}
  \DeclareGraphicsExtensions{.eps,.png}
  \usepackage{pgf}
  \usepackage{tikz}
  \usepackage{pstricks}
\fi

\graphicspath{{./images/}}

\newcommand{\dd}{\textrm{d}}

\makeatletter

\begin{document}

\title{Real-time Cosmology}

\author{Claudia Quercellini}
\email{claudia.quercellini -a.t- uniroma2.it}
\affiliation{Dipartimento di Fisica, Universit\`a di Roma ``Tor Vergata'', via della Ricerca Scientifica 1, 00133 Roma, Italy}
\author{Luca Amendola}
\email{l.amendola -a.t- thphys.uni-heidelberg.de}
\affiliation{Institut für Theoretische Physik, Universität Heidelberg, Philosophenweg
16, 69120 Heidelberg, Germany}
\author{Amedeo Balbi}
\email{balbi -a.t- roma2.infn.it}
\affiliation{Dipartimento di Fisica, Universit\`a di Roma ``Tor Vergata'', via della Ricerca Scientifica 1, 00133 Roma, Italy}
\author{Paolo Cabella}
\email{cabella -a.t- roma2.infn.it}
\affiliation{Dipartimento di Fisica, Universit\`a di Roma ``Tor Vergata'', via della Ricerca Scientifica 1, 00133 Roma, Italy}
\author{Miguel Quartin}
\email{mquartin -a.t- if.ufrj.br}
\affiliation{Instituto de F\'isica, Universidade Federal do Rio de Janeiro, CEP 21941-972, Rio de Janeiro, RJ, Brazil}

\date{\today{}}

\begin{abstract}
   In recent years, improved astrometric and spectroscopic techniques have opened the possibility of measuring the temporal change of radial and transverse position of sources in the sky over relatively short time intervals. This has made at least conceivable to establish a novel research domain, which we dub ``real-time cosmology''. We review for the first time most of the work already done in this field, analysing the theoretical framework as well as some foreseeable observational strategies and their capability to constrain models. We first focus on real-time measurements of the overall redshift drift and angular separation shift in distant sources, which allows the observer to trace the background cosmic expansion and large scale anisotropy, respectively. We then examine the possibility of employing the same kind of observations to probe peculiar and proper accelerations in clustered systems, and therefore their  gravitational potential. The last two sections are devoted to the future change of the cosmic microwave background on ``short'' time scales, as well as to the temporal shift of the temperature anisotropy power spectrum and maps. We conclude revisiting in this context the usefulness of upcoming experiments (like CODEX and Gaia) for real-time observations.
\end{abstract}

\maketitle

\pagenumbering{arabic}

\tableofcontents

\section{Introduction}

In 1920, Willem de Sitter made this remark:  ``The choice between the systems A and B is purely a matter of taste. There is no physical criterion as yet available to decide between them''. He referred to system A and B, respectively, as the Einstein's solution to a quasi-static Universe filled with both matter and a cosmological constant $\Lambda$ (the latter not yet considered as a source term in the Einstein equations, but just as an additional geometrical term which preserved covariance) and his solution to an ``empty'' universe with $\Lambda$. Almost 100 years after de Sitter's comment, cosmology has entered the so-called ``era of precision'': the content of the Universe has been established - at least with respect to the gravitational properties of matter - with very high confidence.

The overall scenario where we, as observers, are supposed to live in turned out to be a universe with nearly flat geometry on large scales, composed by a bit less than 30\% of matter and a bit more than  70\% of a nearly uniform component (called dark energy) with pressure negative enough to drive a late time accelerated expansion. Only a small  percentage of matter ($\sim 4\%$) is made of baryons, while the remaining amount ($\sim 20\%$, called dark matter) does not directly emit electromagnetic radiation. The leading idea behind the formation of cosmic structures is that they arose by gravitational instability from small perturbations in the distribution of dark matter in the early Universe, eventually dragging baryons in the potential wells. This scenario is often called ``concordance model'', since it is consistent with the vast majority of  cosmological observations to date.

The leading cosmological observable is, at present, provided by measurements of the Cosmic Microwave Background (CMB) anistropies: small perturbations in the baryon-photon plasma in the early Universe left an imprint on the photon temperature and polarization anisotropy pattern at the time of decoupling, resulting in a goldmine of crucial information for cosmologists. In particular, the radiation we see as the CMB  appears to come from a spherical surface around the observer such that the radius of the shell is the distance each photon has traveled since it was last scattered at around the epoch of decoupling. The physical scale of the first acoustic peak of the CMB power spectrum is set by the sound horizon at recombination, while its angular scale uncovers the angular diameter distance to last scattering surface, at $z\simeq 1100$ (see \cite{2008arXiv0802.3688H} for an introductory review on the subject).

No less relevant are the recent measurements of baryon acoustic oscillations (BAO). The same scales related to the CMB anisotropy can be spot also in the baryonic matter distribution, leading to acoustic oscillations in the matter power spectrum. The main benefit of BAO comes from the concept of standard ruler: we can infer the distance of an object of known size measuring its angular dimension. In this case the known size is precisely predicted by linear theory of perturbations and corresponds to the size of the sound horizon at decoupling, measured by the CMB. Measuring the radial and tangential size of BAO at a certain redshift one is able to reconstruct the Hubble function $H(z)$ and the angular diameter distance at that redshift (consult Ref. \cite{2009arXiv0910.5224B} for a detailed review on BAO).

Last but not least, cosmologists recently learned to exploit the measurement of luminosity distance of type Ia Supernovae (SNIa) as a function of redshift. After the discovery of a specific correlation between the shape and the height of their light curve, SNIa started to be used as standard candles. The further SNIa appeared dimmer then one would have expected in a Einstein-de-Sitter model, that is in a Universe filled only with matter. This experiments unveiled the presence of a new form of energy (dark energy) that does not emit electromagnetic radiation and whose pressure is so negative that speeds up the expansion of the universe whenever it dominates the density budget (a recent review on the subject can be found in \cite{Amendolabook}).

All these complementary and pivotal measurements (in addition to determination of the Hubble constant \cite{2001ApJ...553...47F}) combined together to shape the previously mentioned concordance scenario. Also, weak lensing observations have recently started to play a prominent role  (see \cite{2010GReGr..42.2177H}). However, these observables offer an information that is primarily focused on distances (angular diameter distance and luminosity distance) and perturbation growth. Reconstructing in an accurate way the background  expansion means reconstructing in an accurate way  $H(z)$. This encodes a great part of the story of cosmological parameters (like the energy density and equation of state parameters, $\Omega_{i}(z)$ and $w_{i}(z)$ respectively) and hence the kinematics and dynamics of expansion. Unfortunately, distances are integrals over $H(z)$, which in turn is an integral over $w_{i}(z)$: therefore, the error propagation of these parameters is quite complex. In other words, the correspondence between distances and $w_i(z)$ is not straightforward. In addition, degeneracies in parameter space and systematic errors might not guarantee that accuracies on the estimates significantly improve with time. Hence the advent of further observables able to test the expansion in a different way and in complementary redshift windows is becoming more and more appealing.

In this context, given the advancement in technology occurred over the last forty years, the idea of measuring temporal variation of astrophysical observable quantities over a few decades, i.e.\ in real time, has become conceivable: hence, the possibility of establishing a novel and appealing research field, which we dubbed ``{\it real-time cosmology}'' \cite{2009PhRvL.102o1302Q}. Real-time observations may be related to variations of radial and transverse position and/or velocity of a given source. In this paper we give, for the first time, an overview of most of the work already done on the topic. In particular, we deemed useful to divide real-time observables into two classes: temporal shifts mainly tracing the background expansion and temporal shift caused by peculiar motions. For each class one can think about two ideal sub-classes, corresponding to drifts in radial and transverse directions with respect to the observer line of sight.

We will start by introducing the first and more ancient real time observable, namely the {\em redshift drift}. First conjectured by Sandage in 1962 \cite{1962ApJ...136..319S}, the redshift drift belongs to the first class of real-time observations and regards the temporal variation of redshift of distant sources as a tracer of the background cosmological expansion (i.e.\ in the radial direction). As we will show in this review, it could be  a very important cosmological probe, since it is a straightforward measure of  the change of $H(z)$ with respect to its present value, and would provide a direct detection of acceleration in the expansion. Today, spectroscopy
has reached already a sensitivity of a few meters per second. Lyman-$\alpha$ clouds along the line of sight of  very distant and bright sources like quasars \cite{1998ApJ...499L.111L} are poorly affected by peculiar motions. By using spectra populated by many sharp lines, some authors have shown that a statistical sensitivity of the required level might already be reached with the next generation of optical telescopes \cite{2008MNRAS.386.1192L}.

By using the same real-time observable and selecting sources for which the cosmological background signal is expected to be tiny (e.g.\ sources closer to us) it would also be possible to track the variation of peculiar velocity of objects in clusters and galaxies over a few decades, allowing the measurement of the acceleration caused by potential wells (which we call {\em peculiar acceleration} \cite{2008PhLB..660...81A}). This procedure opens up the possibility of reconstructing the gravitational potential in a direct way, using this detection to distinguish between different gravity models, like for example Newtonian dynamics and the MOND paradigm \cite{2008MNRAS.391.1308Q}.

Also belonging to the first class of real-time observables is the analogue of redshift drift, but in the transverse direction. In particular it has been shown that the temporal change of the angular separation between distant sources (e.g.\ quasars) can be used to detect a background anisotropic expansion. This is the so-called ``{\em cosmic parallax}'' \cite{2009PhRvL.102o1302Q}.
The standard model of cosmology
rests on two main assumptions: general relativity and a homogeneous
and isotropic metric, the Friedmann-Robertson-Walker metric (henceforth
FRW).
While general relativity has been tested with great precision at least
in laboratory and in the solar system, the issue of large-scale deviations
from homogeneity and isotropy is much less settled. There is by now
abundant literature on possible tests of the FRW metric, and on alternative
models invoked to explain the accelerated expansion by the effect
of strong, large-scale deviations from homogeneity (see
\cite{Celerier:2007} for a review).
In a FRW isotropic expansion the angular separation between sources is  constant in time (except for the effect of peculiar motions) and the cosmic parallax vanishes.
While deviations from homogeneity and
isotropy are constrained to be very small from cosmological observations,
these usually assume the non-existence of anisotropic sources in the
late universe. Conversely,
dark energy with anisotropic pressure may act as a late-time source of
anisotropy. Even if one considers no anisotropic pressure fields,
small departures from isotropy cannot be excluded, and it is interesting
to devise possible strategies to detect them. The anisotropic expansion can be either intrinsically set by the metric itself (like in Bianchi models) or mimicked by an off-center position of the observer in a inhomogeneous Universe (like for example in Lemaître-Tolman-Bondi (LTB) void models).  In the latter case a detection of cosmic parallax would also provide a test of the Copernican Principle stating that we, as observers, do not occupy a favourite position in the Universe.
If again  the signal is dominated by peculiar motions in a bound system, like in our own Galaxy, the angular temporal shift becomes proportional to peculiar acceleration in the transverse direction, which we dub  {\em proper acceleration}.

This review is organized as follows. In the first two sections we focus on the first class of real-time observables, i.e.\ the ones mapping out the background expansion. Speficifically, in Section II we introduce the concept of redshift drift, its derivation and forecasted constraining power for several dark energy FRW models, as well as for less symmetric space-times. Then, in Section III, we examine the cosmic parallax signal in LTB models with an off-center observer, both from an analytical point of view and with a full numerical derivation. We present the cosmic parallax in Bianchi I models as well, where the source of anisotropy is an anisotropically distributed dark energy density.  In the subsequent sections we concentrate on works that examine the second class of real-time observables, that is on signals generated by peculiar motions. In Section IV we derive the real-time peculiar acceleration expression both in linear approximation and in non-linear structures like clusters and galaxies. In Section V proper acceleration detected by temporal drift in the angular position of test particles in our own Galaxy is presented for the first time. Section VI is dedicated to recent papers presenting forecasts of future time variations of CMB temperature, angular power spectrum and maps \cite{2007ApJ...671.1075L, 2007PhRvD..76l3010Z}.
Finally, some details about the observational strategies, instrumental required accuracies (both astrometric and spectroscopic) and capability of already planned mission to measure real-time observables are collected in Section VII. In Section VIII we draw our conclusions.


\section{The redshift drift}
\label{sec:drift}

 The measurement of the expansion rate of the Universe at different redshifts is
crucial to investigate the cause of the accelerated expansion, and to
discriminate candidate models. Until now, a number of cosmological tools have
been successfully used to probe the expansion and the geometry of the Universe.

Depending on the underlying cosmological model, one expects the redshift of any
given object to exhibit a specific variation in time, or {\em redshift drift}. An interesting issue,
then, is to study whether the observation of this variation, performed over a
given time interval, could provide useful information on the physical mechanism
responsible for the acceleration, and be able to constrain specific models. This
is the main goal of this Section. In addition to being a direct probe of the
dynamics of the expansion, the method has the advantage of not relying on a
determination of the absolute luminosity of the observed source, but only on the
identification of stable spectral lines, thus reducing the uncertainties from
systematic or evolutionary effects.

The possibility of using the time variation of the redshift of a source as probe
of cosmological models was first proposed by Sandage~\cite{1962ApJ...136..319S}.
The predicted signal was  less than a cm/s per year and, at the time, deemed
impossible to observe. Although the test was mentioned a few other times in the
literature over the last decades (e.g.
\cite{1962ApJ...136..334M,1981ApJ...247...17L,1980ApJ...240..384R}) it was not
until recently that the feasibility of its observation was reassessed by Loeb
\cite{1998ApJ...499L.111L}  and judged within the scope of future technology
(see also \cite{1997fpc..book.....L}). In particular, the foreseen development
of extremely large observatories, such as the European ELT (E-ELT), the Thirty Meter Telescope (TMT) and the Giant Magellan Telescope (GMT), with diameters in
the range 25-100 m, and the availability of ultra-stable, high-resolution
spectrographs, encouraged new evaluations of the expected signal for the current
standard cosmological model (dominated by a cosmological constant), through the
analysis of realistic simulations. The conclusion of such studies was that the
perspective for the future observation of redshift variations looks quite
promising. For example, the authors in \cite{2005Msngr.122...10P} pointed out
that the CODEX (COsmic Dynamics Experiment) spectrograph should have the right
accuracy to detect the expected signal by monitoring the shift of Lyman-$\alpha$
forest absorption lines of distant ($z \ge 2$) quasars over a period of a few decades. These sources have the advantage of being very stable and basically
immune from peculiar motions. In Section \ref{subsec:driftobs} we will explore
in more detail the observational strategy.

Such new prospects lately prompted renewed interest in the theoretical
predictions of the redshift variation in different scenarios (see
e.g.~\cite{2007PhRvD..75f2001C,2007PhRvD..76f3508L,2007MNRAS.382.1623B,
2007PhRvD..76l3508Z,2010PhLB..691...11Z,2010PhRvD..81d3522Q,2010MNRAS.402..650K,
2009arXiv0910.4825J,2010JCAP...06..017D,2010arXiv1010.0091Y}).

Despite its inherent difficulties, the method has many interesting advantages.
One is that it is a direct probe of the dynamics of the expansion, while other
tools (e.g. those based on the luminosity distance) are essentially geometrical
in nature. This could shed some light on the physical mechanism driving the
acceleration. For example, even if the accuracy of future measurements will turn
out to be insufficient to discriminate among specific models, this test would be
still valuable as a tool to support the accelerated expansion in an independent
way, or to check the dynamical behaviour of the expansion expected in general
relativity compared to alternative scenarios.  It must be noted that radial  BAO surveys can also be used to measure $H(z)$. This is due to the fact that radial BAO are a measure of the comoving distance in a given redshift bin, which, for a narrow enough bin, is inversely proportional to $H(z_{\rm bin})$. Nevertheless, BAO measurements are non-trivial and are subject to their own systematics. On the other hand the redshift drift, despite being observationally challenging, is conceptually extremely simple.  For
example, it does not rely on the calibration of standard candles (as it is the
case of type Ia SNe) or on a standard ruler which originates from the growth of
perturbations (such as the acoustic scale for the CMB) or on effects that depend
on the clustering of matter (except on scales where peculiar accelerations start
to play a significant role). Therefore, the redshift drift will also serve as a useful cross-check for radial BAO. Third, by using distant quasars, it will provide
constraints on the cosmic expansion at redshifts $z>2$, where supernovae and
large scale surveys have difficulties in providing quality data. Finally, it
allows to distinguish between true acceleration, as for dark energy models, and
apparent acceleration, as in void models, as we will discuss in Section
\ref{sec:drift2}.

\subsection{The redshift drift in homogeneous and isotropic universes}

The basic theory behind redshift variation in time is quite simple. One starts
assuming that the metric of the Universe is described by the simplest FRW
metric. The observed redshift of a given source, which emitted its light at a
time $t_s$, is, today (i.e. at time $t_0$),
\begin{equation}
z_s(t_0)=\frac{a(t_0)}{a(t_s)}-1,
\end{equation}
and it becomes, after a time interval $\Delta t_0$ ($\Delta t_s$ for the source)
\begin{equation}
z_s(t_0+\Delta t_0)=\frac{a(t_0+\Delta t_0)}{a(t_s+\Delta t_s)}-1.
\end{equation}
The observed redshift variation of the source is, then,
\begin{equation}
\Delta z_s=\frac{a(t_0+\Delta t_0)}{a(t_s+\Delta t_s)}-\frac{a(t_0)}{a(t_s)},
\end{equation}
which can be re-expressed, after an expansion at first order in $\Delta t/t$,
as:
\begin{equation}
\Delta z_s\simeq\Delta t_0\left(\frac{\dot a(t_0)-\dot a(t_s)}{a(t_s)}\right).
\end{equation}
Clearly, the observable $\Delta z_s$ is directly related to a change in the
expansion rate during the evolution of the Universe, i.e.\ to its acceleration
or deceleration, and it is then a direct probe of the dynamics of the expansion.
It vanishes if the Universe is coasting during a given time interval (i.e.\
neither accelerating nor decelerating). We can rewrite the last expression in
terms of the Hubble parameter $H(z)=\dot a(z)/a(z)$:
\begin{equation}
\label{deltaz}
\Delta z_s=H_0\Delta t\left(1+z_s-\frac{H(z_s)}{H_0}\right),
\end{equation}
where we have dropped the subscript $0$ for simplicity.
The function $H(z)$ contains all the details of the cosmological model under
investigation. Finally, the redshift variation can also be expressed in terms of
an apparent velocity shift of the source, $\Delta v=c\Delta z_s/(1+z_s)$.

\begin{table*}[htdp]
\begin{center}
\begin{tabular}{ l  l }
\\
MODEL & $(H/H_{0})^{2}$\\
\\
\hline \hline
\\
$\Lambda$CDM &
$\frac{\Omega_{k}}{a^{2}}+\frac{\Omega_{m}}{a^{3}}+\Omega_{\Lambda}$\\
\\
\hline
\\
Const. $w$ &
$\frac{\Omega_{k}}{a^{2}}+\frac{\Omega_{m}}{a^{3}}+\frac{\Omega_{DE}}{a^{-3(1+w)
}}$\\
\\
\hline
\\
CPL $w(a)$ &
$\frac{\Omega_{k}}{a^{2}}+\frac{\Omega_{m}}{a^{3}}+\frac{\Omega_{DE}}{e^{3\int
da (1+w(a))/a}}$\\
\\
\hline
\\
INTERACTING  &
$\frac{\Omega_{k}}{a^{2}}+a^{-3}(1-\Omega_{k})(1-\Omega_{DE}(1-a^{\xi}))^{
-3(\frac{w}{\xi})}$\\
\\
\hline
\\
DGP  &
$\frac{\Omega_{k}}{a^{2}}+\Big(\sqrt{\frac{\Omega_{m}}{a^{3}}+\Omega_{r}}+\sqrt{
\Omega_{r}}\Big)^{2}$\\
\\
\hline
\\
CARDASSIAN  &
$\frac{\Omega_{k}}{a^{2}}\Big(1+\frac{(\Omega_{m}^{-q}-1)}{a^{3q(n-1)}}\Big)^{
1/q}$\\
\\
\hline
\\
CHAPLYGIN  &
$\frac{\Omega_{k}}{a^{2}}+(1-\Omega_{k})\Big(A+\frac{(1-A)}{a^{3(1+\gamma)}}
\Big)^{1/(1+\gamma)}$\\
\\
\hline
\\
AFFINE &
$\frac{\Omega_{k}}{a^{2}}+\frac{\tilde{\Omega}_{m}}{a^{3(1+\alpha)}}+\Omega_{
\Lambda}$\\
\\
\hline
\hline
\end{tabular}
\caption{Expansion rate for several dark energy models in the framework of
homogeneous and isotropic cosmologies. The redshift drift evolution
corresponding to these Hubble functions is depicted in
Fig.~\ref{fig:allreddrift}. In the affine model $\tilde{\Omega}_{m}\equiv
(\rho_{0}-\rho_{\Lambda})/\rho_{crit}$ (for more detailed designation of all the
parameters we refer to \cite{2007MNRAS.382.1623B}). }
\label{tab:reddrift}
\end{center}
\end{table*}%
In \cite{2007MNRAS.382.1623B} the authors analysed many currently viable dark
energy cosmological models based on the assumption of homogeneity and isotropy.
Their Hubble expansion rates as a function of the scale factor are collected in
Table~\ref{tab:reddrift}.
These models have often been invoked as candidates to explain the observed
acceleration \cite{2007ApJ...666..716D}, i.e.\ they have not been falsified by
available tests of the background cosmology. For each class  the best fit values
found in \cite{2007ApJ...666..716D} was assumed, and  the parameters were varied
within their 2$\sigma$ uncertainties. Clearly some  models may be preferred with
respect to others based on  the fact that they fit the data well with a smaller
number of parameters.   Nevertheless, it is interesting to explore as many
models as possible, since future observations of the time variation of redshift
could reach a level of accuracy which could allow to better discriminate
competing candidates, and to understand the physical mechanism driving the
expansion.

In order to perform a forecast analysis the predicted accuracy of observations
expected from an experiment like CODEX (see Section \ref{subsec:driftobs}) was adopted. The latter is  entirely
based on the Monte Carlo simulations and discussed by \cite{2005Msngr.122...10P}.
As expected, in its simplest formulation,  the accuracy scales as the square
root of the total number of quasars and is a decreasing function of redshift:
the more the farther the source is from us observer the higher is the number of features
captured in the Lyman-$\alpha$ spectra. Details about the experimental accuracy
and discussions related to it can be found in Section \ref{subsec:driftobs}.

\begin{figure}
\centering
   \includegraphics[width=7.cm]{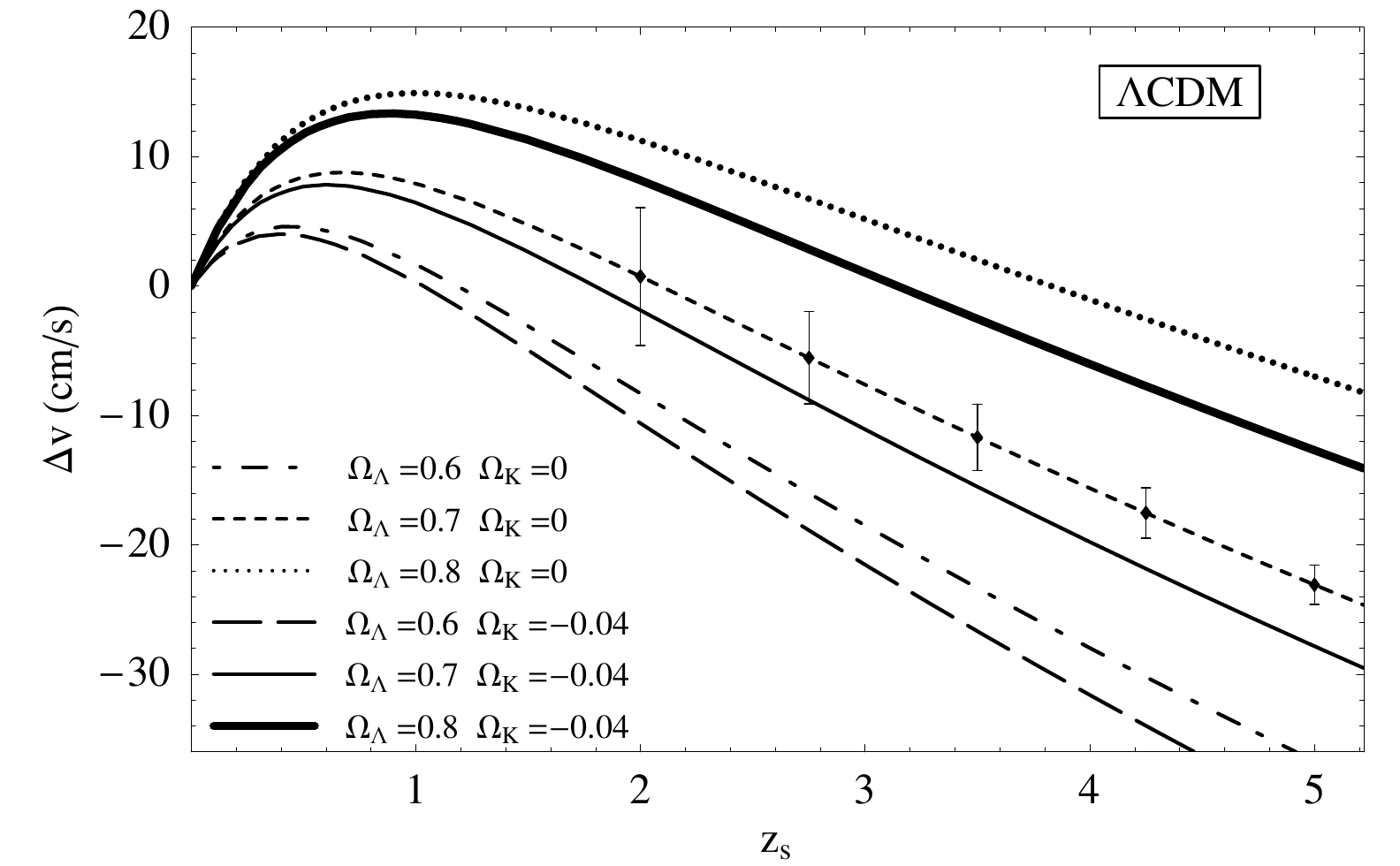}%
   \includegraphics[width=7.cm]{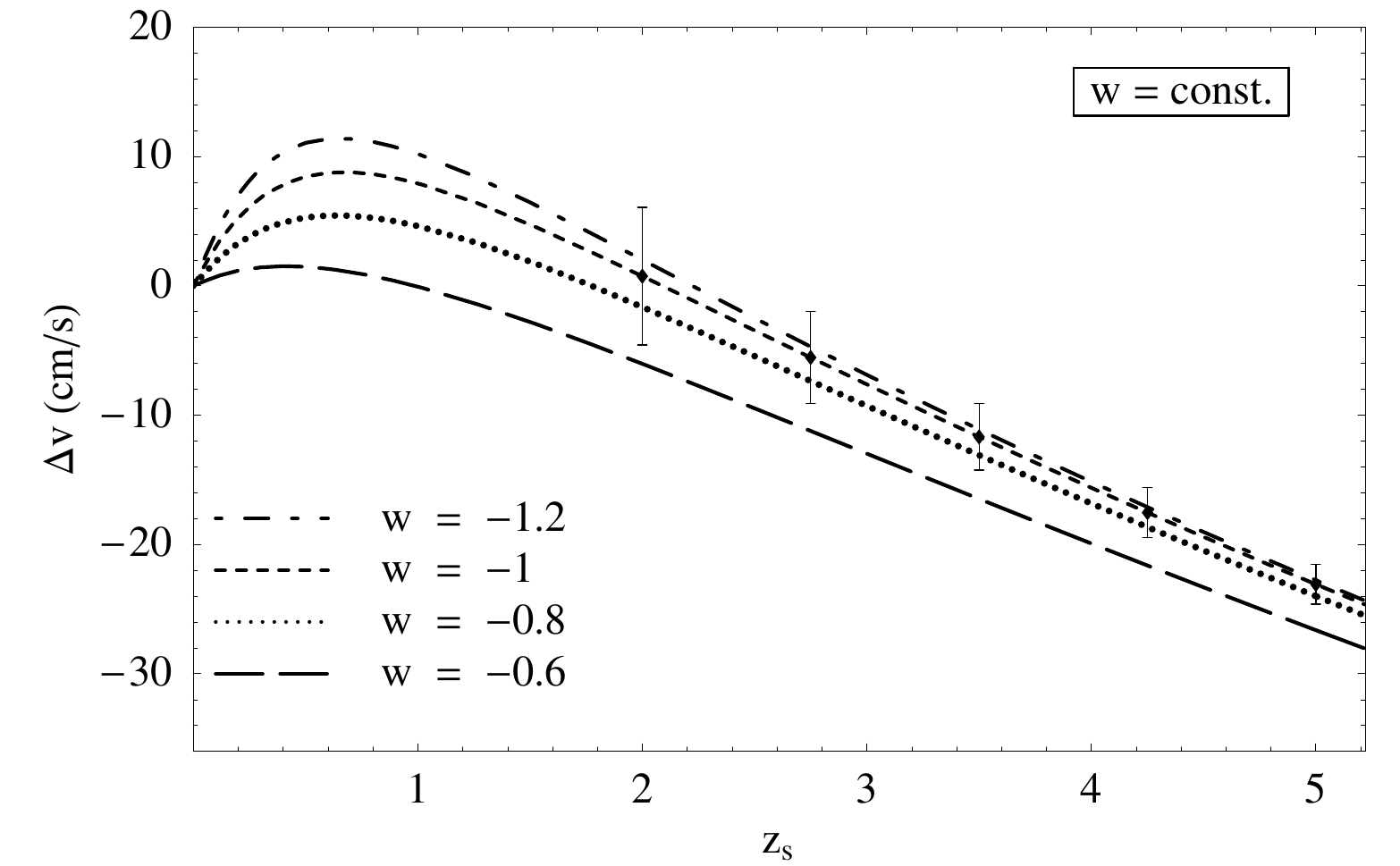}\\
   \includegraphics[width=7.cm]{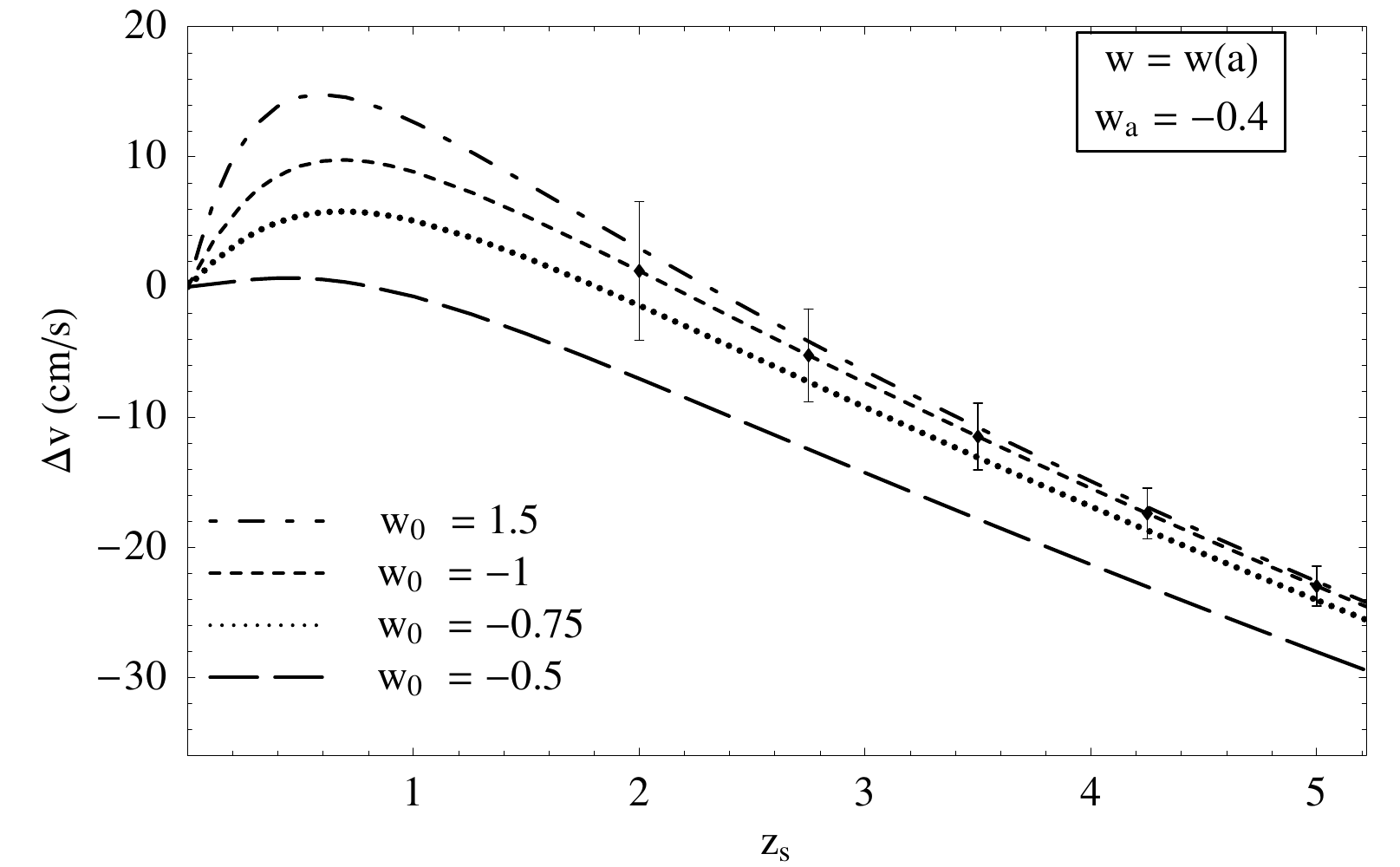}%
   \includegraphics[width=7.cm]{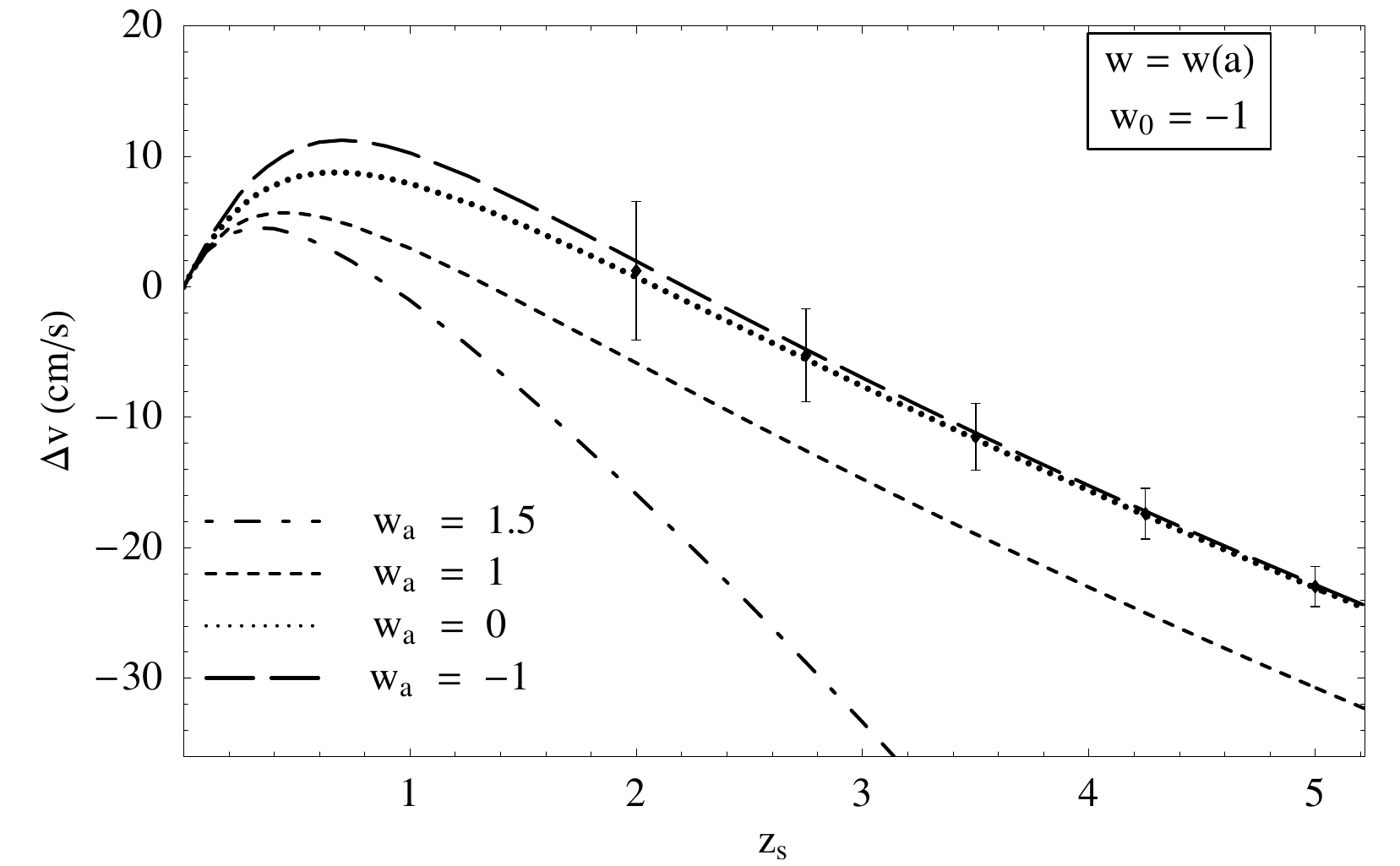}\\
   \includegraphics[width=7.cm]{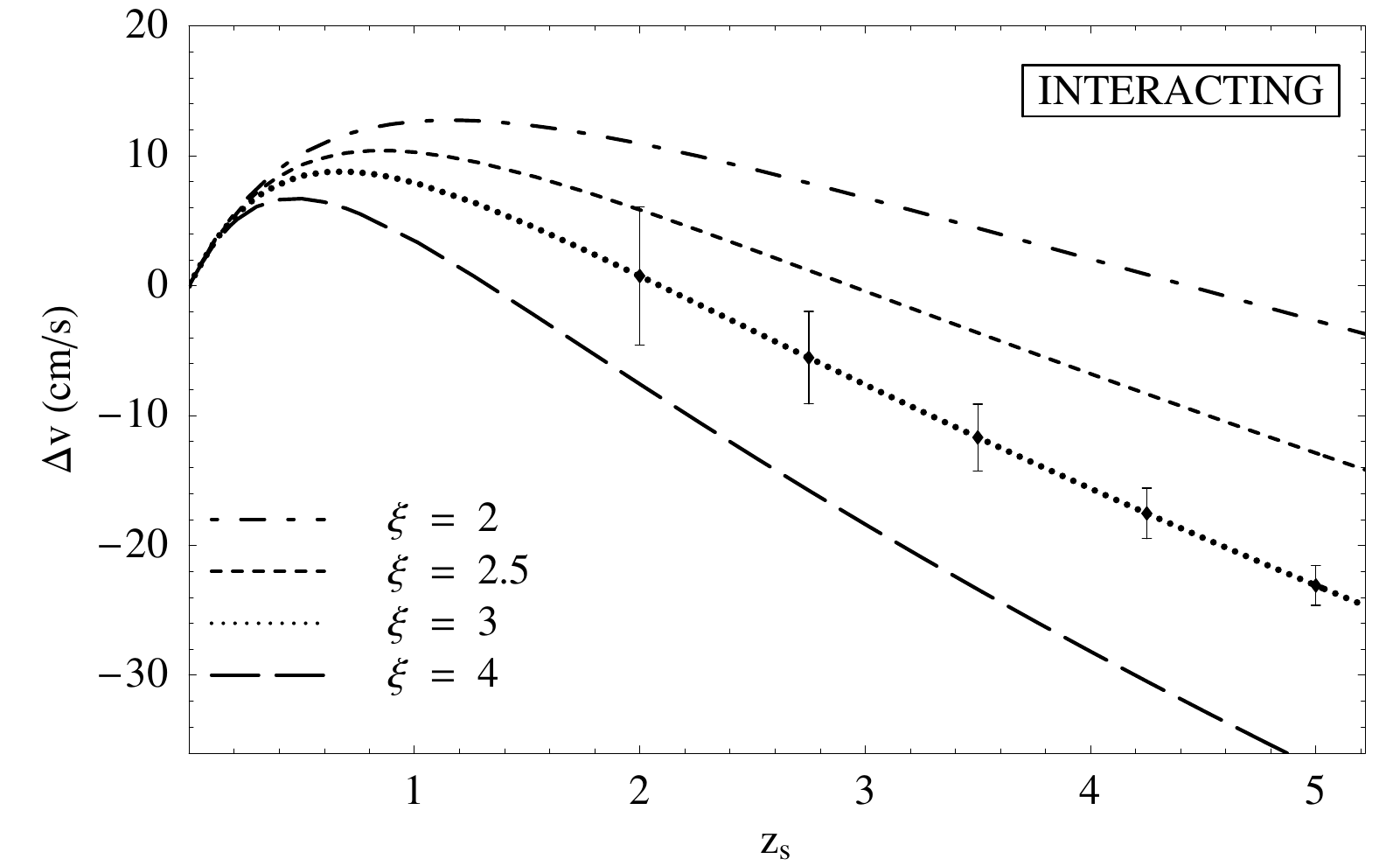}%
   \includegraphics[width=7.cm]{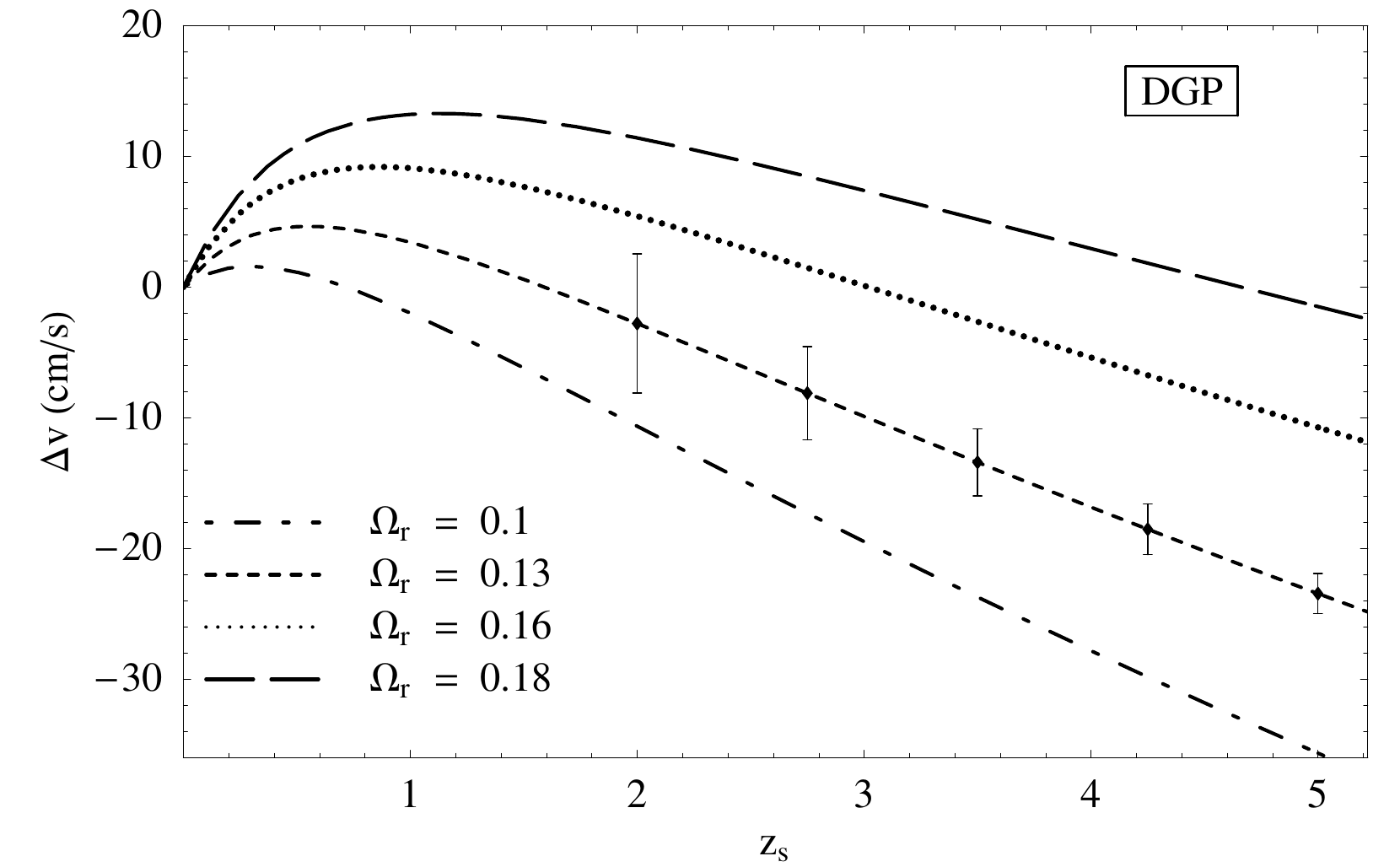}\\
   \includegraphics[width=7.cm]{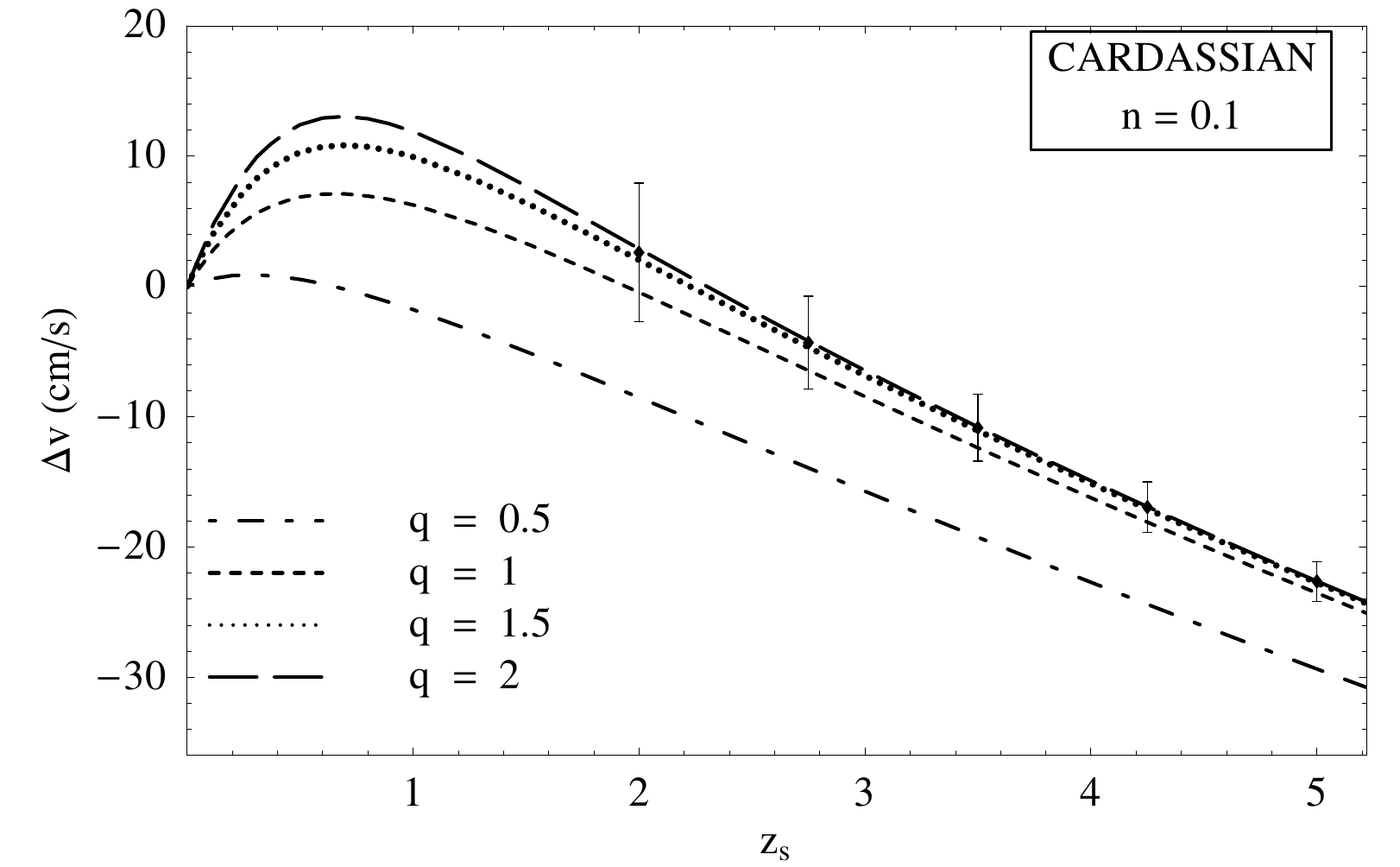}%
   \includegraphics[width=7.cm]{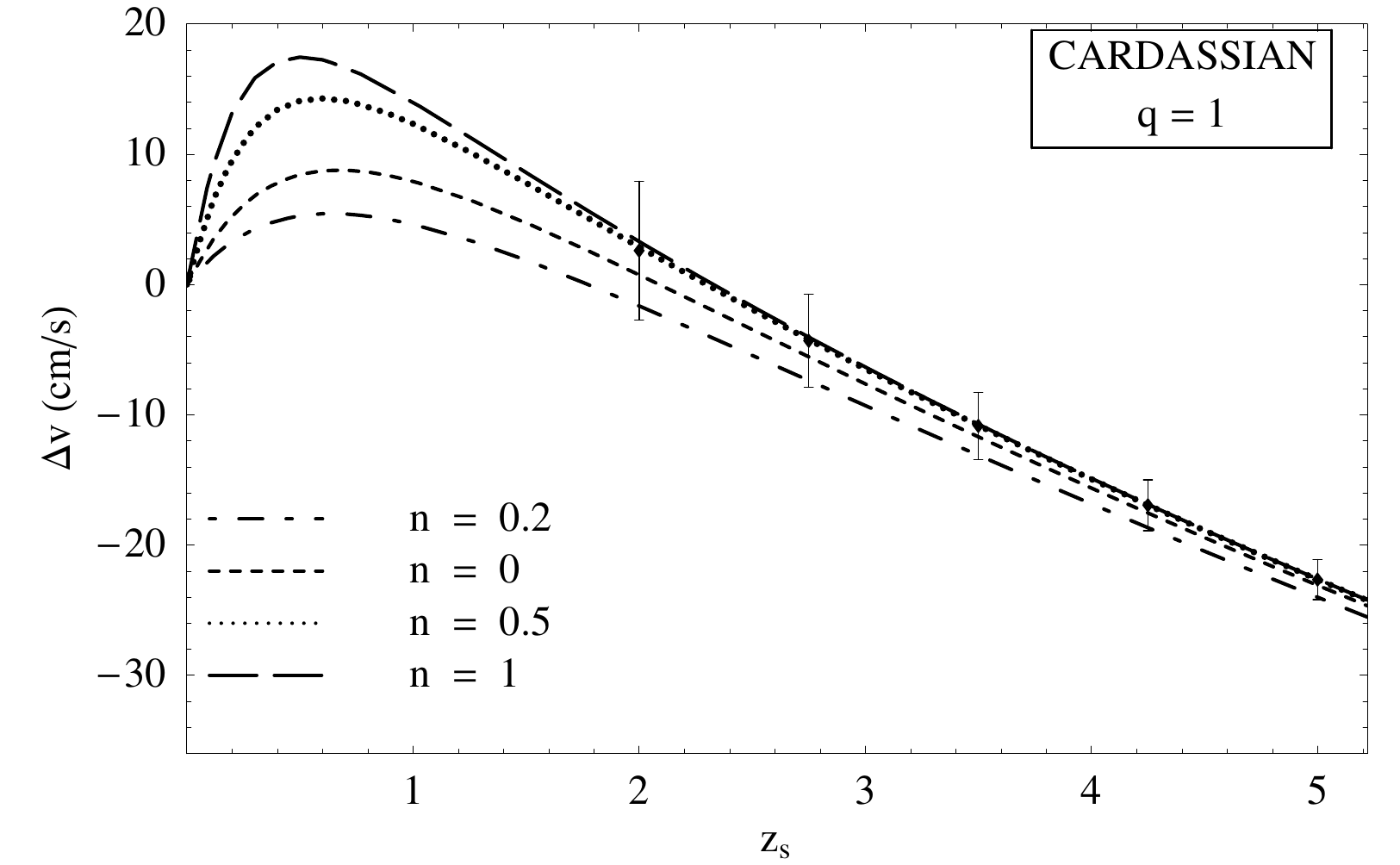}\\
   \includegraphics[width=7.cm]{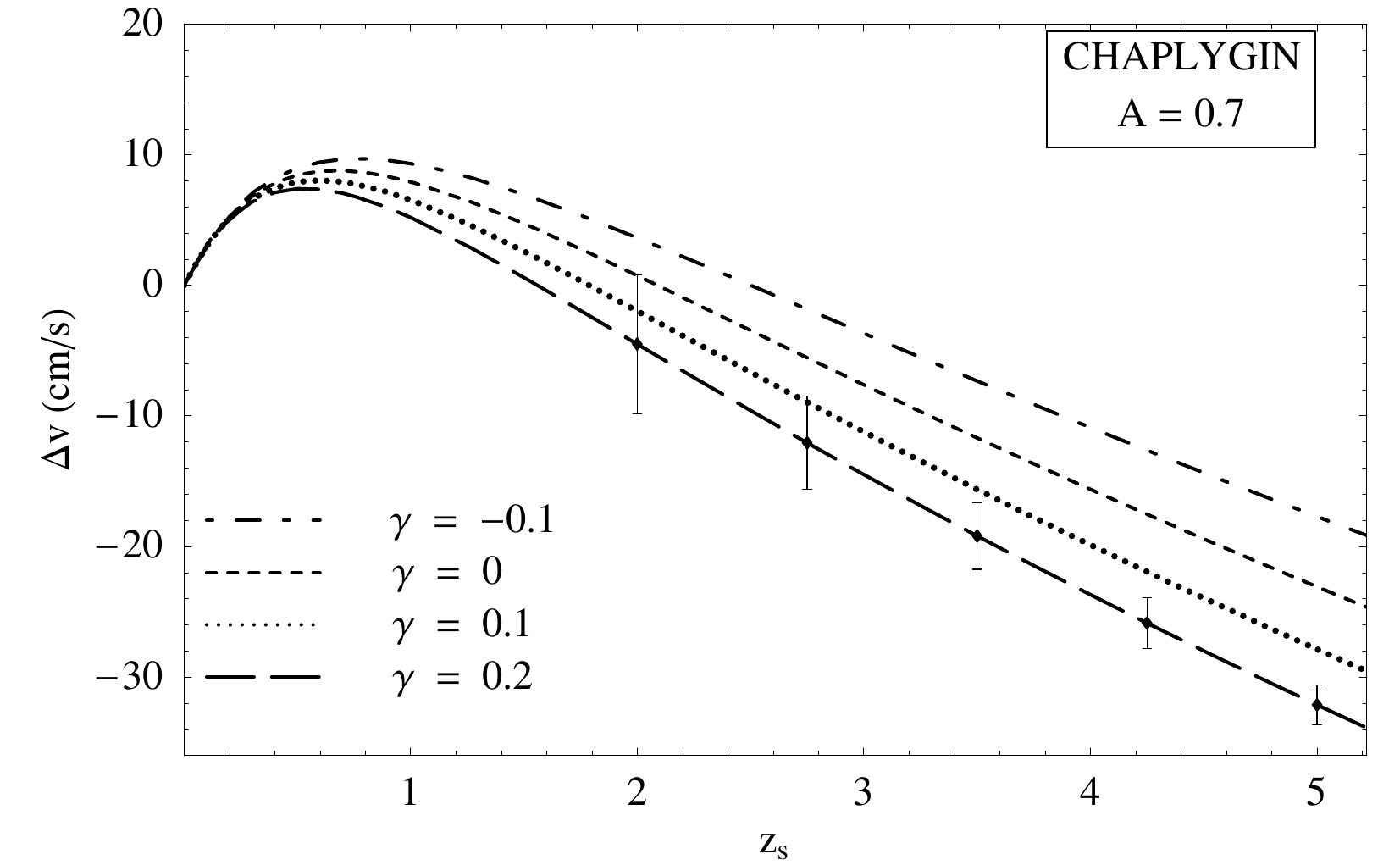}%
   \includegraphics[width=7.cm]{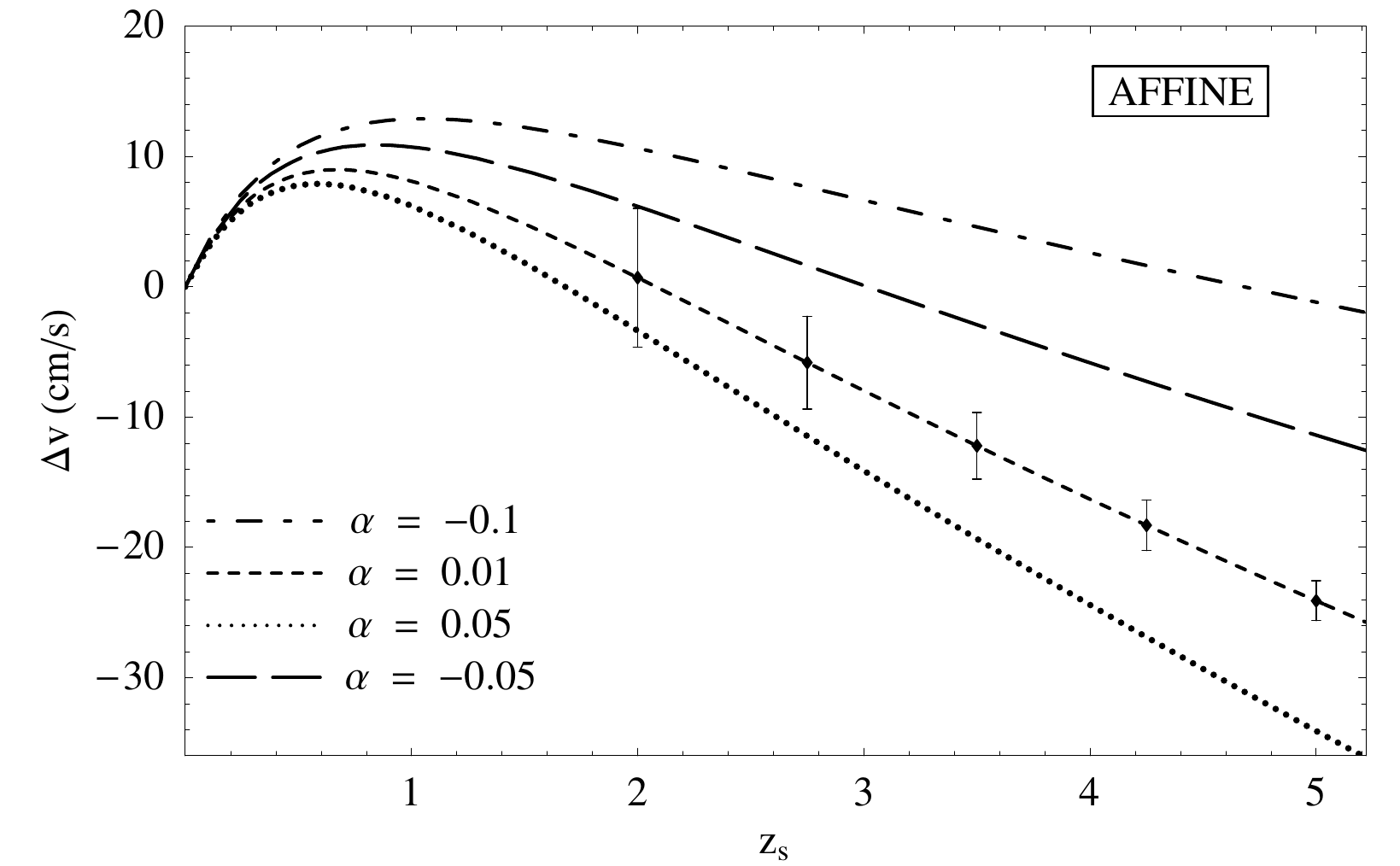}
    \caption{\small The apparent spectroscopic velocity shift over a period $\Delta
t_0=30$ years, for a source at redshift $z_s$, for the models described in
Table~\ref{tab:reddrift}. From Ref.~\cite{2007MNRAS.382.1623B}.}
\label{fig:allreddrift}
\end{figure}

These error bars were used  to construct simulated data and get a feeling of
the possible constraints to viable dark energy models.
Fig~\ref{fig:allreddrift} shows the predicted signal for the models assembled in
Table~\ref{tab:reddrift}. All the predictions  were derived assuming
 $\Delta t=30$ years and a future
dataset containing a total of 40 quasars spectra uniformly distributed over 5
equally spaced redshift bins in the redshift range 2-5 with a S/N=3000, observed
twice over the aforementioned time span. This observational strategy was
properly justified in \cite{2007MNRAS.382.1623B} (details on this assumptions will be encountered again in
Sec.~\ref{subsec:driftobs}).

From Eq.~\ref{deltaz} it is clear that the expected velocity shift signal
increases linearly with $\Delta t$, so that it is straightforward to calculate
the expected signal when a different period of observation is assumed. It is
clear that the observation of velocity shift alone can be affected by the
degeneracies of the parameters that enter $H(z)$, limiting its ability to
constrain cosmological models. The uncertainties on parameter reconstruction
(particularly for non-standard dark energy models with many parameters) can be
rather large unless strong external priors are assumed. When combined with
external inputs, however, the time evolution of redshift could discriminate
among otherwise indistinguishable models.

In Ref.~\cite{2007MNRAS.382.1623B} a Fisher matrix analysis allowed to estimate
the best possible accuracy attainable on the determination of the parameters of
a certain model. Given a set of cosmological parameters $p_i$, $i=1,...,n$, and
the corresponding Fisher matrix $F_{ij}$ (that is easily calculated based on a
theoretical fiducial model and the assumed data errors),
the best possible $1\sigma$ error on $p_i$ is given by $\Delta p_i\equiv
C_{ii}^{1/2}$, where the covariance matrix $C_{ij}$ is simply the inverse of the Fisher
matrix: $C_{ij}=F_{ij}^{-1}$.  The prospect of detecting departures from the
standard  $\Lambda$CDM case could in principle be one of the real assets of
observing the time evolution of redshift, and is thus worthy of closer
investigation. Since the simulated data used in the analysis assume that quasars
are used as a tracer of the redshift evolution, we expect that the more
constrained models will be those that have the largest variability in the
redshift range $2\leq z \leq 5$.
It is at least conceivable that suitable sources at lower redshifts than those
considered in this work could be used to monitor the velocity shift in the
future. This would be extremely valuable, since some non-standard models have a
stronger parameter dependence at low and intermediate redshifts (see Fig.
\ref{fig:allreddrift}), that could be exploited as a discriminating tool. In
\cite{1998ApJ...499L.111L} speculative possibilities of using other sources have
been indicated, like masers in galactic nuclei, extragalactic pulsars or
gravitationally lensed galaxy surveys: this would further extend the lever arm
in redshift space and increase the ability of constraining models. These may
certainly be interesting topics for further studies.

Assuming that the fiducial model has  $\Omega_\Lambda=0.7$ and $\Omega_k=0$ and
that both $\Omega_\Lambda$ and $\Omega_k$ can vary, Ref. \cite{2007MNRAS.382.1623B} found $\Delta\Omega_\Lambda=0.2$
and $\Delta\Omega_k=0.25$ at 1$\sigma$.  Fixing $\Omega_k=0$, the bound on
$\Omega_\Lambda$ becomes 0.007 (at 1$\sigma$). If dark energy is modelled by a
constant equation of state (with a fiducial value $w=-1$) and the flatness
constraint is imposed we find a looser bound on the dark energy density,
$\Delta\Omega_{de}=0.016$, and quite a large error on the equation of state,
$\Delta w=0.58$. This clearly shows that different assumptions on the knowledge
of any parameter has an influence on all the others. The parameter $\Omega_k$
can be much better constrained using external datasets, such as the CMB
anisotropy.

The DGP model is the one for which the tightest constraints have been obtained:
$\Delta\Omega_r=0.0027$ at 1$\sigma$, assuming $\Omega_r=0.13$ as a fiducial
value. This is not only due to the strong dependence of the velocity shift on
$\Omega_r$ (see Fig.~\ref{fig:allreddrift}), but also to the simplicity of the
model, which depends on only one parameter (in this respect, this is the
simplest model, together with the standard flat $\Lambda$CDM). In general, it is
to be expected that models with less parameters perform better.

For what concerns the Chaplygin model, when both $\tilde{A}$ and $\gamma$ vary
freely, no interesting constraint can be obtained observing the velocity shift
with the assumed QSO data: we find $\Delta{\tilde{A}}=0.42$ and
$\Delta\gamma=1.4$ (for the fiducial values $\tilde{A}=0.7$ and $\gamma=0.2$).
Fixing $\tilde{A}$, on the contrary, results in a very tight bound on $\gamma$:
$\Delta\gamma=0.008$.

The interacting dark energy and the affine equation of state models show a large
variability in the redshift range we are exploring: this is to be expected,
since in both models the matter-like component departs from the usual $a^{-3}$
scaling, giving a distinctive signature when one looks at higher and higher
redshifts. If $\Omega_\Lambda$ is known, the affine parameter $\alpha$ can be
reconstructed with an error $\Delta\alpha=0.005$. If, in addition, $w=-1$ we
find $\Delta\xi=0.06$ for the interacting dark energy model (for a fiducial
value $\xi=3$).

The other models do not seem to have very interesting signatures to be
exploited, at least in the redshift range considered in our analysis.

\begin{figure}[ht!]
    \includegraphics[width=12cm]{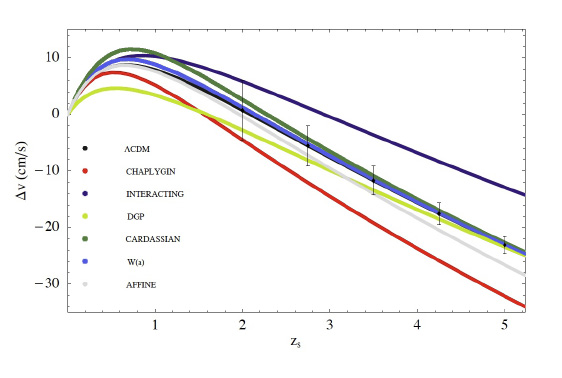} 
    \caption{\small The predicted velocity shift for the models presented in
Table~\ref{tab:reddrift}, compared to simulated data as expected from the CODEX
experiment. The simulated data points and error bars are estimated from Eq.
\eqref{eq:sigma-v}, assuming as a fiducial model the standard $\Lambda$CDM model.
The other curves are obtained assuming, for each non-standard dark energy model,
the parameters which best fit current cosmological observations. From
Ref.~\cite{2007MNRAS.382.1623B}.}
    \label{fig:single}
\end{figure}

Fig.~\ref{fig:single} shows the comparison among the predicted velocity shifts
for all the models described earlier, assuming parameter values that are a good
fit to current cosmological observations (including the peak position of the CMB
anisotropy spectrum, the SNe Ia luminosity distance, and the baryon acoustic
oscillations in the matter power spectrum). In other words, the models shown in
Fig.~\ref{fig:single} cannot be easily discriminated using current cosmological
tests of the background expansion. If we assume that the $\Lambda$CDM model is
the correct one, and simulate the corresponding data points for the velocity
shift, using a $\chi^2$ test we can quantify how well we can exclude the
competing models based on their expected signal. As it is clear from
Fig.~\ref{fig:single}, some models can be excluded with a high confidence level.
In particular, the Chaplygin gas model and the interacting dark energy model
would be excluded at more than $99\%$ confidence level, and that the affine
model would be out of the 1 $\sigma$ region.

All the results above were obtained assuming an equal number (8) of quasars for each of 5
redshift bins centered at $z=\{2,\,2.75,\,3.5,\,4.25,\,5\}$, all of the same redshift width of $0.75$ (such a uniform distribution was also assumed in the simulations performed by~\cite{2005Msngr.122...10P}). A good estimate of the error bars for such binning can be achieved using the brightest quasars from the SDSS DR7 as a benchmark~\cite{2010PhRvD..81d3522Q}. This is explained in more details in Appendix~\ref{app:zdot-errorbars}. Moreover, one should notice that the effective S/N on the measured data points decrease as $\Delta t^{3/2}$, as the signal is linear in time and the noise scales as $\Delta t^{-1/2}$, and can become significantly higher over only a few decades of observations.

\subsection{The redshift drift in less symmetric space-times}
\label{sec:drift2}
FRW universes are based on a maximally symmetric space-time derived from the
assumption of the cosmological principle: the observable universe is homogeneous
and isotropic (in the weaker and matter-of-fact version at least on large
scales). The cosmological principle itself is originated by a combination of the
Corpernican principle, stating that we are not located at a favoured position in
space, and the observed isotropy \cite{1985PhR...124..315E}. The inferred
existence of a late-time accelerated expansion (and consequently the hypothesis
of a dark energy component) has its main foundation on the  conjectured flatness
of space-time justified by CMB observations together with the Copernican
principle.
The verification of the latter has recently attracted attention to test whether
the relaxation of its statement may help explain recent time expansion without
invoking the existence  of a new exotic component or modified gravity theory.

In this framework, less symmetric space-times have been considered; in
particular, since one should expect to observe the redshift drift in any
expanding space-time, in \cite{Uzan:2008} the authors have derived an
expression for this observable in spherically symmetric universes, where the
observer is located at the centre.

One type of possibly useful coordinates that allow for a general treatment are
the observable coordinates $\{w,y,\theta,\varphi\}$, where $w$ marks the past
light-cones of events along the worldline $\mathcal{C}$ of the observer. The
metric is
\begin{equation}
\label{eq:obscoord}
\dd s^{2}=-A^{2}(w,y)\dd w^{2}+2A(w,y)B(w,y)\dd y\dd w+C^{2}(w,y)\dd \Omega^{2},
\end{equation}
spherically symmetric around the world-line $\mathcal{C}$ defined by $y=0$. The
redshift is given by
\begin{equation}
\label{eq:driftgen}
1+z=\frac{(u_{\alpha}k_{\alpha})_{\rm emission}}{(u_{\alpha}k_{\alpha})_{\rm observer}}=\frac{A(w_{0},0)}{
A(w_{0},y)},
\end{equation}
for a given value $w_{0}$. Here the matter velocity and photon wave-vector are
$u^{\alpha}=A^{-1}\delta^{\alpha}_{w}$ and $k^{\alpha}=(AB)^{-1}\delta^{\alpha}_{y}$, respectively.
The isotropic expansion rate around an observer located at the centre is defined
as $H=\nabla_{\alpha}u^{\alpha}/3$ and eventually turns out to be:
\begin{equation}
H(w,y)=\frac{1}{3A}\Big(\frac{\partial_{w}B(w,y)}{B(w,y)}+2\frac{\partial_{w}C(w
,y)}{C(w,y)}\Big).
\end{equation}
In the case of a dust-dominated universe the covariant derivative of $u_{\alpha}$
takes the form $\nabla_{\alpha}u_{\beta}=H(g_{\alpha\beta}+u_{\alpha}u_{\beta})+\sigma_{\alpha\beta}$, where
$\sigma_{\alpha\beta}$ is the traceless and symmetric shear satisfying
$u^{\alpha}\sigma_{\alpha \beta}=0$.

The expression for the redshift drift is then derived from Eq.~\eqref{eq:driftgen}
\begin{equation}
\frac{\delta z(w_{0},y)}{\delta
w}=(1+z)\Big(\frac{\partial_{w}A(w_{0},0)}{A(w_{0},0)}-\frac{\partial_{w}A(w_{0}
,y)}{A(w_{0},y)}\Big).
\end{equation}
Choosing $A(w_{0},0)=1$ and $y$ such that
$\partial_{w}B(w_{0},y)=\partial_{w}A(w_{0},y)$, it follows that
\begin{equation}
\frac{\delta z(w_{0},y)}{\delta
w}=(1+z)H_{0}-H(w_{0},y)-\frac{1}{\sqrt{3}}\sigma(w_{0},y),
\end{equation}
where $\sigma=\sigma^{\alpha\beta}\sigma_{\alpha\beta}/2$ is the scalar shear. Indeed this is the
general form for the redshift drift measured by an observer located at the
centre of a spherically symmetric universe. In the following, we will present
numerical and analytical calculations of the redshift drift in some LTB models also including off-centre observers.

\subsubsection{The redshift drift in Lemaître-Tolman-Bondi void models}
\label{sec:driftLTB}
The most general metric describing a spherically symmetric and inhomogeneous
universe in comoving coordinates is the LTB metric. In particular, the latter is
well designed to describe universes where the observer is located near the
centre of a large void embedded in an Einstein-de Sitter cosmology (see
e.g.~\cite{2008PhRvL.101m1302C,2010MNRAS.405.2231F,Biswas:2010xm,Marra:2011ct}).  Models with void sizes
which, although huge by any means, are ``small'' enough ($z \sim 0.3 - 0.4$) not
to be ruled out due to distortions of the CMB blackbody radiation
spectrum~\cite{Caldwell:2008} are capable of fitting the observed SNIa Hubble diagram and the CMB first peak position and compatible with the COBE results of
the CMB dipole anisotropy, as long as the observer is not too far  from the center~\cite{Alnes:2006a,2010PhRvD..81d3522Q} (although recent works~\cite{Zhang:2010fa,Moss:2011ze,Zibin:2011ma} claim that the blackbody spectrum of the CMB, in combination with the kinetic Sunyaev-Zel'dovich effect due to free electrons effectively rule out these models).
The off-center displacement is limited to $\sim200$ Mpc by supernovae, whereas the CMB dipole limit it to somewhere in between 30 and 60~Mpc depending on some \emph{a priori} assumptions~\cite{Alnes:2006a,2010PhRvD..81d3522Q,Blomqvist:2009}.

In principle, the redshift drift in such models needs an exact treatment where
the full relativistic propagation of light rays is taken into account.
We will begin by introducing the Einstein equations in such a metric and we will
then present the light geodesic equations.

The LTB metric can be written as (primes and dots refer to partial space and
time derivatives, respectively):
\begin{equation}\label{eq:LTB}
    {\textrm{d}}s^{2} = -{\textrm{d}}t^{2} +
\frac{\left[R'(t,r)\right]^{2}}{1+\beta(r)} {\textrm{d}}r^{2} +
R^{2}(t,r){\textrm{d}}\Omega^{2},
\end{equation}
where $\beta(r)$ can be loosely thought of as a position dependent spatial curvature
term. Two distinct Hubble parameters corresponding to the radial and
perpendicular directions of expansion are defined as
\begin{align}
    H_{||} = \dot{R'}/R' \label{eq:def-Hpar} \,,\\
    H_{\perp} = \dot{R}/R \label{eq:def-Hperp} \,.
\end{align}
Note that in a FRW metric $R=ra(t)$ and $H_{||}=H_{\perp}$. This class of models
exhibits implicit analytic solutions of the Einstein equations in the case of a
matter-dominated universe, to wit (in terms of a parameter $\eta$)
\begin{align}
    R=\, & (\cosh\eta-1)\frac{\alpha}{2\beta}+R_{{\rm lss}}\left[\cosh\eta +
\sqrt{\frac{\alpha+\beta R_{\rm lss}}{\beta R_{\rm lss}}} \sinh\eta\right],
\label{eq:sol-R}\\
    \sqrt{\beta}t=\, & (\sinh\eta-\eta)\;\frac{\alpha}{2\beta}\,+ R_{{\rm
lss}}\left[\sinh\eta + \sqrt{\frac{\alpha+\beta R_{\rm lss}}{\beta R_{\rm lss}}}
\left(\cosh\eta-1\right)\right], \label{eq:sol-beta-t}
\end{align}
where $\alpha$, $\beta$ and $R_{{\rm lss}}$ are all functions of $r$. In fact,
$R_{{\rm lss}}(r)$ stands for $R(0,r)$ and we will choose $t=0$ to correspond to
the time of last scattering, while $\alpha(r)$ is an arbitrary function and
$\beta(r)$ is assumed to be positive.  In performing the calculations, it is much simpler to set $R_{\rm lss} = 0$, as $R_{\rm lss} \neq 0$ introduces arduous numerical problems: one only has to note that they will not be valid for $z \gtrsim 10$, as has to be the case anyway.

The full geodesic equations are written in Appendix~\ref{app:ltb-geodesics},
where an algorithm is provided to compute both the redshift drift and the cosmic parallax (see next Section). The former
effect was first calculated in~\cite{2009PhRvL.102o1302Q} (see
Sec.~\ref{sec:parallax})) for two distinct specific LTB models, while the latter
was thoroughly investigated in~\cite{2010PhRvD..81d3522Q} for the same models as well
as for a third one.

\begin{figure}[t!]
    \includegraphics[width=9cm]{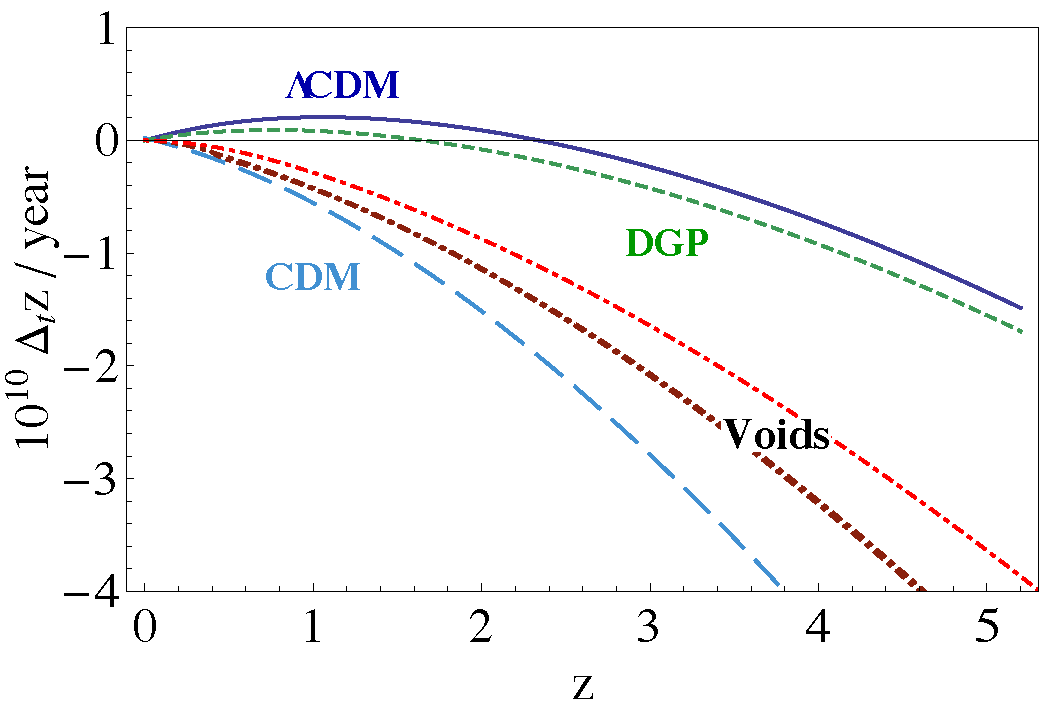}
    \caption{\small The annual redshift drift for different models assuming an observer
at the center. The upper, blue solid lines represent the $\Lambda$CDM model.
The green, dashed line corresponds to a self-accelerating DGP model
with $\Omega_{r_{c}}=0.13$. The dot-dashed lines stand for the 3
void models considered here: the dark brown (indistinguishable) lines
are for Models I and II, while the red line just above correspond
to the cGBH model. The line at the bottom of the plot corresponds to an universe
with only matter in a FRW metric (the CDM model). From
Ref.~\cite{2010PhRvD..81d3522Q}.}
    \label{fig:zdot-per-year}
\end{figure}

Two of these three models (Ref.~\cite{Alnes:2006a,Alnes:2006b}) are
characterized by a smooth transition between an inner void and an outer region
with higher matter density and described by the functions:
\begin{align}
    \alpha(r) & =\left(H_{\perp,0}^{{\rm
out}}\right)^{2}r^{3}\left[1-\frac{\Delta\alpha}{2} \left(1-\tanh\frac{r-r_{{\rm
vo}}}{2\Delta r}\right)\right]\!,\\
    \beta(r) & =\left(H_{\perp,0}^{{\rm out}}\right)^{2}r^{2}
\,\frac{\Delta\alpha}{2} \left(1-\tanh\frac{r-r_{{\rm vo}}}{2\Delta r}\right)\!,
\end{align}
where $\Delta\alpha$, $r_{{\rm vo}}$ and $\Delta r$ are three free parameters and $H_{\perp,0}^{{\rm out}}$ is the Hubble constant at the outer region, set at 51~km~s$^{-1}$~$\mbox{Mpc}^{-1}$. We will dub the two models ``Model I'' and ``Model II'', and define them by the sets $\{\Delta\alpha=0.9,\, r_{\rm vo}=1.46\mbox{ Gpc},\Delta r=0.4\, r_{\rm vo}\}$ and $\{\Delta\alpha=0.78,r_{\rm vo}=1.83$~Gpc$,\Delta
r=0.03\, r_{{\rm vo}}\}$, respectively. These values of $r_{\rm vo}$ correspond, in physical distances (that is, the distance obtained by multiplying the comoving separation by the scale factor), to void sizes of 1.34 and 1.68~Gpc, respectively. The third model under consideration is  the so-called ``constrained'' model proposed in~\cite{Bellido:2008} referred to as the ``cGBH'' model. For this model, the parameters were chosen  to  maximize the likelihood as obtained in~\cite{Bellido:2008}; this model can be written  in terms of $\alpha$ and $\beta$ (\cite{2010PhRvD..81d3522Q}). The main difference between the three models is that Model~II features a much sharper transition from the void and that the cGBH model is almost twice as large. Although in principle the redshift drift will depend on the source position in the sky for off-center observers, it
was shown in~\cite{2010PhRvD..81d3522Q} that, unless the LTB models violate the CMB dipole measurements, the differences across the sky are less than 5\%.

\begin{figure}[t!]
    \includegraphics[width=7.5cm]{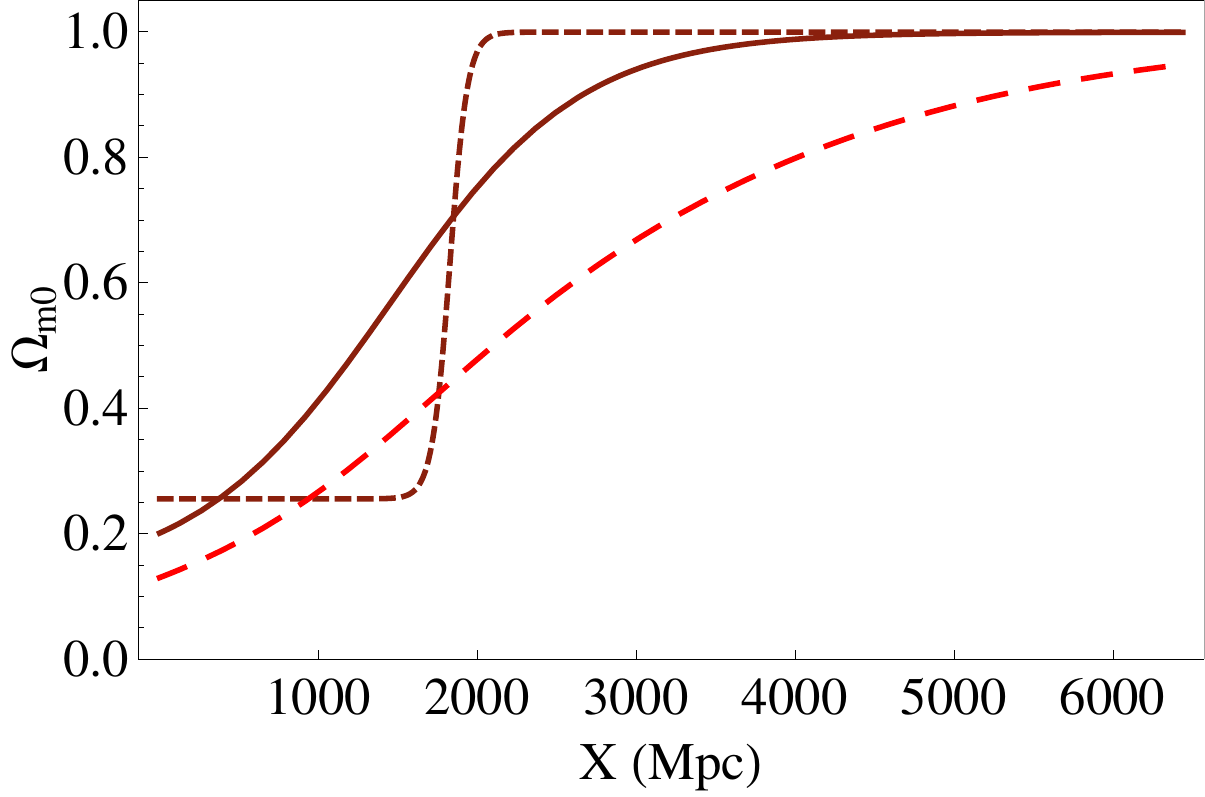} \qquad
    \includegraphics[width=7.5cm]{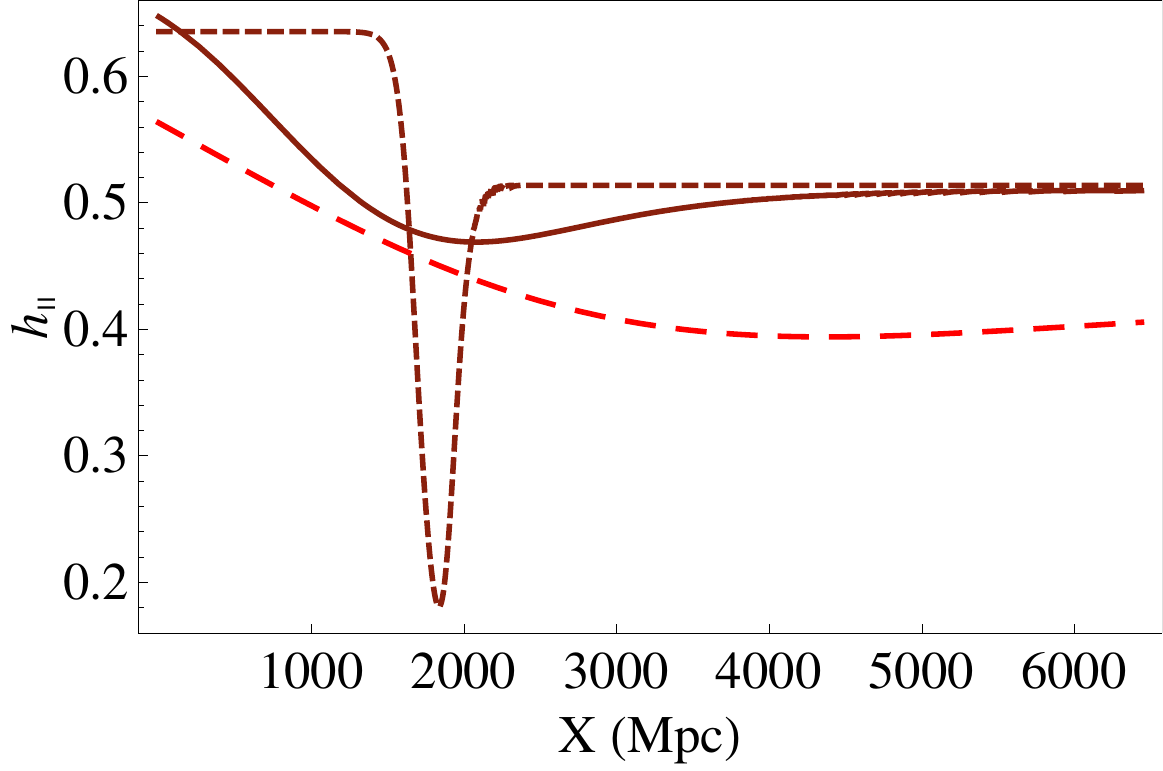}
    \caption{$\Omega_{m0}\,$ and $\,h_{||} \equiv H_{||}/H_0\,$ for Model I (solid), Model II (dashed curve) and the cGBH model (red, long-dashed) as a function of the physical (as opposed to comoving) distance $X$. Note that the definition for $\Omega_{m0}$ we use differ from  the one in~\cite{Alnes:2006a,Alnes:2006b} and we do not get their
    characteristic over-density bump (or shell) surrounding the void.}
    \label{fig:ltb-models}
\end{figure}

The 3 LTB models we shall investigate in this and in the following section are depicted in Fig.~\ref{fig:ltb-models}.

Fig.~\ref{fig:zdot-per-year} illustrates the redshift drift as a function of
redshift for $\Lambda$CDM, the DGP model, the old matter dominated model (CDM)
and the 3 different void models. As could be expected, the void models predict a
curve which is in between CDM and  $\Lambda$CDM.

As can be seen from Fig.~\ref{fig:single}, in many dark energy models the
redshift drift is positive at small redshift, but becomes negative for $z\gtrsim
2$. On the other hand, a giant void mimicking dark energy  produces a very
distinct $z$ dependance of this drift, and in fact one has that $\dd z/\dd t$
is always negative (see Fig.~\ref{fig:zdot-per-year}). This property and its
potential as discriminator between LTB voids and $\Lambda$CDM was first pointed
out in~\cite{2008PThPh.120..937Y}. Moreover, it was recently proven in \cite{2010arXiv1010.0091Y} that under certain general conditions the property $\dd z/\dd t<0$ always holds in void models.

\begin{figure}[t!]
    \includegraphics[width=8.3cm]{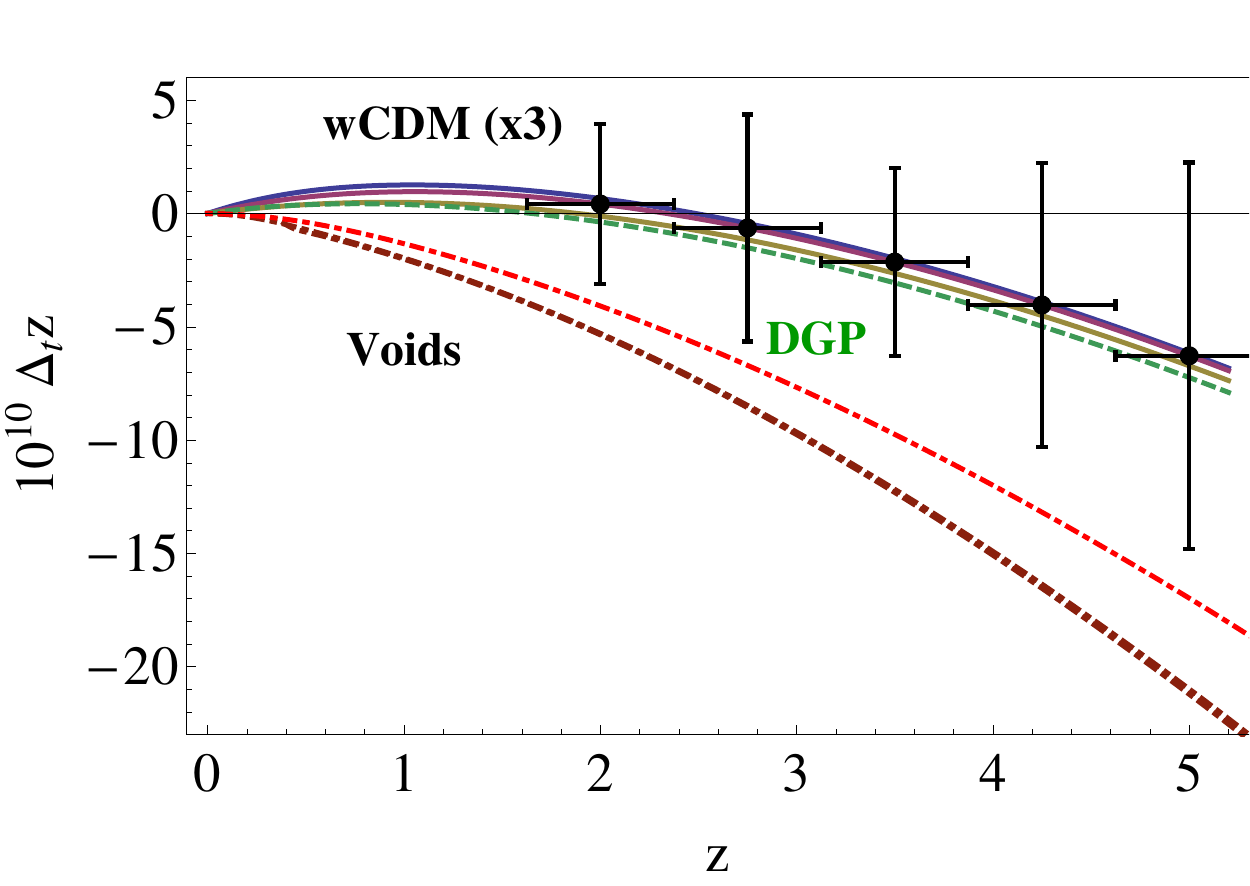} \qquad
    \includegraphics[width=8.3cm]{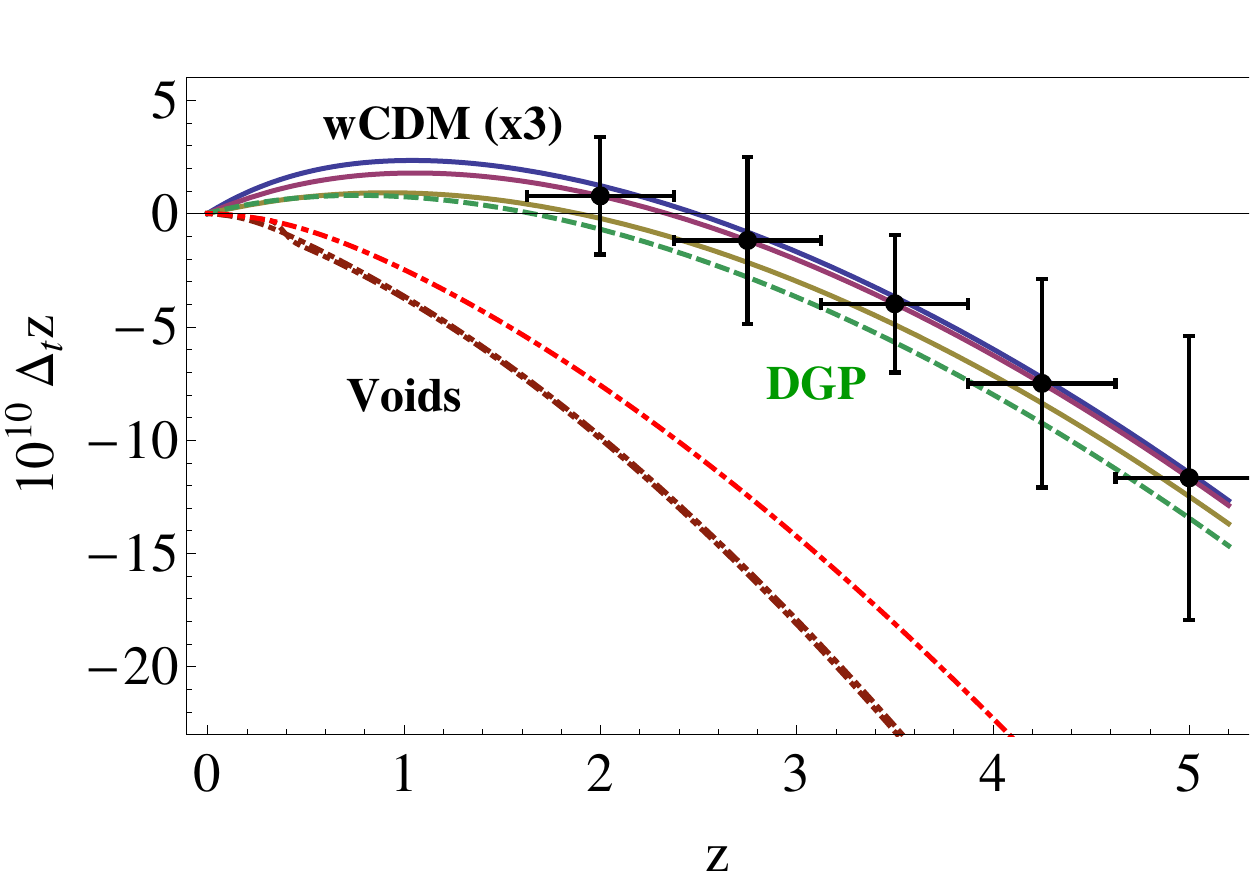}
    \caption{\small Redshift drift for different dark energy models for a
    total mission duration of $7$ (left) and 13 (right)  years and CODEX forecast error bars. In each plot, the upper 3, solid lines represent const. wCDM models for $w=-1.25$ (uppermost), $w=-1$ (second) and $w=-0.75$ (third uppermost). The green, dashed line corresponds to a self-accelerating DGP model with $\Omega_{r_{c}}=0.13$. The three dot-dashed lines at the bottom of the plot stand for the 3 void models considered here: the dark brown  (indistinguishable) lines are for Models I and II, while the red line   just above correspond to the cGBH model. Similar to figures from  Ref.~\cite{2010PhRvD..81d3522Q}, except for the inclusion of horizontal error bars.}
    \label{fig:zdot-ltb-codex}
\end{figure}

Fig.~\ref{fig:zdot-ltb-codex} depicts the redshift drift for both 7 and 13 years time span (\cite{2010PhRvD..81d3522Q}). Also plotted are the forecasted error bars
obtainable by CODEX at the E-ELT. The correspondent possible achievable significance of detection is listed in Table~\ref{tab:zdot-codex}. As can be seen from this table, void models can be ruled out with a $\sim 5\sigma$ confidence-level within a decade of observations.

\begin{table}[t]
\begin{tabular}{|c|c|c|c|}
    \hline
    \parbox[c][0.6cm]{0.5cm}{}\hspace{.5cm} \textbf{Model} \hspace{.5cm}  &  \textbf{ 7 years }   &  \textbf{ 10 years }   &  \textbf{ 13 years }  \tabularnewline
    \hline
    Models I / II  & \parbox[c][0.7cm]{0.5cm}{} $3.0\sigma$  & $6.2\sigma$  & $10\sigma$ \tabularnewline
    \hline
    cGBH Model  & \parbox[c][0.7cm]{0.5cm}{} 2.2$\sigma$  & 4.3$\sigma$  & $7.1\sigma$ \tabularnewline
    \hline
\end{tabular}
\caption{Estimated achievable confidence levels by the CODEX mission in 7,
10 and 13 years.}
\label{tab:zdot-codex}
\end{table}

The  predicted accuracy assumed in~\cite{2010PhRvD..81d3522Q} although adopting the same expression as in Eq.~\eqref{eq:sigma-v} (except for the redshift power law exponent, which is not substantially relevant), leads to an error bar that is somewhat higher at higher
redshift. This is the outcome of two factors: quasars which are at high redshift are usually dimmer, so their S/N  is lower, and that may win over the power law factor. Moreover, the time spent observing each quasar is assumed to be the same, and this accounts for larger error bars at high redshift due to the lower apparent magnitudes of the corresponding quasars. On the other hand, in~\cite{2007MNRAS.382.1623B} the relative integration time for these sources was assumed to be increased in order to achieve the same average signal-to-noise ratio at all redshift bins. The observational strategy will be better analyzed in Sec.~\ref{sec:observations}. In order to estimate the error bars in the left panel of Fig.~\ref{fig:zdot-ltb-codex}, Ref.~\cite{2010PhRvD..81d3522Q} made use of available SDSS quasars, selecting the brightest quasars in each redshift bin using the appropriate band for such bin. These details are covered in the Appendix~\ref{app:zdot-errorbars}. In the right panel of the same figure a similar plot is shown from Ref.~\cite{2010JCAP...06..017D}, where the authors also included redshift errors.

\section{Cosmic Parallax}
\label{sec:parallax}

In every anisotropic expansion the angular separation between any two sources
varies in time (except for particular sources aligned on symmetry axes), thereby
inducing a \emph{cosmic parallax}  effect~\cite{2009PhRvL.102o1302Q}. This is
totally analogous to the classical stellar parallax, except here the parallax is
induced by a differential cosmic expansion rather than by the observer's own
movement. In other words, cosmic parallax will be present whenever one has
\emph{shear} in cosmology. An anisotropic expansion can either be experienced by
an {\rm off-centre} observer in an inhomogeneous and isotropic universe or a
centered observer embedded in an intrinsically anisotropic expansion. LTB void
models belong to the former class, and cosmic parallax was extensively analysed
in \cite{2009PhRvL.102o1302Q,2010PhRvD..81d3522Q,2009arXiv0905.3727F}, while
Bianchi I models are specific example of the latter class and were studied
in~\cite{2009PhRvD..80f3527Q}.

Measuring the cosmic parallax is therefore an independent test of late-time
anisotropies in the Universe. More rigorously, it is a test of late-time
anisotropic expansion, or shear. Competing tests include direct reconstruction
of the Hubble diagrams using Supernovae~\cite{Alnes:2006b,Blomqvist:2009} and
the CMB dipole~\cite{2010arXiv1009.0273F} and quadrupole~\cite{Appleby:2009}.

\subsection{Cosmic parallax in Lemaitre-Tolman-Bondi void models}
\label{sec:parallaxLTB}

As discussed in Section~\ref{sec:driftLTB}, LTB universes have two different
Hubble parameters and therefore appear anisotropic to any observer except the
central one. Therefore any such observer will see a sky affected by cosmic
parallax. In fact, the amount of anisotropy is at first order directly
proportional to this off-center distance.  In void models, the cosmic parallax
is also an independent test of the Copernican principle but without the
degeneracy with our peculiar velocity that afflicts the constraints from the
cosmic microwave background
dipole~\cite{Alnes:2006a,Caldwell:2008,2010arXiv1009.0273F}. This fact was used
in~\cite{2009PhRvL.102o1302Q,2010PhRvD..81d3522Q} to evaluate the feasibility of
detecting the cosmic parallax in dark energy motivated void models.

Due to their large distances and point-like properties, quasars are the obvious
choice for observing the cosmic parallax.

Figure~\ref{fig:CP-overview} depicts the overall scheme describing a possible
time-variation of the  position of a pair of sources that expand radially with respect to the center but anisotropically with respect to the observer. We label the two sources $a$ and $b$, and the two observation times 1 and 2. In what follows, we will refer to ($t$, $r$, $\theta$, $\phi$) as the comoving coordinates with origin on the center of a
spherically symmetric model. Peculiar velocities aside, the symmetry of such a
model forces objects to expand radially outwards, keeping $r$, $\theta$ and
$\phi$ constant.

\subsubsection{Estimating the parallax}
\label{sec:estimate}

Following Fig.~\ref{fig:CP-overview}, let us assume an expansion in a flat FRW space from a ``center'' $C$ observed by an off-center observer $O$ at a distance
$X_{\rm obs}$ from $C$. Since we are assuming FRW it is clear that any point in space could be considered a {}``center'' of expansion: it is only when we will consider a LTB universe that the center acquires an absolute meaning. The relation between the observer line-of-sight angle $\xi$ and the coordinates of a source located at a physical radial distance $X$ (corresponding to a comoving radial distance $r$) and angle $\theta$ in the $C$-frame is
\begin{equation}
    \cos\xi=\frac{X\cos\theta-X_{{\rm obs}}}{(X^{2}+X_{{\rm obs}}^{2}-2\,
X_{{\rm obs}}X\cos\theta)^{1/2}},
\end{equation}
where all angles are measured with respect to the $CO$ axis. We follow the
approach of Ref.~\cite{2009PhRvL.102o1302Q}  and  assume for simplicity (and
clarity) that both sources share the same $\phi$ coordinate.

\begin{figure}[t!]
    \includegraphics[width=9cm]{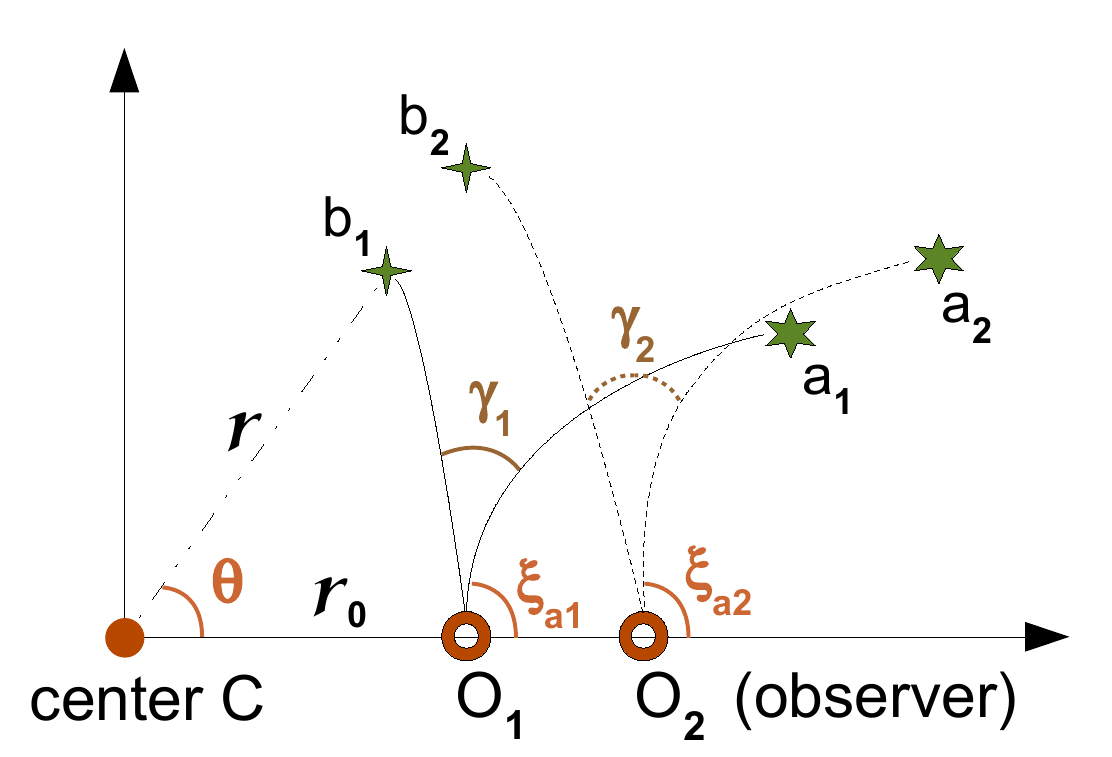}
    \caption{\small Cosmic parallax in LTB models. $C$ stands for the center of symmetry, $O$ for the off-centre observer, a and b for two distinct distant light sources, such as quasars. For these latter three, the subindex $1$ and $2$ refer to two different times of observation. For clarity purposes  we assumed here that the points $C,\,O,\,$a$_{1},\,$b$_{1}$ all lie on the same plane. By symmetry, points a$_{2},\,$b$_{2}$ remain on this plane as well. Comoving coordinates $r$ and $r_{0}$ correspond to physical coordinates $X$ and $X_{0}$. The difference between the angular separation of sources, $\,\Delta_t\gamma\equiv\gamma_1 - \gamma_2$, is the cosmic parallax. The angular separation $\gamma_t$, in turn, is calculated as the difference between the angle $\xi$ of the incoming geodesics coming from a and b at time $t$ ($\gamma_1 \equiv \xi_{a1} - \xi_{b1}$).  From Ref.~\cite{2009PhRvL.102o1302Q}.}
    \label{fig:CP-overview}
\end{figure}

Consider first two sources at location $a_{1},b_{1}$ on the same
plane that includes the $CO$ axis with an angular separation $\gamma_{1}$
as seen from $O$, both at distance $X$ from $C$. After some time
$\Delta t$, the sources move to positions $a_{2},b_{2}$ and the
distances $X$ and $X_{{\rm obs}}$ will have increased by $\Delta_{t}X$
and $\Delta_{t}X_{{\rm obs}}$ respectively, so that the sources subtend
an angle $\gamma_{2}$ (see Fig.~\ref{fig:CP-overview}). In a FRW universe, these
increments are such
that they keep the overall separation $\gamma$ constant. However,
if for a moment we allow ourselves the liberty of assigning to the
scale factor $a(t)$ and the $H$ function a spatial dependence, a
time-variation of $\gamma$ is induced. The variation
\begin{equation}
    \Delta_{t}\gamma\;\equiv\;\gamma_{1}-\gamma_{2}
\end{equation}
is the cosmic parallax effect and can be easily estimated if we suppose that the
Hubble law is just generalized to
\begin{equation}
    \Delta_{t}X=XH(t_{0},X)\Delta t\equiv XH_{X}\Delta t\,,
\end{equation}
where
\begin{equation}
    X(r)\equiv\int^{r}g_{rr}^{1/2}\dd r'=\int^{r}a(t_{0},r')\dd
r'\,,\label{eq:phys-distance}
\end{equation}
generalizes the FRW relation $\, X_{{\rm FRW}}=a(t_{0})r$ in a metric
whose radial coefficient is $g_{rr}$.

For two sources $a$ and $b$ at distances much larger than $X_{{\rm obs}}$  (which in practice in usual models corresponds to $z_{a,b} \gtrsim 0.1$),
after straightforward geometry we arrive at
\begin{equation}
    \Delta_{t}\gamma\,=\,\Delta tX_{{\rm obs}}\left[(H_{{\rm obs}}-H_{a})\frac{\sin\theta_{a}}{X_{a}}-(H_{{\rm obs}}-H_{b})\frac{\sin\theta_{b}}{X_{b}}\right] .\label{eq:deltagamma-full}
\end{equation}
It is important to note that this simple analytical estimate have been verified numerically, and the angular dependence of the cosmic parallax for sources at similar distances has been verified to hold to very high precision. As can be seen above, the signal $\Delta_{t}\gamma$ in \eqref{eq:deltagamma-full} depends
both on the sources' positions on the sky (the angles $\theta_{a,b}$) and on their radial distances to the center ($X_a$ and $X_b$). In what follows we will consider two simplified scenarios for which the sources lie either: $(i)$ on approximately the same redshift but different positions; $(ii)$ on approximately the same line-of-sight but different redshifts.

For case $(i)$, we can average over $\theta_{a,b}$ to obtain the average cosmic parallax for two arbitrary sources in the sky (still assuming they lie on the same plane that contains $CO$). If both sources are at the similar redshifts $z_a\simeq z_b \equiv z$ (corresponding to a physical distance $X$), then the average cosmic parallax effect is given by
\begin{align}
    \langle\Delta_{t}\gamma\rangle_{{\rm perp}} \,\simeq\,\frac{s\,\Delta t\,(H_{{\rm obs}}-H_{X})}{4\pi^{2}}\int_{0}^{2\pi}\!\!\int_{0}^{2\pi}\!\!|\sin\theta_{a} -\sin\theta_{b}| \,\dd\theta_{a}\dd\theta_{b}\,  \,=\,\frac{8}{\pi^{2}}s\,\Delta t\,(H_{{\rm obs}}-H_{X})\,.\label{eq:average-cp}
\end{align}
where we defined a convenient dimensionless parameter $s$ such that
\begin{equation}
    s\equiv\frac{X_{{\rm obs}}}{X}\ll1\,.\label{eq:s-definition}
\end{equation}
Note that at this order  the difference between the observed angle $\xi$ and $\theta$ can be neglected \cite{2009PhRvL.102o1302Q}. We can also convert the above intervals $\Delta X$ into the redshift interval $\Delta z$ by using the relation $\;r=\int\dd z/H(z)$. Using~\eqref{eq:phys-distance} we can write $\;\Delta X=a(t_{0},X)\Delta z/H(z)\sim\Delta z/H(z)\,$ (we impose the normalization $\,a(t_{0},X_{{\rm obs}})=1$), where $\,H(z)\equiv H(t(z),X)$. One should note that in a non-FRW metric, one has $\,s\neq r_{0}/r$.

In a FRW metric, $H$ does not depend on $r$ and the parallax vanishes. On the other hand, any deviation from FRW entails such spatial dependence and the emergence of cosmic parallax, except possibly for special observers (such as the center of LTB). A constraint on $\Delta_{t}\gamma$ is therefore a constraint on cosmic anisotropy.

Rigorously, the use of the above equations is inconsistent outside a flat FRW scenario; one actually needs to perform a full integration of light-ray geodesics in the new metric. Nevertheless, following~\cite{2009PhRvL.102o1302Q} we shall assume  that for an order of magnitude estimate we can simply replace $H$ with its space-dependent counterpart given by LTB models. In order for an alternative LTB cosmology to have any substantial effect (e.g., explaining the SNIa Hubble diagram) it is reasonable to assume a difference between the local $H_{{\rm obs}}$ and the distant $H_{X}$ of order $H_{{\rm obs}}$~\cite{Alnes:2006a}. More precisely, putting $\, H_{obs}-H_{X}=H_{obs}\Delta h\,$ then using \eqref{eq:average-cp} one has that the average $\Delta_{t}\gamma$
is of order
\begin{equation}
    \langle\Delta_{t}\gamma\rangle\Big|_{\rm perp}\sim\,20\, s\,\Delta h \;\mu  {\rm as/year}  \label{eq:perp-cp-estimate}
\end{equation}
for two sources at the same redshift.

Similarly, for case $(ii)$ we haver source pairs at approximately same position $\theta$ but different (yet similar) redshifts, and one has (using~\eqref{eq:deltagamma-full})
\begin{equation}
\begin{aligned}
    \Delta_{t}\gamma\Big|_{{\rm rad}}  \;\sim\; s\,\sin\theta\Delta h\Delta
t\,\Delta z/X\;\mu\mbox{as/year}  \;\sim\;\,20\, s\sin{\theta}\,\Delta
h\frac{\Delta z}{z}\;\mu\mbox{as/year}\,,
\end{aligned}
\end{equation}
 where it was assumed that $X\sim zH(z)^{-1}$. The average radial
cosmic parallax for sources between 10 and 200 times $X_{{\rm obs}}$ can be
obtained
numerically to be
\begin{align}
    \langle\Delta_{t}\gamma\rangle_{{\rm rad}} \;\simeq\;\frac{\,\Delta
t\,(H_{{\rm obs}}-H_{X})\sin\theta}{190^{2}} \int_{10}^{200}\!\int_{10}^{200}
\left|\frac{1}{s_{a}}-\frac{1}{s_{b}}\right| \,\dd (1/s_{a})\,\dd (1/s_{b})
    \;=\;0.014\,\sin\theta\Delta t\,(H_{{\rm obs}}-H_{X}) \,.
\label{eq:deltagamma}
\end{align}
Therefore, one can estimate for the radial signal
\begin{equation}
    \langle\Delta_{t}\gamma\rangle\Big|_{{\rm rad}}\sim\,0.3\sin\theta\Delta h
\mu {\rm as/year}\,,
    \label{eq:rad-cp-estimate}
\end{equation}
which is very similar to its same-shell counterpart~\eqref{eq:perp-cp-estimate},
except for the $\sin\theta$ modulation.

Moreover, one has to address the main expected source of noise, to wit the
intrinsic peculiar velocities of the sources. The variation in angular
separation for sources at angular diameter distance $D_{A}$ (measured by the
observer) and peculiar velocity $v_{{\rm pec}}$ can be estimated
as~\cite{2009PhRvL.102o1302Q}
\begin{equation}
    \Delta_{t}\gamma_{{\rm pec}}=\left(\frac{v_{{\rm
pec}}}{500\,\frac{\mbox{km}}{\mbox{s}}}\right)\left(\frac{D_{A}}{1\,\mbox{Gpc}}
\right)^{-1}\!\left(\frac{\Delta
t}{10\,\mbox{years}}\right)\mu\mbox{as}.\label{eq:pecvel}
\end{equation}
This velocity field noise is therefore typically smaller than the experimental
uncertainty (especially for large distances) and again will be averaged out for
many sources. The above relation was further investigated
in~\cite{2009arXiv0903.3402D}, where it was proposed to estimate $D_{A}$ via
observations of $\Delta_{t}\gamma_{\rm pec}$ due not to voids but by our motion
with respect to the CMB.

Finally, two competing effects induce similar dipolar parallaxes: one is due to
our own peculiar velocity and the other by a change in aberration of the sky due
to the acceleration of the Solar System in the the Milky Way. Both effects have
nonetheless distinct redshift dependence, which can be used to tell all three
apart. Figure~\ref{fig:dgammath-vs-z} depicts the three dipolar effects; we will
come back to this issue in Section~\ref{subsec:parobs}.

\subsubsection{Numerical derivation} \label{sec:geodesic}

As suggestive as the above estimates be, they need confirmation from
an exact treatment where the full relativistic propagation of light
rays is taken into account.  We will thus consider in what follows the three
LTB void models introduced in Sec.~\ref{sec:driftLTB}, dubbed Models I, II and
cGBH~\cite{2010PhRvD..81d3522Q}. In all three cases the off-center (physical)
distance is set to 30~Mpc. Assuming no peculiar velocity of the observer, this
distance was shown to be compatible with both the CMB dipole
\cite{Alnes:2006a,2010PhRvD..81d3522Q,2010arXiv1009.0273F} and supernovae data
\cite{Alnes:2006b,Blomqvist:2009}.

\begin{figure}[t!]
    \includegraphics[width=9cm]{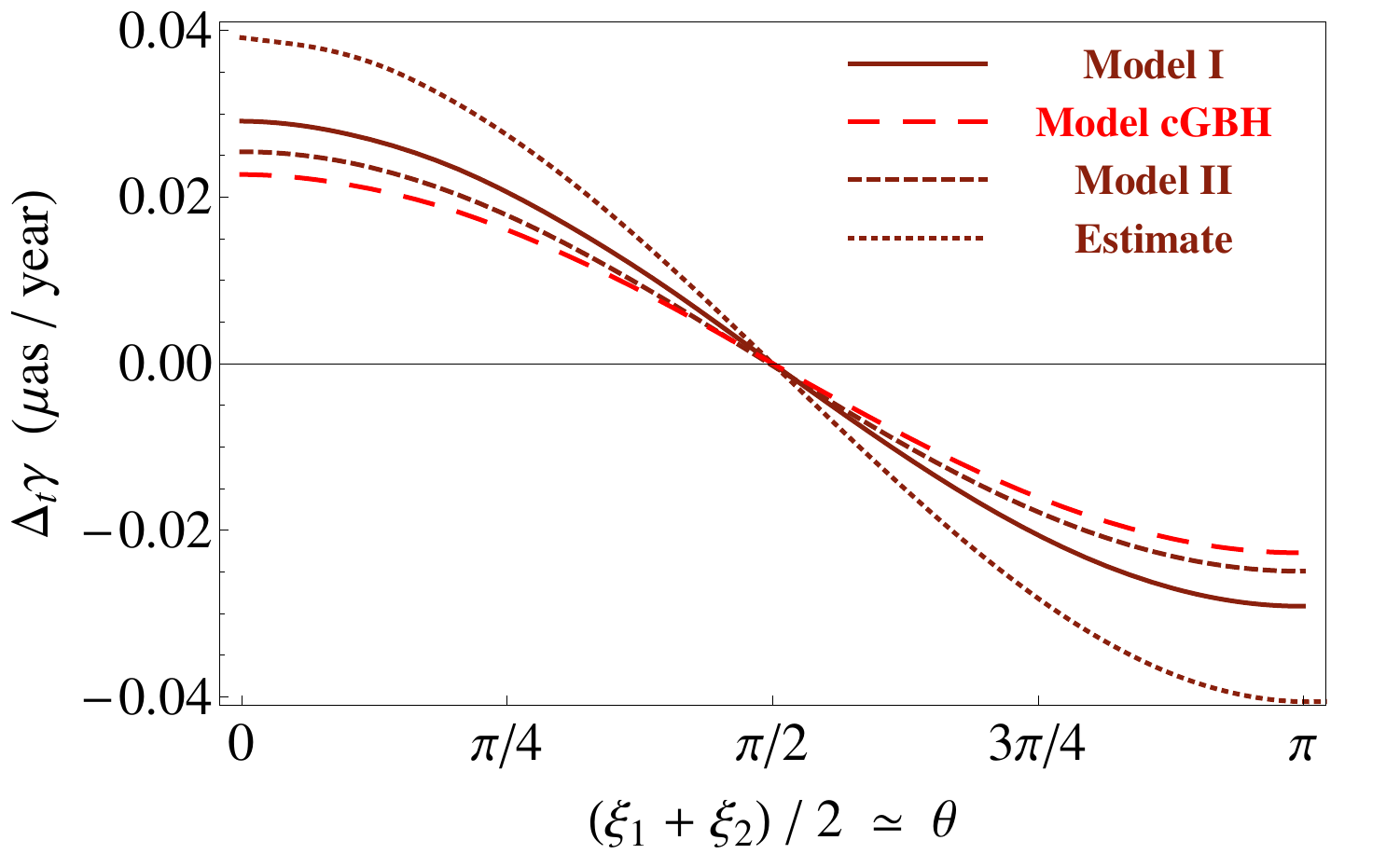}
    \caption{\small $\Delta_{t}\gamma$ for two sources at the same shell, at $z=1$,
    for Model I (full lines), Model II (dashed), the cGBH model (red,
    long-dashed lines) and the FRW-like estimate (dotted). The horizontal axis is the time-average angular position $\xi$ of the source in the sky, and can be approximated by the inclination angle $\theta$ to good precision. The plotted lines correspond to a separation of $90^{\circ}$ in the sky between the sources. The off-center distance is assumed to be $30$ Mpc.  From Ref.~\cite{2010PhRvD..81d3522Q}.}
    \label{fig:dgammath}
\end{figure}

To compute numerically the cosmic parallax effect in LTB models, an algorithm
was laid down in~\cite{2009PhRvL.102o1302Q}; it can be found in
Appendix~\ref{app:ltb-geodesics}. Using this algorithm, we plot in
Fig.~\ref{fig:dgammath} $\Delta_{t}\gamma$ for three sources at $z=1$, for
models I and II as well as for the cGBH model and the FRW-like estimate. One can
see that the results do not depend sensitively on the details of the shell
transition and that in both cases the FRW-like estimate gives a reasonable idea
of the true LTB behavior.

\begin{figure}[t!]
\includegraphics[width=9cm]{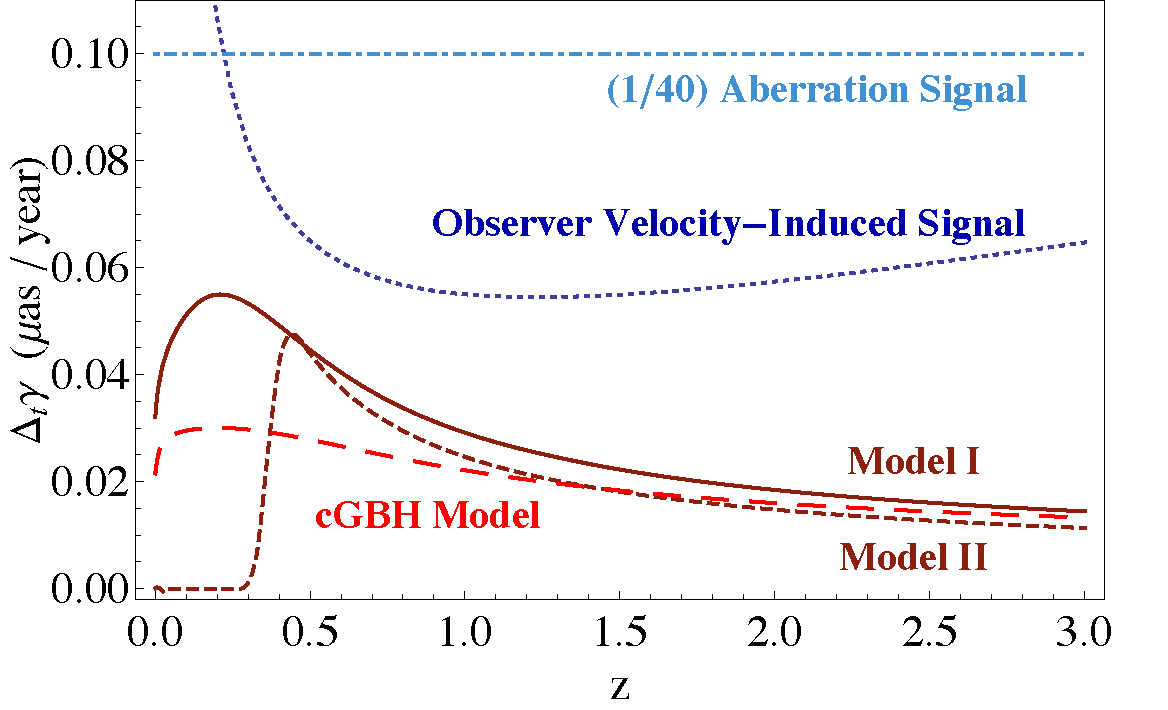}
\caption{\small $\Delta_{t}\gamma$ for two sources at the same shell but separated
by $90^{\circ}$ as a function of redshift assuming a $30$ Mpc
off-center distance. The dark, brown lines correspond to the cosmic
parallax in Models I (full lines) and II (dashed); the red long-dashed
lines to the cGBH model; the light, blue dotted lines represent
$1/40$ of the aberration-induced signal (see text), which does not depend
on redshift; the dark dotted lines stand for the parallax induced
by our own peculiar velocity (assumed to be 400 km/s).
Since all effects are dipolar, the curves plotted here are proportional
to the amplitude of such dipoles. The actual amount of noise depend
on the angle between the center of the void and the directions of
acceleration and peculiar velocity of the measuring instrument. Notice
that in Model II the cosmic parallax is zero inside the
void, which is expected as $H_{||}$ is constant inside the void in that model [see Fig.~\ref{fig:ltb-models} and~\eqref{eq:average-cp}]. The vanishing cosmic parallax in Model II inside the void is a well-understood peculiar feature of that model. From Ref.~\cite{2010PhRvD..81d3522Q}.}
\label{fig:dgammath-vs-z}
\end{figure}

Fig.~\ref{fig:dgammath-vs-z} illustrates the redshift dependence of
the cosmic parallax effect for two sources at the same shell (i.e.,
same redshift) but separated in the sky by $90^{\circ}$ (which is
the average separation between two sources in an all-sky survey):
one source is located at $\xi=-45^{\circ}$, the other at $\xi=+45^{\circ}$.
Also plotted are the two major sources of systematic noise, which
will be discussed in Section~\ref{subsec:parobs}: our own peculiar
velocity and the change in the aberration due to the acceleration
of the observer. As will be shown, all the effects we are considering
are dipolar and the lines in Fig.~\ref{fig:dgammath-vs-z} are
proportional to the amplitudes of such dipoles. Note that both systematics
have different $z$-dependence than the cosmic parallax produce in void models,
and in principle all three effects can be separated.

\subsection{ Cosmic parallax in homogeneous and anisotropic models}
\label{sec:paralaxBianchi}
The Bianchi solutions describing the anisotropic line element were treated as
small perturbations to a FRW background and its effect on the CMB pattern was
studied by
\cite{1973MNRAS.162..307C,1985MNRAS.213..917B,1995A&A...300..346M,
1996A&A...309L...7M,1996PhRvL..77.2883B,1997PhRvD..55.1901K}. These anisotropic
models face two main drawbacks: in order to fit large scale CMB patterns they
sometimes require unrealistic choice of the cosmological parameters; the early
time  inflationary phase isotropises the universe very efficiently, leaving a
hardly detectable residual anisotropy. However, all these difficulties vanish
if the anisotropic expansion is  generated only at late time, excited by a dominating anisotropically stressed dark energy component. In
\cite{2009PhRvD..80f3527Q} a general treatment of the cosmic parallax in Bianchi
I models has been derived, and subsequently connected to simple phenomenological
anisotropic dark energy model \cite{2008ApJ...679....1K}.

\subsubsection{Cosmic parallax in Bianchi I models}
\begin{figure}[t!]
    \includegraphics[width=8cm]{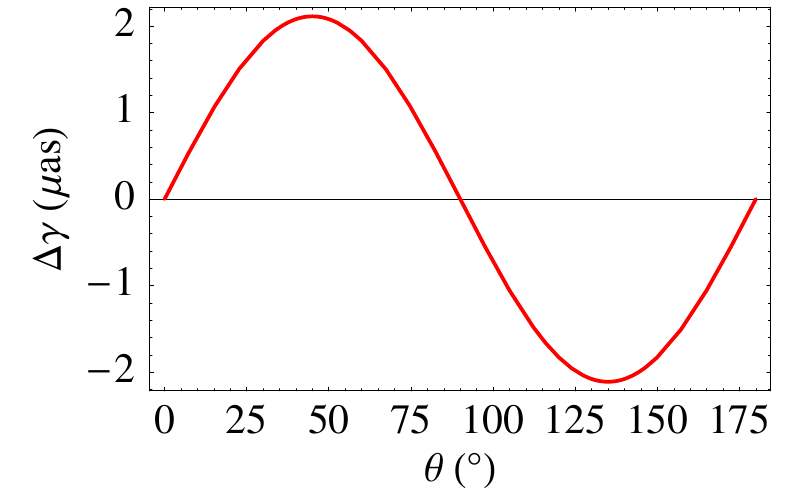}
        \includegraphics[width=8cm]{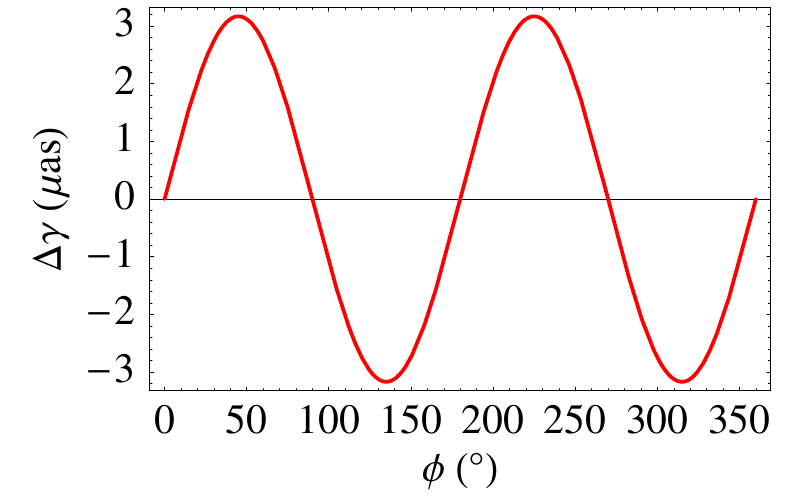}
    \caption{\small Cosmic parallax in Bianchi I models for $h_{x}=0.71$,
$h_{y}=0.725$, $h_{z}=0.72$, i.e. $\Sigma_{0X}=-0.012$ and $\Sigma_{0Y}=0.009$).
The time interval is $\Delta t=10$yrs as: \emph{(Left)} a function of $\theta$
for
    $\phi=\Delta\phi=0$ and $\Delta\theta=90^{o}$, which corresponds
    to the (X,Z) plane; and \emph{(Right)} a function of $\phi$ for
$\theta=90^{o}$,  $\Delta\theta=0$ and $\Delta\phi=90^{o}$ which corresponds
    to the (X,Y) plane.  From Ref.~\cite{2009PhRvD..80f3527Q}.}
    \label{fig:bianthe-phi}
\end{figure}

The unperturbed metric in Bianchi I models can be written in Cartesian
coordinates as:
\begin{equation}
ds^{2}=-dt^{2}+a^{2}(t)dx^{2}+b^{2}(t)dy^{2}+c^{2}(t)dz^{2},
\label{metric}
\end{equation}
where the three expansion rates are defined as $H_{X}=\dot{a}/a$,
$H_{Y}=\dot{b}/b$ and
$H_{Z}=\dot{c}/c$.
Here the derivatives are taken with respect to the coordinate time. Bianchi I
models exhibit no overall vorticity but shear components  $\Sigma_{XYZ}=
H_{X,Y,Z}/H-1$, where H is an
effective expansion rate, $H=\dot{A}/A$, with $A=(abc)^{1/3}$. In homogeneous
and anisotropic models like  Bianchi I models we expect a different signal with
respect to
the dipolar LTB one derived in  Sec.~\ref{sec:estimate}, being less contaminated
by systematics (e.g. observer's velocity and acceleration) that could mimic the
cosmological signal and hence be even more predictive.
Let us consider two sources $A$ and $B$ in the sky located at physical
distance from us observers $\mathbf{O}_{[A,B]}=(X,Y,Z)_{[A,B]} =
(R\sin{\theta}\cos{\phi},R\sin{\theta}\sin{\phi},R\cos{\theta})_{[A,B]}\,$,
where $R=\sqrt{X^{2}+Y^{2}+Z^{2}}$ and $(\theta,\phi)$ are spherical
angular coordinates. Their angular separation on the celestial sphere
reads
\begin{eqnarray}
\mathbf{OA}\cdot\mathbf{OB}=\cos{\gamma}=\cos{\theta_{A}}\cos{\theta_{B}}+\sin{
\theta_{A}}\sin{\theta_{B}}\cos{\Delta\phi},\label{gamma}
\end{eqnarray}
 with $\Delta\phi=(\phi_{A}-\phi_{B})$.If during evolution of the Universe
$\Delta_t\gamma$ is different from zero, then a cosmic parallax arises.
If the expansion  is homogeneous but anisotropic, the angular separation between
two points, in a $\Delta t$ interval, changes as:
\begin{eqnarray}
-\sin{\gamma}\Delta_{t}\gamma & = &
\sin{\theta_{A}}\cos{\theta_{B}}(\Delta_{t}\theta_{B}\cos{\Delta\phi}-\Delta_{t}
\theta_{A})+\cos{\theta_{A}}\sin{\theta_{B}}(\Delta_{t}\theta_{A}\cos{\Delta\phi
}-\Delta_{t}\theta_{B})\label{dgamma}\\
 & + &
\sin{\theta_{A}}\sin{\theta_{B}}\sin{\Delta\phi}(\Delta_{t}\phi_{B}-\Delta_{t}
\phi_{A}).\nonumber \end{eqnarray}
Looking at the equation above, some intuitive solutions can be derived. For
instance, in the limit
of $\Delta_{t}\phi_{A}=\Delta_{t}\phi_{B}=\phi_{A}=\phi_{B}=0$ the motion is
confined on the
(X,Z) plane and the cosmic parallax reduces to
$(\Delta_{t}\theta_{A}-\Delta_{t}\theta_{B})$
(see left panel in Fig.~\ref{fig:bianthe-phi}). Similarly, on the (X,Y) plane
the signal
is $(\Delta_{t}\phi_{A}-\Delta_{t}\phi_{B})$ (see right panel
Fig.~\ref{fig:bianthe-phi}).
In figure~\ref{fig:bianthe-phi}   the shear parameters at present are allowed to
appreciably deviate from 0. This explains why the cosmic parallax is a few orders
of magnitude larger than the one in~\cite{2009arXiv0905.3727F}. The main
motivation for this will be presented later on.

In general, the signal will be given by the combination of the anisotropic
expansion
of the sources and the change in curvature induced by the shear on the photon
path from the emission to the observer.
As for the LTB models, the photon trajectory in a Bianchi I Universe will not be
radial.
However, while in LTB models this effect on the cosmic parallax is enhanced by
inhomogeneity
(although a FRW description of null geodesic has been shown to give a fairly good
approximation, see Fig~\ref{fig:dgammath}),
in Bianchi I models the geodesic bending for a single source has been shown to
amount at most to about
7$\%$ \cite{2009PhRvD..80f3527Q}, which allows to adopt the straight geodesics
approximation.
With this approximation and considering the relations $\phi=\arctan{(Y/X)}$ and
$\theta=\arccos{(Z/\sqrt{X^{2}+Y^{2}+Z^{2}})}$ one can write down the temporal
evolution equations
for the angular coordinates as:

\begin{eqnarray}
\Delta_{t}\phi & = & \frac{XY}{X^{2}+Y^{2}}(H_{0Y}-H_{0X})\Delta t =
\frac{\sin{2\phi}}{2}(\Sigma_{0Y}-\Sigma_{0X})H_{0}\Delta t\label{dang}\\
\Delta_{t}\theta & = & \frac{Z\,
R^{-2}}{\sqrt{X^{2}+Y^{2}}}\Big[X^{2}(H_{0X}-H_{0Z})+Y^{2}(H_{0Y}-H_{0Z})\Big]
\Delta t \nonumber \\
&=&
\frac{\sin{2\theta}}{4}\Big[(3(\Sigma_{0X}+\Sigma_{0Y})+\cos{2\phi}(\Sigma_{0X}
-\Sigma_{0Y})\Big]H_{0}\Delta t,
\label{dangb}
\end{eqnarray}
where the shear components at present satisfy the transverse condition
$\Sigma_{0X}+\Sigma_{0Y}+\Sigma_{0Z}=0$.

Equations~\eqref{dang}-\eqref{dangb} show the quadrupolar behaviour of the cosmic
parallax for Bianchi I
models in the $\phi$ and $\theta$ coordinate, respectively. This functional form
recovers the one expected for the first non-vanishing multipole expansion of the
CMB large scale
relative temperature anisotropies in Bianchi I model~\cite{1995A&A...300..346M}.
In general, the pattern on the sky of the cosmic parallax signal will be given
by plugging the above equations into~(\ref{dgamma}), to wit
$\Delta_{t}\gamma=\Delta_{t}\gamma(\theta,\phi,\Delta\theta,\Delta\phi,\Sigma_{
0X},\Sigma_{0Y},H_{0},\Delta t)$, where the only further conjecture is that $H$
does not appreciably vary in $\Delta t$ \cite{2009PhRvD..80f3527Q}.
At first order, this seems reasonable for the time intervals under
consideration.
The signal is independent
on the redshift, which means for instance that source pairs along the same line
of sight undergo
the same temporal change in their angular separation or that aligned quasars
would stay aligned.

\begin{figure}[t!]
\includegraphics[width=16cm]{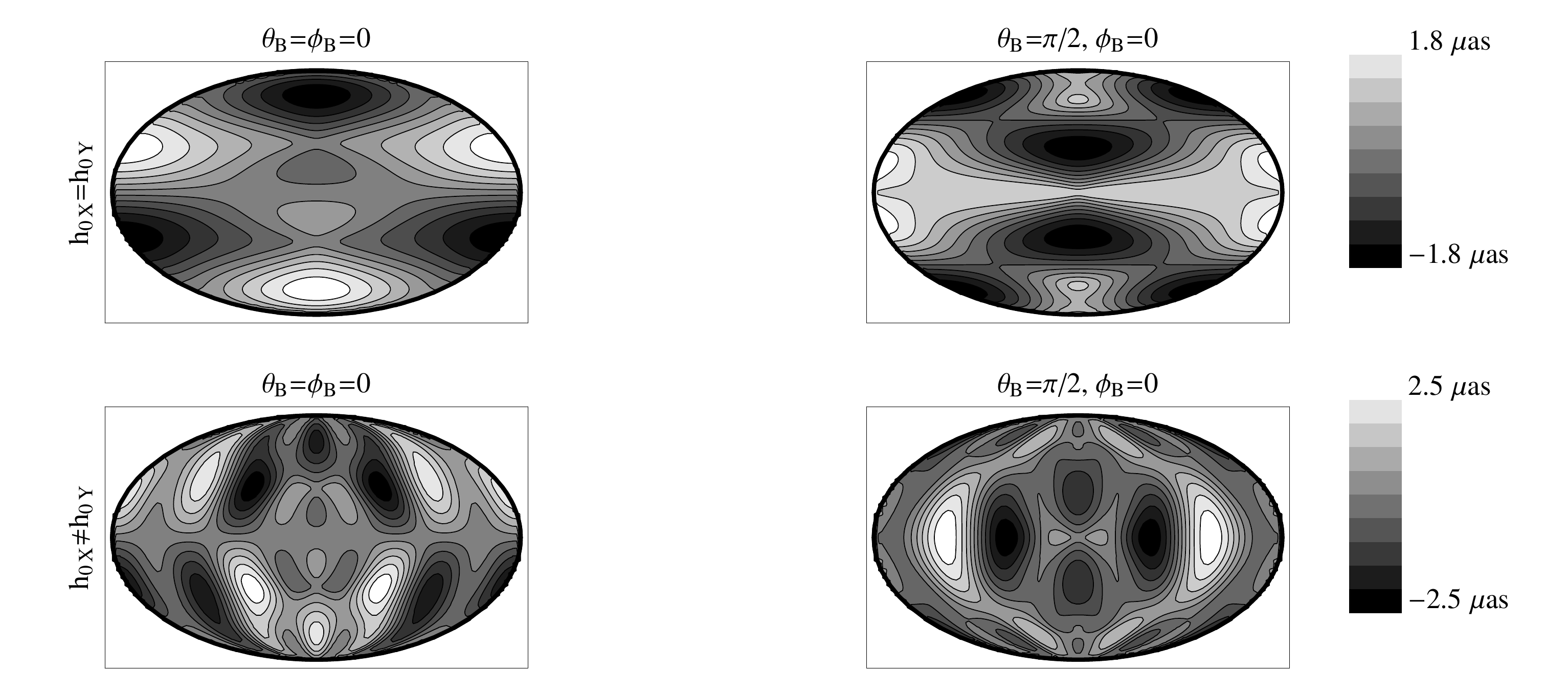}
\caption{\small Mollweide contour plot for cosmic parallax in Bianchi I models for
one source fixed at two different location in the sky. Upper panels
show the signal for ellipsoidal models ($h_{0Z}=0.72$ and $h_{0X}=h_{0Y}=0.71$),
while in lower panels $h_{0Y}=0.725$. Lighter colours correspond
to higher signal and on the horizontal and vertical axes angular coordinates
vary in the range $\phi:[0,2\pi]$ and $\theta:[0,\pi]$, respectively.
The time interval is $\Delta t=10$yrs. From Ref.~\cite{2009PhRvD..80f3527Q}.}
\label{moll}
\end{figure}

In Fig.~\ref{moll}  a Mollweide projection of the isocontours of the cosmic
parallax for peculiar values of the angles
$\theta$ and $\phi$ with the same choice of parameters of
Fig.~\ref{fig:bianthe-phi}  is shown.
In the first case, one of the sources is fixed at the north pole and, in the
other, one the sources
lives in the plane (X,Y).  As expected, when the source is at an equatorial
position the symmetry
with respect to the (X,Y) plane is preserved, while when the source is at the
north pole a symmetry
with respect to the (X,Z) plane a cosmic parallax emerges. In a FRW universe the
components
of the shear simultaneously vanish and so does the cosmic parallax.

\subsubsection{Cosmic parallax induced by dark energy}
\label{sec:parallaxDE}
CMB quadrupole has been used to put strong constraints on Bianchi
models~\cite{1996PhRvL..77.2883B,1997PhRvD..55.1901K,1995A&A...300..346M}. In a
LCDM Universe, the anisotropy parameters scale as the inverse of comoving
volume
leading so to a natural isotropization of the expansion from the recombination
up to present with typical limits of the shear parameters of the order
$\sim10^{-9}\div10^{-10}$ (resulting in a cosmic parallax signal of order
$10^{-4}\mu$as).
However, this constraints must be relaxed in the cases where the anisotropic
expansion takes place after decoupling,
 due to, for instance, vector fields representing anisotropic dark energy
~\cite{2008ApJ...679....1K}.
In \cite{2009PhRvD..80f3527Q}  the authors analysed the cosmic parallax applied
to a specific anisotropic
phenomenological dark energy model in the framework of Bianchi I
models~\cite{2008ApJ...679....1K,2008JCAP...06..018K}
(we refer to these papers for details). The anisotropic expansion is
caused by the anisotropically stressed dark energy fluid whenever
its energy density contributes to the global energy budget.

Let us consider a physical model where the  Universe expansion is driven by the
anisotropically stressed
dark energy fluid. After recombination, the energy momentum-stress tensor is
dominated by the
dark matter and dark energy components and can be written as:

\begin{eqnarray}
T_{(m)\nu}^{\mu} & = &
\mbox{diag}(-1,w_{m},w_{m},w_{m})\rho_{m}\label{eq:Tmunu-de}\\
T_{({\rm DE)\nu}}^{\mu} & = & \mbox{diag}(-1,w,w+3\delta,w+3\gamma)\rho_{{\rm
DE}},
\label{eq:Tmunu-de2}
\end{eqnarray}
where $w_{m}$ and $w$ are the equation of state parameters of matter and dark energy and the skewness parameters $\delta$ and $\gamma$ can be interpreted as the difference of pressure along the x and y and z axis. Note that the energy-momentum tensor ~\eqref{eq:Tmunu-de} is the most general one compatible with the metric~\eqref{metric}~\cite{2008ApJ...679....1K}. The reason why limits from cosmic parallax could be more sensitive with respect to the ones coming from CMB~\cite{2009arXiv0905.3727F} is that the parameters $\delta$ and $\gamma$ are allowed to grow after the decoupling. Assuming constant anisotropy parameters, an experimental
constraint $\delta=-0.1$ is not completely excluded by supernovae data, since it lies on
the 2$\sigma$ contours of the $\gamma-\delta$ plane, if a prior on $w$ and $\Omega_{m}$ is assumed \cite{2008ApJ...679....1K}. More phantom equation of state parameters and/or larger matter densities allow
for larger value of delta. In addition, and more generally, time dependent $\delta$ and $\gamma$ functions, mimicking for example specific minimally coupled vector field with double power law potential, can escape these constraints. It can be shown~\cite{2009PhRvD..80f3527Q} that the dynamical solutions for the quantities of interest can be found by expanding around the critical points the generalized Friedman equations and the  continuity equations for matter and dark energy~\cite{2008JCAP...06..018K,2008ApJ...679....1K}:

\begin{equation}
\begin{aligned}
U'= & U(U-1)[\gamma(3+R-2S)\;+\,\delta(3-2R+S)\,+\,3(w-w_{m})]\\
S'= &
\frac{1}{6}(9-R^{2}+RS-S^{2})\big\{S[U(\delta+\gamma+w-w_{m})+w_{m}-1]-6\,
\gamma\, U\big\}\\
R'= &
\frac{1}{6}(9-R^{2}+RS-S^{2})\big\{R[U(\delta+\gamma+w-w_{m})+w_{m}-1]-6\,
\delta\, U\big\},
\end{aligned}
\label{sys}
\end{equation}
where $U\equiv\rho_{{\rm DE}}/(\rho_{{\rm DE}}+\rho_{m})$ and the
derivatives are taken with respect to $\log(A)/3$. In the above equations we
have introduced:
\begin{equation}
\begin{aligned}
R & \,\equiv\,(\dot{a}/a-\dot{b}/b)/H\;=\;\Sigma_{X}-\Sigma_{Y}\,\\
S & \,\equiv\,(\dot{a}/a-\dot{c}/c)/H\;=\;2\Sigma_{X}+\Sigma_{Y}\,,
\end{aligned}
\label{dom}
\end{equation}
that naturally define the degree of anisotropy. Since at present the dark energy
contribution
to the total density is about 74$\%$ , among the several solutions of the linear
system (\ref{sys}) beside the Einstein-de
Sitter case ($R_{*}=S_{*}=U_{*}=0$), one is interested
in the fixed points where the contribution of the dark energy is either
dominant:

\begin{equation}
R_{*}\,=\,\frac{6\delta}{\delta+\gamma+w-1},\;\;\;
S_{*}\,=\,\frac{6\gamma}{\delta+\gamma+w-1},\;\;\; U_{*}=1,
\label{eq:de-domination}
\end{equation}
or in its scaling critical stage:
\begin{equation}
\begin{aligned}
&
R_{*}\,=\,\frac{3\delta(\delta+\gamma+w)}{2(\delta^{2}-\delta\gamma+\gamma^{2})}
,
\quad
S_{*}\,=\,\frac{3\gamma(\delta+\gamma+w)}{2(\delta^{2}-\delta\gamma+\gamma^{2})}
,\;\;\;
U_{*}\,=\,\frac{w+\gamma+\delta}{w^{2}-3(\gamma-\delta)^{2}+2w(\gamma+\delta)},
\end{aligned}
\label{scal}
\end{equation}

where $\rho_{{\rm DE}}/\rho_{m}=const.$, i.e., the fractional
dark energy contribution to the total energy density is constant. In this case,
one needs to ensure that the scaling regime in not too much extended in the
past, in order to avoid a too long accelerated epoch overwhelming the structure
formation era \cite{2009PhRvD..80f3527Q}.

\subsubsection{Forecastings}
\label{sec:parallaxfore}

\begin{table}[t!]
\centering
\begin{tabular}{@{}lccc|ccr@{}}
\hline
 &  &  & \tabularnewline
Experiment  & $N_{s}$  & $\sigma_{acc}$  & $\Delta t$ & $\sigma_{\Sigma_0}$ & LTB $\sigma_{X_{\rm obs}}$ & LTB detection (30 Mpc) \tabularnewline
 &  &  &  & & & \tabularnewline
\hline
 &  &  &  & & &\tabularnewline
Gaia  & 500,000  & $\;90\;\mu$as$\;$  & 2.5yrs  &  $\;1.5 \cdot 10^{-3}$  & $\;190-310$ Mpc & $< 0.001\sigma$ \tabularnewline
&  &  & \tabularnewline
Gaia+  & 500,000  & $\;50\;\mu$as$\;$  & 5yrs  & $\;8.3 \cdot 10^{-4}$ & $\;46-70$ Mpc & $0.03\sigma - 0.24 \sigma$ \tabularnewline
 &  &  & \tabularnewline
Gaia++  & 1,000,000  & $5\;\mu$as  & $\;$10yrs$\;$ & $\;6 \cdot 10^{-5}$ & $\;1.6-2.5$ Mpc & $32\sigma - 49 \sigma$ \tabularnewline
 &  &  & \tabularnewline
\hline
\end{tabular}
\caption{Specifications adopted for Gaia-like and Gaia++ experiments, where $N_{s}$ is the total number of sources, $\sigma_{acc}$ is the experimental astrometric accuracy, $\Delta t$ is the time interval between two measurements, $\sigma_{\Sigma_0}$ is the 1$\sigma$ error bar on the present shear in the Bianchi I models here considered and the last 2 columns give estimates on detection of different LTB dark energy models (we used Models I, II and cGBH, discussed in Section~\ref{sec:geodesic}). The first LTB column gives the 1$\sigma$ limit on the off-center distance $X_{\rm obs}$ if one assumes a FRW metric; the second LTB column shows with how many $\sigma$ one would detect a LTB model assuming in all cases $X_{\rm obs} = 30$ Mpc. Model I is the easiest to detect or rule out, Model II the hardest.
Note that the error $\sigma_{\Sigma0}$ does not depend directly on the interval $\Delta t$ as it has been factored out in~\eqref{dang}-\eqref{dangb}; instead, it depends only on the average positional accuracy $\sigma_{acc}$ and the number of sources $N_{s}$. \label{tab:gaia}}
\end{table}

Detecting a cosmic parallax for a Bianchi I Universe requires an astrometric instrument with the highest performance in detecting quasar positions like the next generation astrometric experiment Gaia (described in Section~\ref{subsec:parobs}). In this section we present forecastings using the Fisher matrix formalism and assuming the instrumental specifications of Gaia and 2 other enhanced Gaia-like missions (dubbed Gaia+ and Gaia++) with an average time difference between observations $\Delta t$ ranging between $2.5$ and $10$ years~\cite{2009PhRvD..80f3527Q,2010PhRvD..81d3522Q} and different positional accuracy. Table~\ref{tab:gaia} lists the details of all 3 experiments (we refer to Section~\ref{subsec:parobs} for more accuracy about the observational strategy).

Since the average time separation between 2 observations is half the mission duration, Gaia (with nominal 5 year mission duration) should provide an average $\Delta t = 2.5$ years. Moreover, since accuracy depend strongly on quasar magnitude, it is important to estimate how will Gaia's quasar catalogue be distributed in magnitude. Using SDSS catalogue as a baseline, a detailed estimate of what the Gaia mission as planned will provide was carried out in Ref.~\cite{2010PhRvD..81d3522Q}, and it was found that the average accuracy would be $\sigma_{acc}=90 \;\mu$as (see \cite{Lindegren:2008} and Section~\ref{subsec:parobs} for details). We also include here estimates for two enhanced Gaia-like missions, following Ref.~\cite{2009PhRvD..80f3527Q}: one with 500,000 quasars uniformly distributed on the sphere with a constant average accuracy of $\sigma_{acc}=50\;\mu$as and with average time separation between 2 observations of 5 years (here dubbed Gaia+) and another one, with 1 million quasars, average accuracy $\sigma_{acc}=5\;\mu$as and average $\Delta t$ of 10 years (here dubbed Gaia++). Furthermore using average accuracy numbers, it is trivial to re-scale the final errors to a different accuracy of any Gaia-like misiion.

In general Bianchi I models, the cosmic parallax signal depends on four parameters: the average Hubble function at present, the time span and the two Hubble normalized anisotropy parameters at present. However, for the allowed range of values, contours in the $(\Sigma_{0X},\Sigma_{0Y})$ frame  do not depend on the value of $H_{0}$. Moreover, due to the linearity of our equations,
stretching the time interval between the two measurements or improving the instrumental accuracy would result in a trivial scaling on the final constraints presented here.

\begin{figure}[ht!]
 \includegraphics[width=9cm]{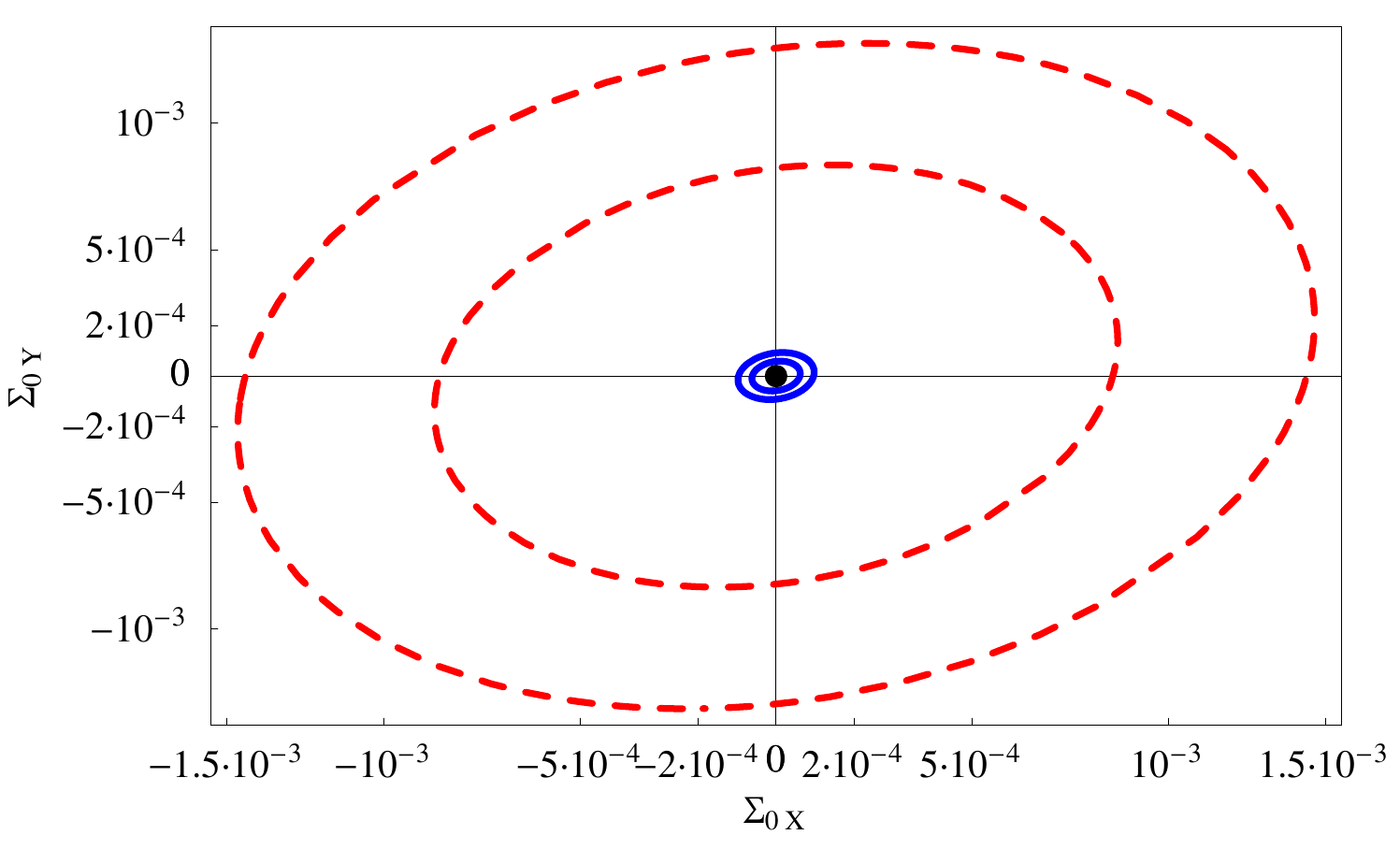}
\caption{\small Fisher contours of Cosmic parallax for Gaia+ and Gaia++ specifications (dashed and solid lines, respectively). The double contours identify $1\sigma$ and $2\sigma$ regions for $\Delta t=10$yrs.  From Ref.~\cite{2009PhRvD..80f3527Q}.}
\label{cont1}
\end{figure}

\begin{figure}[ht!]
 \includegraphics[width=17cm]{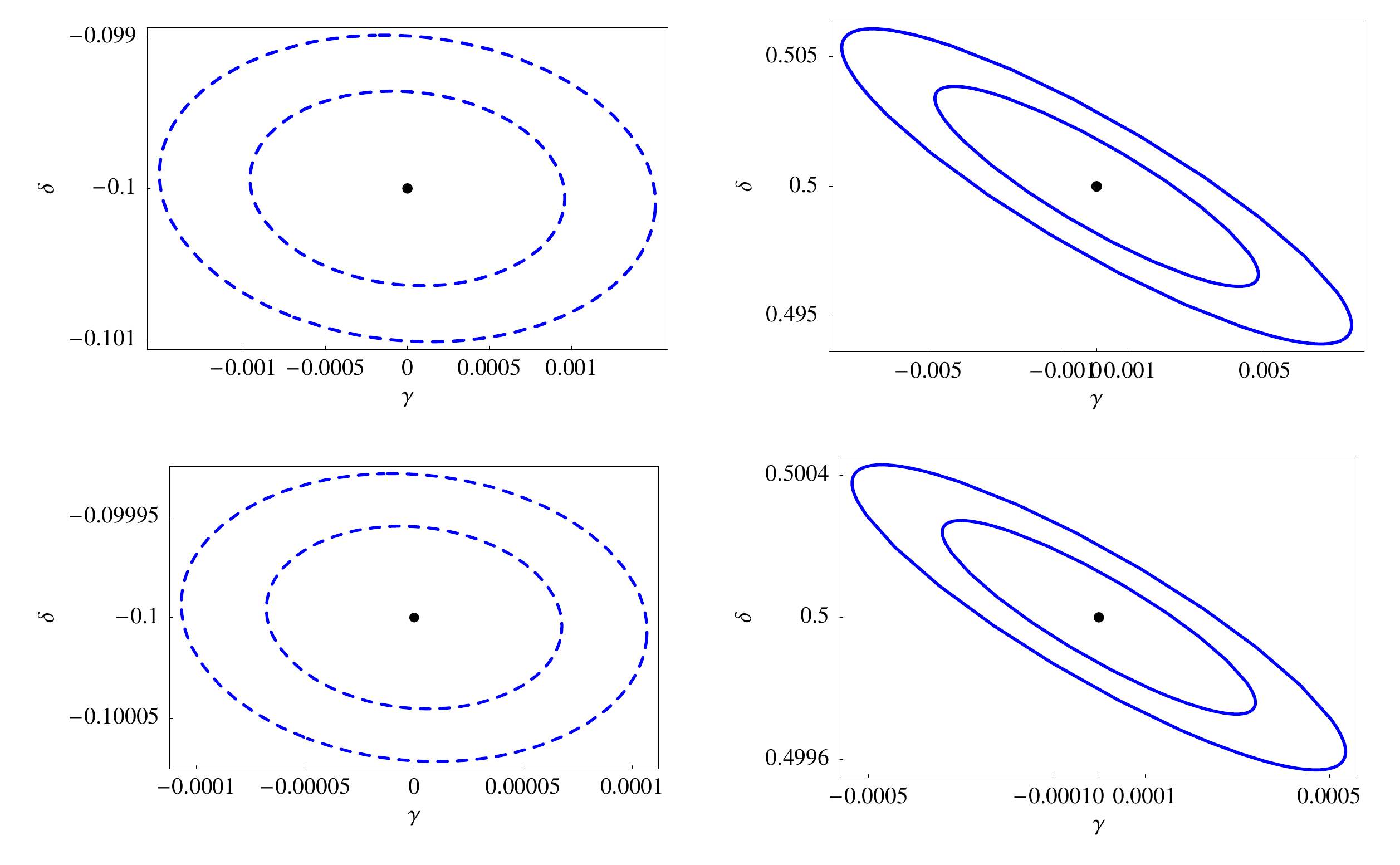}
\caption{\small Projected Fisher contours for the skewness dark energy parameters
for Gaia+ (upper panels) and Gaia++ specifications (lower panels). The double contours identify $1\sigma$ and $2\sigma$ regions for $\Delta t=10$yrs.
The dashed lines represent the case of an ellipsoidal universe with
$\, w=-1$, $\, U_{0}=0.74\,$ and $\,\delta=-0.1\,$ ($R_{0}\simeq0.2$)
approaching the dark energy dominated critical point (where $\, U=1\,$
and $\, R_{*}\simeq0.3$), while the solid lines represent an ellipsoidal
universe that has just entered the scaling regime, with $\, w=-1$,
$\, U_{0}=0.74\,$ and $\,\delta=0.5\,$ ($R_{*}\simeq-0.5$).  From
Ref.~\cite{2009PhRvD..80f3527Q}.}
\label{DEcont}
\end{figure}

The Fisher matrix is defined as:
\begin{equation}
F_{i,j}=\sum_{l}\frac{\partial\Delta_{t}\gamma_{(l)}}{\partial\Sigma_{0i}}\frac{
1}{\sigma_{acc}^{2}}\frac{\partial\Delta_{t}\gamma_{(l)}}{\partial\Sigma_{0j}},
\label{fish}
\end{equation}
where all separations are taken with respect to a reference source
and index $l$ runs from $2$ up to the number of quasars $N_{s}$
to take into account the spherical distances to all other sources.
In fact, one should notice that the Gaia accuracy positional errors  are obtained having already averaged over 2$N_{s}$ coordinates.
The Fisher error ellipses are shown in Fig~\ref{cont1}; the constraints turn out to be of the same order of magnitude of the CMB limits on the shear at decoupling. The  1$\sigma$ errors on $\Sigma_{0X}$ and $\Sigma_{0Y}$ turn out to be $8.3\cdot10^{-4}$ and $6\cdot10^{-5}$ for Gaia+
and Gaia++, respectively. The calculation was performed assuming the null
hypothesis (i.e. a Friedmann Robertson Walker isotropic expansion). However, due to the linearity of the cosmic parallax equations, the effect of considering alternative hypothesis is merely the shift of the ellipses' centers to the values of the new fiducial model.\footnote{This is not generally the case with Fisher Matrices, where the area of the ellipses can change appreciably with different fiducial models.}
The constraints on the skewness dark energy parameters of
equations~(\ref{eq:Tmunu-de})-(\ref{eq:Tmunu-de2})
 can be provided by mapping out the Fisher matrix into the new parameter space
${\bf p}=(\delta,\gamma)$ via $F'=A^{T}FA$, where
$A_{ij}=\partial\Sigma_{0i}/\partial p_{j}$ \cite{2009PhRvD..80f3527Q}.
 For the scaling solution~(\ref{scal})
the error
contours are shown in the right panels of Fig.~\ref{DEcont}, for
both Gaia+ and Gaia++ configurations.
Conversely, if the expansion is driven towards a future dark energy
 dominated solution, equations~(\ref{eq:de-domination}) do not represent
the anisotropy parameters at present (see  \cite{2009PhRvD..80f3527Q}). In order
to derive a more appropriate functional form for them, the
linearized system~(\ref{sys}) was solved around
solution~(\ref{eq:de-domination})
where  $\log(A)$, fixed to 0,   selects the present time.
For this second case, results are shown in the left panels of Fig.~\ref{DEcont}.

The derived constraints are of the order $10^{-3}\div10^{-4}$ with a net
improvement of about 2 or 3 order of magnitude with respect to the limits coming
from SNIa data ~\cite{2008ApJ...679....1K}. It is worth noticing that even though
the available number of supernovae will drastically increase in the future, it  might
be hard to improve the constraints at such a high level because of
the integral dependence of the luminosity distance on the skewness
parameters. Therefore the cosmic parallax seems to be an ideal candidate for
testing the anisotropically stressed dark energy.
In this analysis two main experimental approximations were assumed. The first
one was to consider the experimental covariance matrix as diagonal. The
off-diagonal correlations is still under investigation by Gaia collaboration and
once these are provided the Fisher formalism would naturally include them in the
final limits. The second approximation was that only the statistical errors were
considered in our Fisher analysis. We refer to Section.~\ref{subsec:parobs} for
discussion on possible systematic effects such as peculiar velocity of the
objects or aberration change induced by our own motion.

Before concluding we comment briefly on the possibility of testing anisotropy
through the accumulated effect on distant source (galaxies, quasars,
supernovae) distribution. The number density of quasars will change so that
the number counts should show some level of anisotropy.
If sources shifts by  0.1$\mu$as/year during the dark energy
dominated regime, then the accumulated shift will be of the order
of 1 arcmin in $10^{9}$years and up to fraction of a degree in the
time from the beginning of acceleration to now. If the initial distribution
is isotropic, this implies that sources in one direction will be denser
than in a perpendicular direction by roughly $1/90\approx10^{-2}$ \cite{2009PhRvD..80f3527Q}.
This anisotropy  might be seen as a large-scale feature on the angular
correlation function of distant sources, where we expect any intrinsic
correlation to be negligible. The Poisson noise become negligible
for $N\gg10^{4}$: for instance, a million quasars could be sufficient
to detect the signal. Although the impact of the selection procedure
and galactic extinction is uncertain, this crude calculation
shows that the real-time effect could be complemented by standard
large-scale angular correlation methods.%

\section{Peculiar Acceleration }
\label{sec:pec-accel}

In Sections~\ref{sec:drift} and \ref{sec:parallax}  we reviewed the time
evolution of the cosmological redshift signal (the redshift drift) and of  the overall  angular
separations between distant sources (the cosmic parallax), respectively. These are both cosmic
signatures of a background  expansion: the first one is a perfect tool to
probe late time acceleration, while the second one tracks a possible anisotropic
expansion.

Many spurious effects might spoil these cosmological
signals: Earth rotation, relativistic corrections, the
acceleration of the Sun in the Galaxy and peculiar motion of the source.
Of particular interest is the redshift variation induced by linear perturbations, which -- in analogy with the peculiar velocities arising from the same physical mechanism -- we dub {\em peculiar acceleration}. This signal may contain additional information on the matter clustering amplitude. Also worth investigating is the acceleration field in bound structures (galaxies or clusters), which can be used as a probe of the local
gravitational field and can be a measure of the mass inside structures and
help distinguishing between competing gravity theories
\cite{2008PhLB..660...81A,2008MNRAS.391.1308Q}.

To be fair, the practical observability of these effects is still to be assessed in any detail. The main question is to select  convenient targets  to measure
 the high resolution redshifts. A possible candidate is the 21cm lines of neutral hydrogen clouds:
future experiments like SKA plan in fact \cite{2004NewAR..48.1095C} to measure redshifts with an error of the order of $10^{-6}$.
Whether this precision, combined with  SKA's  high angular resolution, and possibly  other atomic or molecular lines,  is enough to reach the detection threshold remains to be seen.

\subsection{Peculiar redshift drift in linear approximation}

\label{linear}

The redshift  is by definition the ratio of the frequency of photons measured
at the observer position (O) and the emission position (E). We suppose
that the emitter has world line E with unit four velocity $u_{O}^{a}$
and the observer has world line O with four velocity $u_{E}^{a}$.
If the light ray is an affinely parameterised null geodesic with tangent
vector $l^{a}$, then the expression for the redshift can be written
as
\begin{equation}
1+z=\frac{g_{\alpha\beta}u_{E}^{\alpha}k^{\beta}}{g_{\alpha\beta}u_{O}^{\alpha}k^{\beta}},\label{redshift}
\end{equation}
where $k^\beta$ is the photon 4-momentum.
In \cite{2008PhRvD..77b1301U} the authors expressed the redshift
at first order in perturbations assuming a Newtonian gauge
$g_{\alpha}=-a^{2}(1+2\phi)d\eta^{2}+a^{2}(1-2\psi)\delta_{ij}dx^{i}dx^{j}$
and neglecting the effect of gravity waves. The resulting expression
for the redshift is
\begin{equation}
(1+z)=\frac{a(\eta_{O})}{a(\eta_{E})}\Big[1+[\phi+e^{i}v_{i}]_{E}^{O}-\int_{E}^{
O}(\phi'+\psi')d\eta\Big],\label{redshift2}
\end{equation}
 where $e^{i}$ and $v^{i}$ represent the spatial components of $l^{a}$
and $u^{a}$, respectively, and prime refers to a derivative with
respect to the conformal time. Eq.~\ref{redshift2} is clearly composed
by adding up the cosmological redshift to the gravitational redshift
expressed in this gauge. The variation of conformal time at E is associated
to the variation of conformal time at O with a relation that takes
into account the difference between cosmic and proper time due to
the motion of the observer and the emitter, and it reads:
$\delta\eta_{E}\simeq[1+\mathbf{e}\cdot(\mathbf{v_{E}}-\mathbf{v_{O}})]
\delta\eta_{O}$.
The final expression for the redshift drift in this coordinate system
is the following
\begin{equation}
\dot{z}=\dot{\bar{z}}(\eta_{O},z)+\zeta(\eta_{O},z,\mathbf{e},\mathbf{x_{O}}),
\label{driftpert}
\end{equation}
 where $(\mathbf{x_{O}},\eta_{O})$ is the location of the observer,
the dot refers to derivative with respect to observer proper time
and the first order redshift drift in metric perturbations and $v$
is
\begin{equation}
\zeta(\eta_{O},z,\mathbf{e},\mathbf{x_{O}})=-\phi_{O}\dot{\bar{z}}(\eta_{O},
z)+(1+z)\big[\mathbf{e}\cdot\mathbf{\dot{v}}-\dot{\psi}\big]_{E}^{O}.
\label{driftpert2}
\end{equation}
 Notice that the term $\mathbf{e}\cdot\mathbf{\dot{v}}$ is the line
of sight peculiar acceleration of the source under consideration.
The first term in Eq.~\ref{driftpert2} is clearly generated by the
gravitational potential at the observer location, while the second
involves respectively the variation of the Doppler effect due to the
relative motion of the observer and the source
($\zeta_{\dot{v}}=(1+z)[\mathbf{e}\cdot\mathbf{\dot{v}}]_{E}^{O}$)
as well as the equivalent of the integrated Sachs-Wolfe effect of
CMB due to temporal variation of $\psi$ between E and O
($\zeta_{\dot{\psi}}=(1+z)[\dot{\psi}]_{E}^{O}$).

On sub-horizon scales $\psi=\phi$ and
$\nabla^{2}\phi=\frac{3}{2}H^{2}\Omega_{m}(z)a^{2}\delta$,
where $\Omega_{m}(a)$ is the time dependent matter density parameter
and the matter density contrast $\delta$ is proportional to the growth
factor $D(t)$. Picking up the growing mode $D_{+}(t)$, the time
evolution of the gravitational potential reads
\begin{equation}
\dot{\phi}=H\phi[f(t)-1],\label{potevol}
\end{equation}
 where $f(t)=d\log D_{+}/d\log a$. Using this expression one can
easily couple the root mean square fluctuations of $\dot{\phi}$ to
$\sigma_{\phi}$, which in turn are related to the matter density
fluctuations $\sigma_{\delta}(z)=\int\frac{d^{3}k}{2\pi^{3}}P_{\delta}(k,z)$,
where $P_{\delta}(k,z)$ is the matter power spectrum. Assuming that
$f\simeq1$ at the time of the emission and $t=0$ is the time of
observation, the root mean square redshift drift fluctuations caused
by the integrated Sachs-Wolfe effect is the following \cite{2008PhRvD..77b1301U}
\begin{equation}
<\zeta_{\dot{\phi}}^{2}>^{1/2}=(1+z)H_{0}[f(0)-1]\sigma_{\phi}(0).\label{sw}
\end{equation}
 In cosmological linear theory it is also possible to express the
velocity gradients $\theta(x,t)=\partial_{i}v_{i}/aH$ in terms of
the density fluctuations, such that $\theta(x,t)=-f(t)\delta(x)$.
Since the local linear peculiar acceleration is
$\dot{v_{i}}=-Hv_{i}-\partial_{i}\phi/a$
and the Fourier components of the velocity density contrast satisfy
$k^{2}Hv_{i}=-f(t)H^{2}ak_{i}\delta_{k}$, then the total root mean square
fluctuations on the
redshift drift due to the variation of the Doppler is the combination
of two terms, at the observer and at the emission location respectively
\begin{eqnarray}
<\zeta_{\dot{v}}^{2}>^{1/2}(z) & = &
[<\zeta_{\dot{v},O}^{2}>(z)+<\zeta_{\dot{v},E}^{2}>(z)]^{1/2}\label{doppler}\\
<\zeta_{\dot{v},O}^{2}>^{1/2}(z) & = &
(1+z)\Big[\frac{3}{2}\Omega_{m0}-f(0)\Big]H_{0}^{2}\hat{\sigma}_{\delta}
(0)\nonumber \\
<\zeta_{\dot{v},E}^{2}>^{1/2}(z) & = &
\Big[\frac{3}{2}\Omega_{m}(z)-f(z)\Big]H(z)^{2}\hat{\sigma}_{\delta}(z),
\nonumber
\end{eqnarray}
 with
$\hat{\sigma}_{\delta}(z)=\int\frac{d^{3}k}{2\pi^{3}}P_{\delta}(k,z)/k^{2}$.

\begin{figure}[ht!]
 \includegraphics[width=9cm]{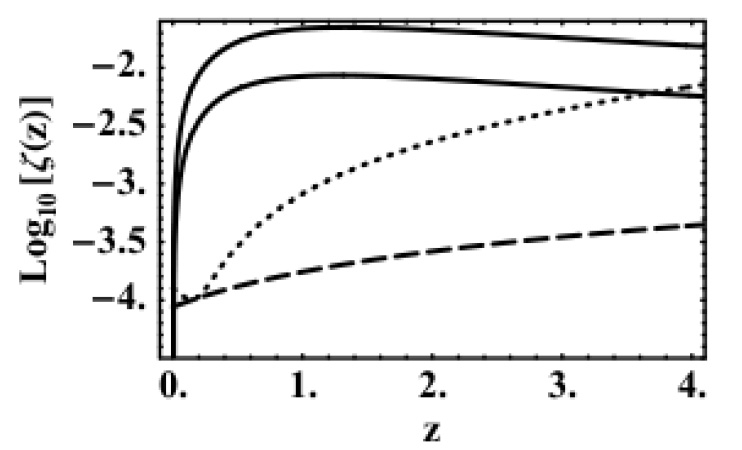}
\caption{\small Amplitude of the r.m.s. of the systematic errors
$\zeta_{\dot{v}}$ due to cosmic acceleration effects. The
contribution of $\zeta_{\dot{v},O}$ (dashed line) is subdominant compared
to the one of $\zeta_{\dot{v},E}$ (dotted line). The solid lines
represents the difference between a standard $\Lambda$CDM model
and cosmological models with either $w = -0.95$ (upper solid line)
or $w = -0.98$ (lower solid line). From Ref.~\cite{2008PhRvD..77b1301U}.}
 \label{fig:linear}
\end{figure}

Assuming the explicit function $f(t)$ for a simple flat $\Lambda$CDM
model and the prescription by \citet{1986ApJ...304...15B}
to estimate the matter power spectrum (and
$\hat{\sigma}_{\delta}(z)=\hat{\sigma}_{\delta}(0)$),
in \cite{2008PhRvD..77b1301U} the authors presented the dependencies
of the peculiar acceleration contribution to the redshift drift fluctuations
compared to differences in cosmological signal between a $\Lambda$CDM
model and a model with dark energy constant equation of state
(Fig.~\ref{fig:linear}).
In this particular models the dominant component is $\zeta_{\dot{v},E}$
which rises at a percent level up to $z=4$. In contrast, they find
that the Sachs-Wolfe redshift variance is much smaller, of order
$<\zeta_{\dot{\phi}}^{2}>^{1/2}\simeq(1+z)H_{0}\cdot10^{-5}$
(cfr Eq.~\ref{sw}). This means that the ratio of $\zeta_{\dot{v},E}$ and $\zeta_{\dot{\phi}}$ to the cosmological signal is  about $10^{-2}$ and $10^{-5}$, respectively. Obviously, this might not hold for other cosmological
models.

\subsection{Peculiar redshift drift in non-linear structures}

\label{nonlinear}

At linear level the peculiar acceleration signal acts as a noise over
the cosmological signal and one needs a large number of sources (and
spectral features) to average the latter out. At small distances however
the peculiar field dominates over the cosmological one and can be observed directly in objects
where the matter density contrast has turned non-linear. In two
papers \cite{2008PhLB..660...81A,2008MNRAS.391.1308Q} it was shown
that the peculiar acceleration in nearby clusters and galaxies is
in fact of the same order of magnitude of the cosmological signal
at larger distances and could be measured with the same instrumentation.

Moreover, the peculiar acceleration field can help
solving one of the most outstanding issues in astrophysics and cosmology,
namely the measure of the total mass of clustered structures. The most frequently used methods rely on kinematic measurements,
where the velocity dispersion of some suitable class of test particles
is used to infer the virial mass of the object. (e.g. the rotation
curves of spiral galaxies and the velocity dispersion of elliptical
galaxies). This of course requires the assumption of virialization. The same is true
for methods based on $X$-ray emission from the hot intra-cluster gas
confined by the gravitational potential.
On the other hand, as we will show in the next two subsections,
 the measurement of peculiar acceleration acceleration
(i.e. the local peculiar redshift drift) is a direct probe
of the acceleration field and does not assume virialization.

\begin{figure*}[h]
 \includegraphics[width=6.cm]{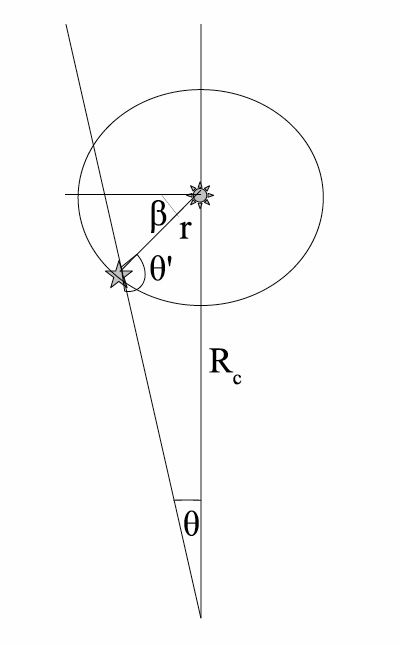}
  \includegraphics[width=10cm]{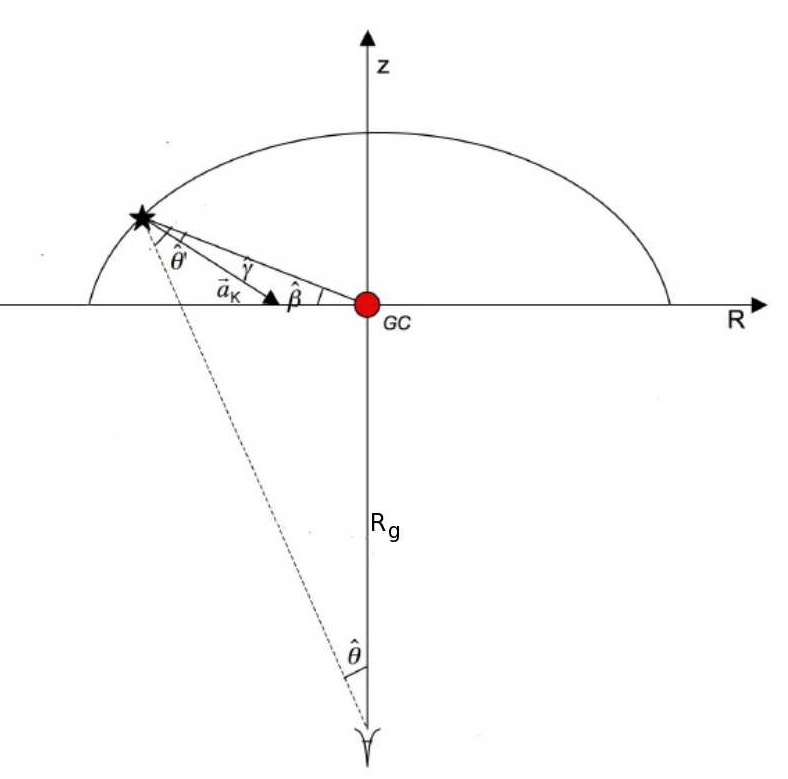}
\caption{{\small Definition of geometric quantities used in the text. {\it Left}: a schematic diagram of a galaxy cluster geometry. $R_c$ is the distance of the cluster center from the observer, while $r$ is the distance of a test particle from the cluster center. {\it Right}: Kuzmin potential in cylindric coordinates for a disc galaxy seen from an observer at distance $R_g$ from the galaxy center (GC). The disc lies edge-on on the axis $R$. The curved contour shows an equipotential curve and $\vec{a}_{K}$ represents the acceleration experienced by a test particle having galactic latitude $\hat\beta$. Note that the acceleration does not point towards the galaxy center. }}
 \label{fig:geom}
\end{figure*}

\subsubsection{Galaxy clusters}
\label{sec:clusters}
Let us consider a particle at the spherical-coordinate position
$(r,\alpha,\gamma)$
in a system centered on a cluster. For a schematic representation of angle and
distances we refer to the left diagram of Fig.~\ref{fig:geom}. The peculiar
acceleration of a particle
reads
\begin{equation}
\vec{a}=\vec{\nabla}\Phi=\Phi_{,r}\hat{r}+\frac{1}{r\sin\gamma}\Phi_{,\alpha}
\hat{\alpha}+\frac{1}{r}\Phi_{,\gamma}\hat{\gamma}
\label{pecacc}
\end{equation}
where the tilded quantities are unit versors and $\Phi$ is the gravitational
potential (Earth's local acceleration is assumed to be properly subtracted). The
acceleration along the line of sight
unit vector $\hat{s}$ is then
\begin{equation}
a_{s}=\hat{s}\cdot\vec{\nabla}\Phi.
\label{pecacc2}
\end{equation}
For simplicity  the potential is assumed to be spherically symmetric,
namely $\Phi=\Phi(r)$. Therefore the line of sight acceleration is
as follows:
\begin{equation}
a_{s}=\cos\theta'\Phi_{,r},
\end{equation}
 where $\theta'$ is as in Fig.~\ref{fig:geom}. In the same figure
 $R_{c}$ is defined as the cluster distance from observer, $r$ as
the particle distance from cluster's center and $\theta$ the viewing
angle. For small viewing angles $\theta$ (i.e. for $r\ll R_{c}$)
we have that the peculiar acceleration along the line of sight is
\cite{2008PhLB..660...81A}
\begin{equation}
a_{s}=\cos\theta'\Phi_{,r}\approx\sin\beta\Phi_{,r}|_{r=R_{c}\theta/\cos\beta}
\label{accapprox}
\end{equation}
 and under the asumption of  spherical symmetry it follows that
 \begin{equation}
a_{s}\approx\sin\beta\frac{GM(r)}{r^{2}},
\end{equation}
 where $r\equiv R_{c}\theta/\cos\beta$. The assumption of spherical symmetry simplifies the treatment and allows to use the
NFW profile for the cluster density but it is by no means crucial to the
argument and does not imply any assumption on the dynamics, i.e. on the degree of virialization.

Obviously,  in nearby clusters of galaxies the peculiar acceleration
signal-to-noise should be higher, if we consider the cosmological redshift drift
as noise. In
a cluster the dark matter halo gives the main contribution to the gravitational potential . In Ref.~\cite{2008PhLB..660...81A} a
Navarro-Frenk-White matter density profile is assumed
\begin{equation}
\rho(r)=\frac{\delta_{c}\rho_{cr}}{\frac{r}{r_{s}}(1+\frac{r}{r_{s}})^{2}},
\label{NFW}\end{equation}
 where $r_{s}=r_{v}/c$ sets the transition scale from $r^{-3}$ to
$r^{-1}$, $c$ is a dimensionless parameter called the concentration
parameter, $\rho_{cr}=3H_{0}^{2}/8\pi G$ is the critical density
at the redshift of the halo,
$r_{v}$ is the virial radius inside which the mass density equals
$\Delta_{c}\rho_{cr}$ and $\delta_{c}$ is the characteristic overdensity
for the halo given by
\begin{equation}
\delta_{c}=\frac{C\Delta_{c}c^{3}}{3},\label{deltac}
\end{equation}
 where
 \begin{equation}
C=\Big[\log(1+c)-\frac{c}{1+c}\Big]^{-1}.
\end{equation}
In addition, $\Delta_{c}$ is the nonlinear density contrast for
a virialized object and enters the expression for the mass $M_{v}=4/3\pi
r_{v}^{3}\Delta_{c}\rho_{cr}$.
Its value depends on the cosmological model and assuming a $\Lambda$CDM
was set to $\Delta_{c}=102$ \cite{1996MNRAS.282..263E}.

Then the mass associated with the radius $r$ is
\begin{equation}
M(r)=M_{v}C\Big(\log{(1+\frac{r}{r_{s}})}-\frac{\frac{r}{r_{s}}}{1+\frac{r}{r_{s
}}}\Big)
\label{mass}
\end{equation}
 and consequently
 \begin{equation}
\Phi(r)_{,r}=\frac{GM_{v}}{r_{s}^{2}}C\Big(\frac{\log{(1+\frac{r}{r_{s}})}}{
(\frac{r}{r_{s}})^{2}}-\frac{1}{\frac{r}{r_{s}}(1+\frac{r}{r_{s}})}\Big).
\label{pot}
\end{equation}
 Considering a time interval $\Delta t$ the velocity shift of the
particle test due to the peculiar acceleration along the line of sight
is $\Delta v=a_{s}\Delta t$ and turns out to be
\begin{eqnarray}
\Delta v&=&\frac{GM_{v}}{r_{s}^{2}}C\Delta
t\sin{\beta}\Big(\frac{\log{(1+\frac{r}{r_{s}})}}{(\frac{r}{r_{s}})^{2}}-\frac{1
}{\frac{r}{r_{s}}(1+\frac{r}{r_{s}})}\Big)\nonumber \\
\label{pot}
&=&0.44\frac{\textnormal{cm}}{\textnormal{sec}}\sin\beta\frac{\Delta
t}{10yr}\frac{M_{v}}{10^{14}M_{\odot}}(\frac{r_{s}}{1\textnormal{Mpc}})^{2}
C\Big(\frac{\log{(1+\frac{r}{r_{s}})}}{(\frac{r}{r_{s}})^{2}}-\frac{1}{\frac{r}{
r_{s}}(1+\frac{r}{r_{s}})}\Big)
\label{eq:shift}
\end{eqnarray}
with $r=R_{c}\theta/\cos\beta$. For a typical cluster value of $C\approx1$,
it turns out therefore that the typical shift for a galaxy cluster
is of the order of 1 cm/sec , similar to the cosmological value at
$z\approx1$ \cite{2007MNRAS.382.1623B} . The maximal value
depends in  general on the separation angle $\theta$ between cluster center and galaxy;
for $\theta\to 0$ we obtain
\begin{equation}
\Delta v_{max}=0.44\frac{\textnormal{cm}}{\textnormal{sec}}\frac{\Delta
t}{10y}\frac{M_{v}}{10^{14}M_{\odot}}(\frac{r_{s}}{1\textnormal{Mpc}})^{2}\frac{
C}{2}.
\label{eq:shift2}
\end{equation}
If we can observe $\Delta v_{max}$ as a function of $\theta$ for a given cluster then
we may measure a combination of the cluster parameters $M_v,C,r_s$ and compare this with
the expected results from numerical simulations.

\begin{figure}[ht!]
\centering \includegraphics[width=8.5cm]{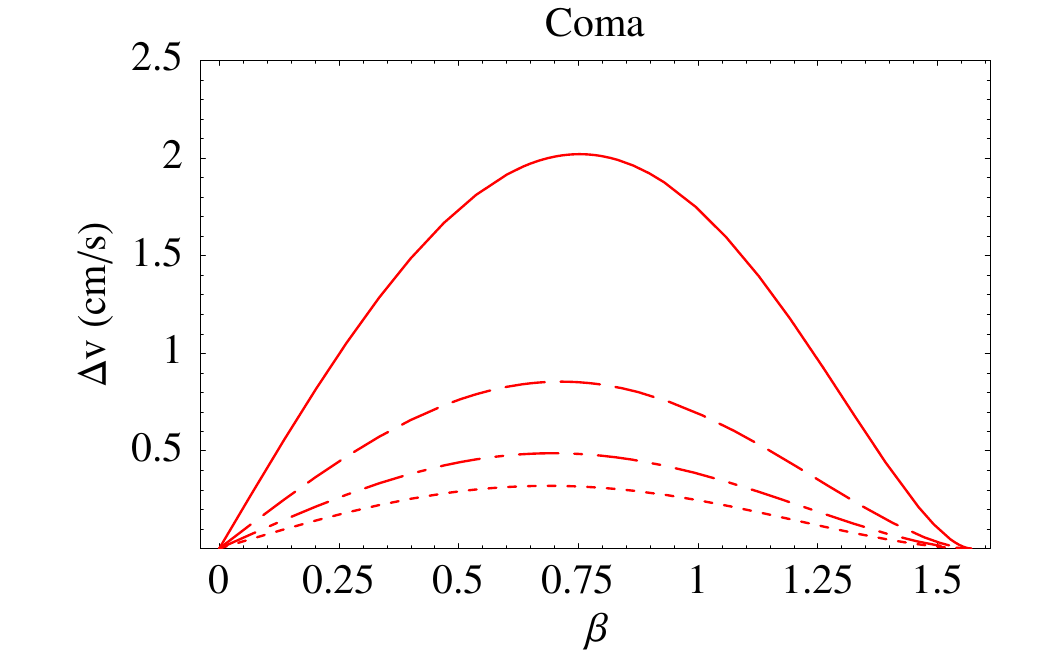}
\includegraphics[width=8.5cm]{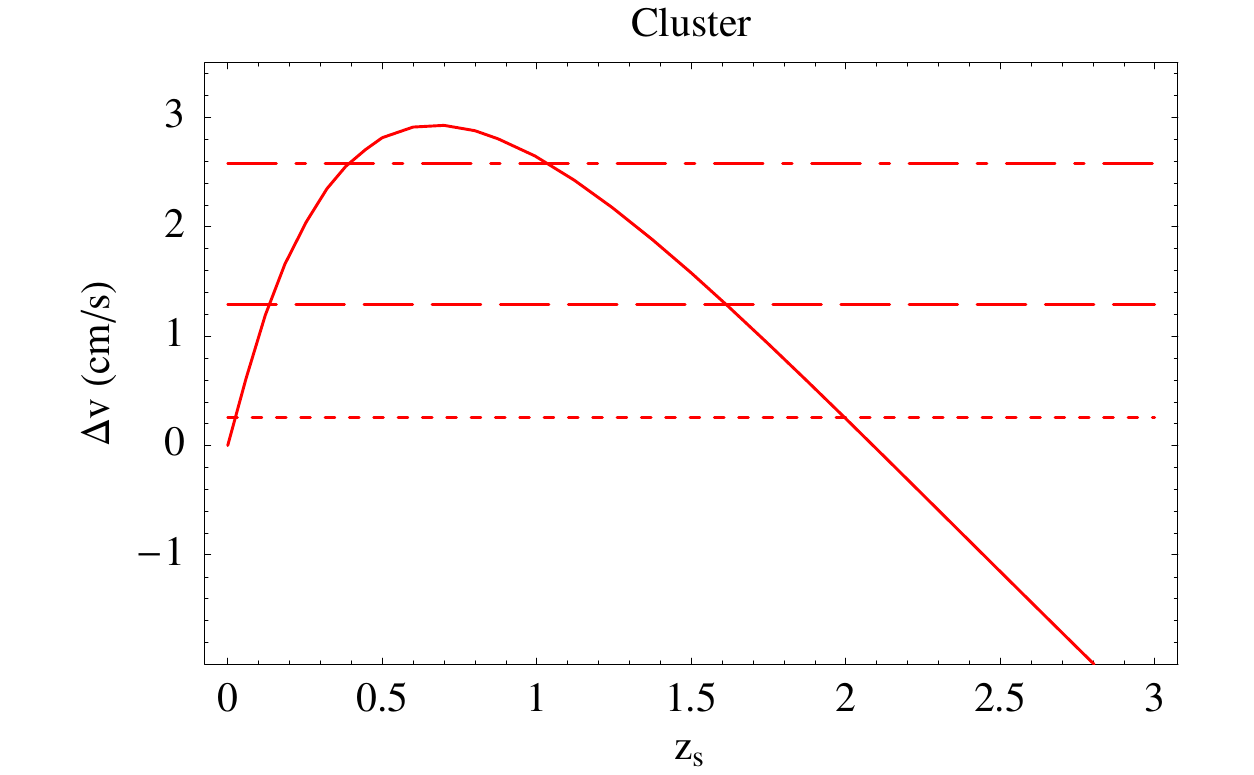}
\caption{{\small {\it Left}: predictions for $\Delta v$ in cm/sec for Coma
($R_{c}=100$Mpc,
$r_{v}=2.7$Mpc , $c=9.4$) for four values of $\theta$ up to $\theta_{max}=0.027$
equally spaced, starting from $\theta_{min}=\theta_{max}/4$ (top
to bottom). This value of $\theta_{min}$ corresponds to a radius
of $0.675$Mpc. {\it Right}: The cosmological velocity shift as a function of
redshift for a $\Lambda$CDM
model (solid line) and the maximum of the velocity shift due to the
peculiar acceleration for $\theta=\theta_{max}/4$ and three different
value of the mass of a rich cluster: $M=10^{14}M_\odot$ (short-dashed
line), $M=5\cdot10^{14}M_\odot$ (long-dashed line), $M=10^{15}M_\odot$
(long-short-dashed line). From Ref.~\cite{2008PhLB..660...81A}}}
\label{fig:coma}
\end{figure}

The maximum velocity shift of a cluster is plotted in the right-hand panel of
Fig.~\ref{fig:coma} together with the cosmological signal for a $\Lambda$CDM
with $\Omega_{\Lambda}=0.7$. The redshift at which the two signals are similar depends on the mass of the cluster
and goes from  very small
($z\simeq 0.02$) to $\approx 0.4$. In addition, there is a
second redshift, around $z\simeq 2$, at which
the cosmological signal is negligible. However observations at such
a distance could not be reliable, not only because of the difficulty
of precise measurements at this redshift, but also because the zero-crossing
of the function is cosmology-dependent. For massive and nearby clusters like Coma, the cluster peculiar
acceleration is therefore  the dominating effect. In principle of course the
cosmological signal could also be subtracted from the peculiar one
after averaging over a large number of galaxies.

Among the clusters,  Coma seems a very suitable candidate:
it is the most studied and best known cluster of galaxies,  it is almost
perfectly spherically symmetric, very massive ($M_{v}=1.2\times10^{15}M_{\odot}$),
and close to our Local Group ($R_{c}=100$Mpc, $z=0.02$ ; the cosmological
velocity shift is of the order of $10^{-1}$cm/sec at this redshift
\cite{2008PhLB..660...81A}).
Assuming, as above, a NFW density distribution, the other parameters
are: $r_{v}=2.7$ Mpc, $c=9.4$ and $r_{s}=0.29$ Mpc \cite{2003MNRAS.343..401L}.
The velocity shift predicted for Coma is shown in Fig.~\ref{fig:coma}.

\begin{figure}[ht]
 \centering
  \includegraphics[width=8cm]{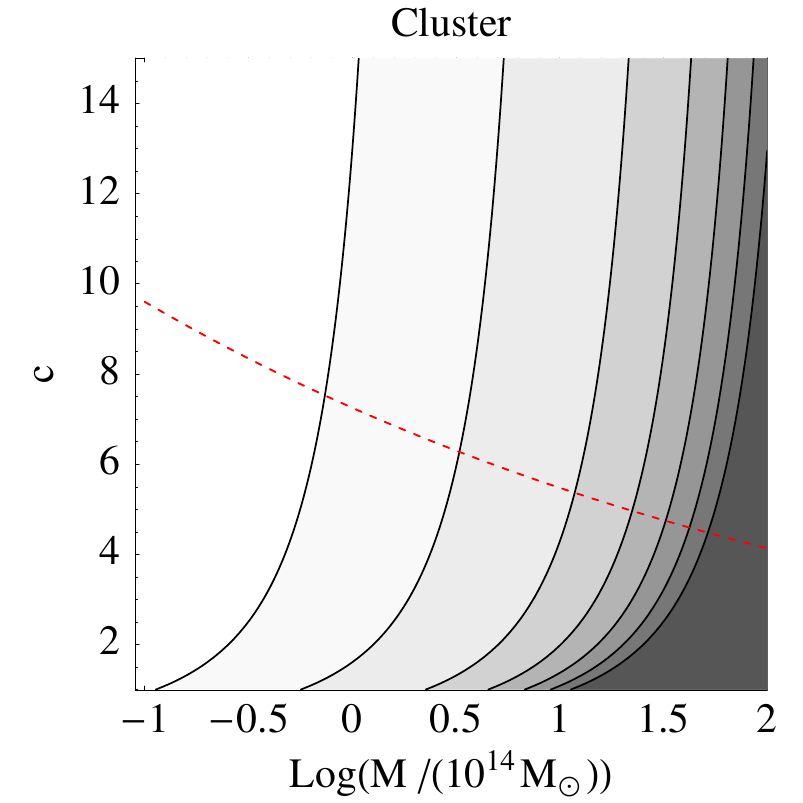}
\caption{{\small Contour plot for the velocity shift of a cluster at
$\theta_{max}/4$
as a function of the concentration $c$ and the mass $M$. Higher
values correspond to darker regions and the contours levels are (left
to right) $\Delta v=(0.1,0.5,2,4,6,8,10,15)$cm/s. The short-dashed
line is the fitting formula by \cite{2001MNRAS.321..559B}. From
Ref.~\cite{2008PhLB..660...81A}.}}

\label{fig:distr}
\end{figure}

In Fig~\ref{fig:distr} the contour curves corresponding
to the same $\Delta v_{max}$ for $\theta=\theta_{max}/4$ (and different
combinations of $c$ and the mass, fixing $r_s$ to the Coma value) are overplotted on a fitting formula  for the
concentration parameter as a function of the mass derived from N-body
simulations \cite{2001MNRAS.321..559B}. A
measurement of $\Delta v_{max}$ at different mass scales would provide then a
test of the concentration parameter fitting formula \cite{2008PhLB..660...81A}, i.e. of the
relation between $C$ and $M_v$.

Another observable is  the density distribution
of galaxies along a line of sight $\theta$
\begin{equation}
N(R_{c},\theta,s)=2\pi\rho_{g}\left(\frac{R_{c}\theta}{\cos\beta(\Delta v)}\right)\frac
{R_{c}^{3}\theta^{2}}{\cos^{2}\beta(\Delta v)}\frac{d\beta}{d\Delta v}d\Delta v,
\label{eq:obs3}
\end{equation}
where  $\rho_{g}$ is the radial distribution of galaxies
(possibly derived by the projected number density).
Comparing $N$ with the observed numbers one could
reconstruct $\rho (r)$ \cite{2008PhLB..660...81A}.

\subsubsection{Galaxy}
\label{sec:galaxies}
A similar analysis can be also applied to our own galaxy
\cite{2008PhLB..660...81A}. Here the effect of the peculiar acceleration is even stronger than for distant clusters
and,  following Ref. \cite{2008MNRAS.391.1308Q}, we can ask ourselves whether we can use  the Galaxy redshift drift as a test of gravity.
Different gravity theories, in fact,
produce a different acceleration field and this could in principle be compared to
observations.
 In Ref. \cite{2008MNRAS.391.1308Q}  the authors
modelled the galaxy following two competing gravity theories: a baryon disc
embedded in a dark matter halo (we call this model disk+CDM halo) and  the same disc subject to a different
Poisson equation beyond a certain scale (the MOND model \cite{1983ApJ...270..371M}, to be denoted as disk+MOND). The question
is whether  these two completely different
descriptions could be distinguished by the velocity shift as an observable,
given the same rotation curves. Of course the best target to do so is our own
Galaxy, where the known globular clusters in the stellar halo might be good test
particles.
The bulge component will be neglected in both scenarios, since its spherical
symmetry allows one to treat it simply as an additional contribution to the
total mass for scales outside the bulge. In this section for a schematic
representation of angles and distances in the galaxy we refer to the left
diagram of Fig. \ref{fig:geom}. The velocity shift is in both scenarios $\Delta v=a_{s}\cdot \Delta t$, where $a_{s}$ is the total line of sight acceleration caused by disk+CDM halo and disk+MOND, respectively.

\paragraph{Disk+CDM halo}
If a test particle (e.g.\ stars, gas) orbits on the disc, its peculiar acceleration
is affected only by the mass embedded within its distance from the centre, due to the symmetry of the distribution, and
it is the same as if this mass was totally concentrated in the centre.

The disc component can be modeled as the so-called Kuzmin disc, namely a disc
with superficial density
\begin{equation}
 \label{densi}
 \Sigma(r)=\frac{hM}{2\pi(R^{2}+h^{2})^{3/2}},
\end{equation}
where M is the total disc mass, $R$ and $z$ are cylindrical coordinates and $h$
is the scale length of the disc.
The two-parameter Newtonian gravitational potential outside the disc is
\begin{equation}
 \label{kuzpot}
 \phi_{K}=-\frac{MG}{[R^{2}+(|z|+h)^{2}]^{1/2}},
\end{equation}
and the equipotential surfaces are concentric spheres centered at $\pm h$ (see
left diagram of Fig.~\ref{fig:geom}). The test particle acceleration then reads
\begin{equation}
 \label{acc2}
 a_{K}=-\frac{MG}{[R^{2}+(|z|+h)^{2}]},
\end{equation}
with $|z|$ accounting for the acceleration field above and below the disc.  The
force field of a Kuzmin disc no longer converges towards the origin of axis.
Consequently, the projection angle $\theta'$ must be corrected by $\gamma$,
namely the angle between the acceleration and the radial direction $r$.
On the other hand the peculiar acceleration of a test particle outside the disc,
e.g.\ a globular cluster, is not only affected by the  gravitational potential
generated by the disc, but also by the possible presence of a dark matter halo
needed to explain the observed rotation curves.
In standard Newtonian mechanics  the halo potential is modelled as a logarithmic function
\begin{equation}
\label{halo}
\phi_{H}=\frac{1}{2}v_{0}^{2}\log{\Big(R_{c}^{2}+R^{2}+\frac{z^{2}}{q}\Big)},
 \end{equation}
where $R_{c}$ is the scalelength, $q$ is the halo  flattening ($q=1$ recovers
spherical symmetry) and $v_{0}$ is the asymptotic value of the velocity at large
radii.

The Kuzmin and the halo accelerations must be projected along the line of sight
and then added together. While for spherical symmetric logarithmic potential the
acceleration is radial and the angle between the line of sight and $r$ is
simply $\theta'$ (as in Fig.~\ref{fig:geom}), as mentioned before, the
projection angle for the  Kuzmin acceleration does not point towards the origin.
One clearly has $\cos\theta'\approx \sin\beta$. Using $r\cos\beta\simeq R_g
\theta$, $z\simeq\pm\sqrt{r^2-R^2}$ and $R\simeq R_g\theta$, the two line of
sight accelerations read \cite{2008PhLB..660...81A}
\begin{eqnarray}
a_{s,K}&=&\frac{MG}{R_{g}^{2}\theta^{2}\Big[1+\Big(|\tan{\beta}|+\frac{h}{R_{g}
\theta}\Big)^{2}\Big]}\sin{(\beta\mp\gamma)}\\
a_{s,H}&=&\frac{v_{0}^{2}R_{g}\theta\sqrt{1+\frac{\tan^{2}{\beta}}{q^{2}}}}{R_{c
}^{2}+R_{g}^{2}\theta^{2}(1+\frac{\tan^{2}{\beta}}{q^{2}})}\sin{\beta}.
\end{eqnarray}

\paragraph{Disk+MOND}

The Modified Newtonian Dynamics (MOND) paradigm was  first proposed by Milgrom
to reconcile discrepancies between general relativity and galaxy scale dynamics
\cite{1983ApJ...270..371M}.
As presented in its  first version, MOND by itself
violates conservation of momentum and energy. The subsequent Bekenstein-Milgrom
formulation
of MOND \cite{1984ApJ...286....7B} leaves the Newtonian law of motion intact and
modifies the standard Poisson equation for the
Newtonian gravitational potential as follows:
\begin{equation}
 \label{mond_pois}
 \nabla \Big[ \mu \Big( \frac{|\nabla \psi|}{a_{0}}\Big)\nabla \psi\Big]=4\pi G
\rho,
\end{equation}
where $\psi$  is the MOND gravitational potential, $\nabla^{2}\phi_{N}=4\pi G
\rho$, and $a_{0}$ is a scale that was estimated by Milgrom to be $a_{0} = 1.2
\cdot10^{-10}$m/s$^{2}$. The original shape of $\mu$ that helps rendering the
right profile of rotational velocities is $\mu(|a|/a_{0}) = 1$ for $|a|\gg
|a_0|$ and $\mu(|a|/a_{0}) = |a|/a_{0}$ for $|a| \ll  |a_0|$.

Eq.~\ref{mond_pois} is a non-linear equation which might be hard to solve
analytically, except for a class of symmetric configurations. In
\cite{Brada:1995}, a class of disc-galaxy models has been introduced for which
exact solutions of the MOND field equation exist.
Subtracting the usual Poisson equation from the MOND equation (\ref{mond_pois})
one has
\begin{equation}
 \label{mond_pois2}
 \nabla \Big[ \mu \Big( \frac{|\nabla \psi|}{a_{0}}\Big)\nabla \psi - \nabla
\phi_{N}\Big]=0.
 \end{equation}
For configurations with spherical, cylindrical or plane symmetry the relation
between the MOND  field and the Newtonian  field becomes
 \begin{equation}
 \label{mond_pois3}
 \mu \Big( \frac{|\nabla \psi|}{a_{0}}\Big)\nabla \psi = \nabla \phi_{N}.
 \end{equation}
 This equation permits a straightforward relation between the two potentials,
such that by assuming a matter density
distribution, one can solve the Poisson equation for the Newtonian potential and
then invert it to get the MOND one.

The function $\mu$ defined in MOND has been reformulated during the years,
moving from a step function as the
aforementioned type to, e.g., $\mu(x) = x/\sqrt{1+x^{2}}$. As pointed out in
\cite{Brada:1995}, the function $I(x) = x\mu(x)$ is therefore
monotonic and invertible, and the inverse is related to the function  $\nu(y) =
I^{-1}(y)/y$. Eq.~\ref{mond_pois3} now reads as
\begin{equation}
 \label{out1}
 \nabla \psi=\nu\Big( \frac{|\nabla \phi_{N}|}{a_{0}}\Big)\nabla \phi_{N}.
 \end{equation}

 The exact MOND solution is then given by Eq.~(\ref{out1}), with
 \begin{equation}
 \label{acc}
 \vec{a}_M=-\nabla\psi=a_{0}I^{-1}\Big(\frac{a_{N}}{a_{0}}\Big)\frac{\vec{a}_{N}
}{a_{N}}
   \end{equation}
where $a_M$ and $a_N$ are the MOND and Newtonian acceleration, respectively
\cite{2008MNRAS.391.1308Q}.

For $\mu(x) = x/\sqrt{1+x^{2}}$, then  $\nu(y) =[1/2 +\sqrt{y^{-2} + 1/4}]$ and
consequently
\begin{equation}
 \label{accmond}
 a_{M}=a_{K}\frac{\Big(1+\sqrt{1+\frac{4a_{0}^{2}}{|a_{K}|^{2}}}\Big)^{1/2}}{
\sqrt{2}}.
 \end{equation}

\begin{figure}[ht!]
  \centering
    \includegraphics[width=8.7truecm]{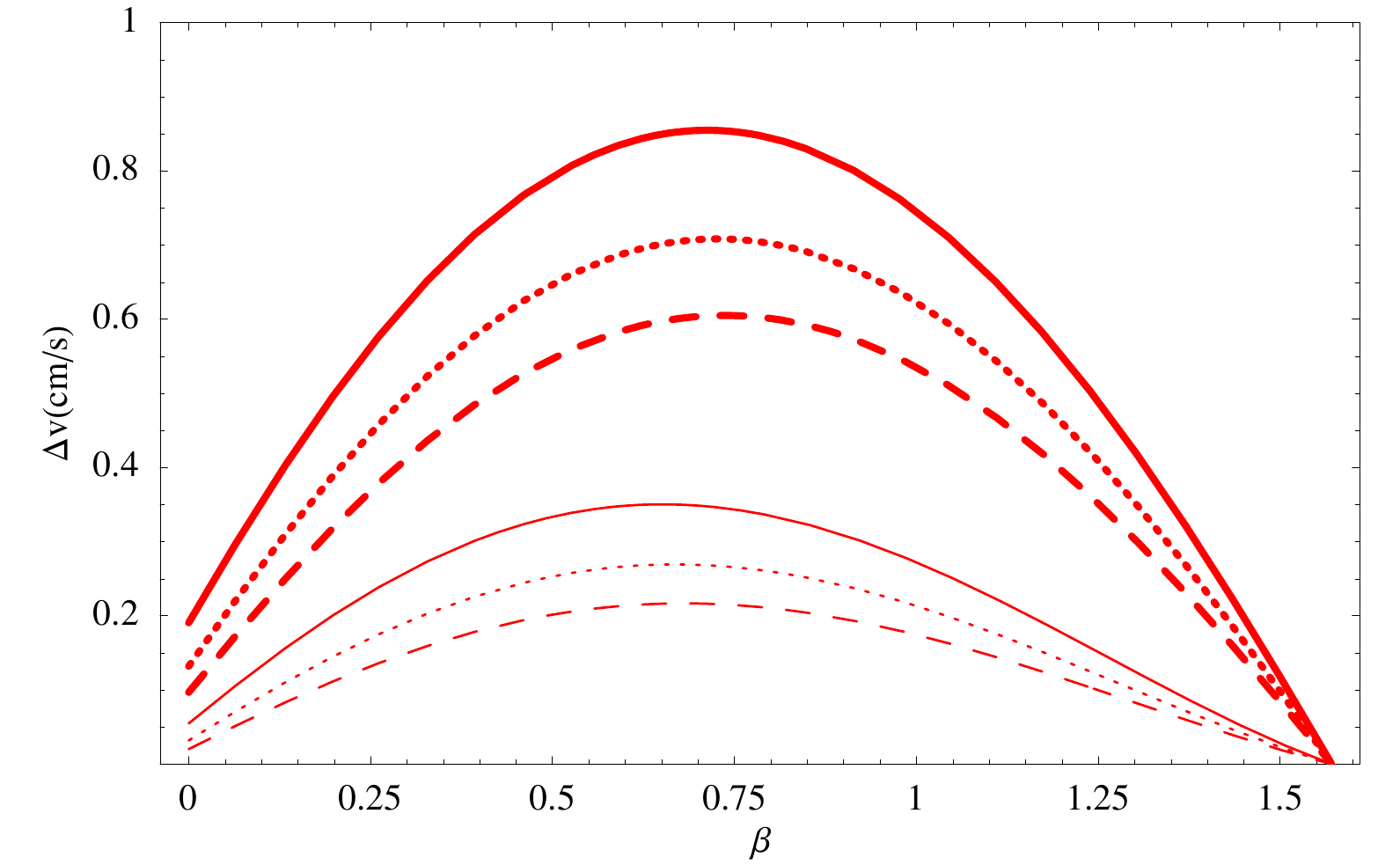}
       \includegraphics[width=8.7truecm]{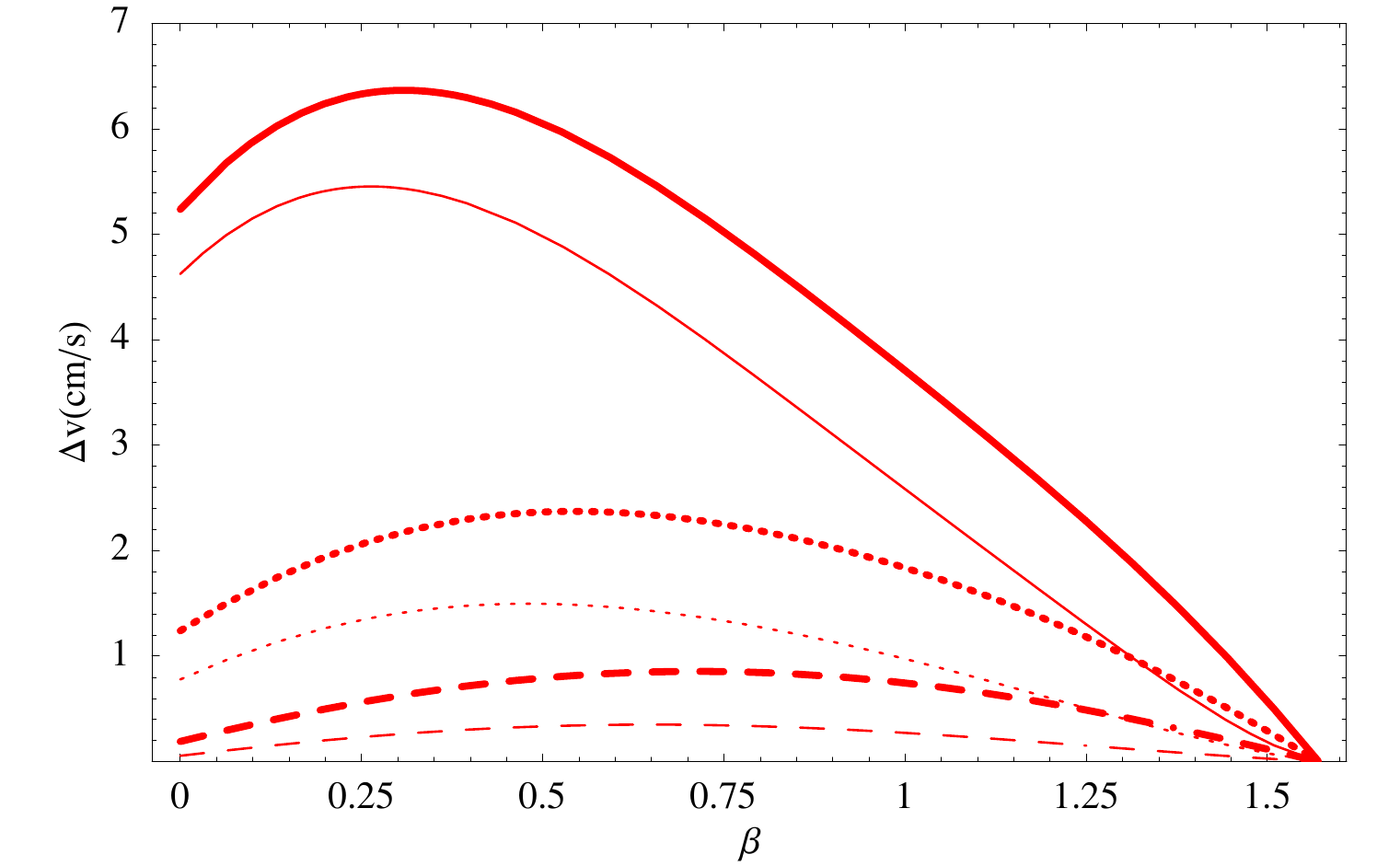}
    \caption{ \small  {\it Left}: velocity shift curves caused by peculiar
accelerations as a function of $\beta$ for Newtonian configuration (thin lines,
CDM halo+Kuzmin disc) and MOND (thick lines). For both configurations curves are
drawn for three different values of $\theta$, from top to bottom:
$\{0.05,0.06,0.07\}$, corresponding to distances from the centre $\{40,48,56\}$
kpc. {\it :Right}: same as the left panel, but  for three different values of
$\theta$, from top to bottom: $\{0.01,0.03,0.05\}$, corresponding to distances
from the centre $\{8,24,40\}$ kpc. From Ref.~\cite{2008MNRAS.391.1308Q}.}
    \label{fig:pec1}
\end{figure}

The expression for the peculiar acceleration in MOND, outside the disc, is as
follows:
\begin{eqnarray}
\label{accmond2}
 a_{s,M}= a_{M}\cdot \sin{(\beta \mp \gamma)},
 \end{eqnarray}
 with
 \begin{eqnarray}
 a_{M}=
\frac{MG\Big(1+\sqrt{1+\frac{4a_{0}^{2}}{M^{2}G^{2}}R_{g}^{4}\theta^{4}[1+(|\tan
{\beta}|+\frac{h}{R_{g}\theta})^{2}]^{2}}\Big)^{1/2}}{\sqrt{2}R_{g}^{2}\theta^{2
}\Big[1+\Big(|\tan{\beta}|+\frac{h}{R_{g}\theta}\Big)\Big]},
 \end{eqnarray}
 and
\begin{equation}
\label{gam}
\gamma=\arccos{\frac{r^{2}+x^{2}-h^{2}}{2 r x}};\qquad
x^{2}=r^{2}+h^{2}-2rh\sin{\beta}.
 \end{equation}
The different shape of the peculiar acceleration between Newtonian and MOND
configurations are
shown in Fig.~\ref{fig:pec1}. The predicted velocity shift reaches a maximum
value along the line of sight,  $\Delta v_{max}$, that depends on the
gravitational configuration. The maximum also depends on the distance from the
galactic centre, increasing with decreasing distance.  By estimating  $\Delta
v_{max}$ from a sample of test particles (e.g.\ globular clusters) outside the
disc of the galaxy along each line of sight, one might therefore hope to
distinguish the Newtonian and the MOND gravitational potentials.  In our Galaxy,
knowing precisely the coordinates of test particles one might trace out the
distribution of the signal.

In this case we clearly are internal observers. The projected acceleration of
the Sun on the disc, which has its maximun value on the order of a few cm/s, is
assumed to be substracted, and the cluster-centred acceleration of single stars
in globular clusters is assumed to be averaged out. The time scales are
short enough that the motion of the Milky way  within the Local Group and the
Local Group relative to Virgo, the Great Attractor  and other distant cosmic
structures can be considered  approximatively constant. In
\cite{2008MNRAS.391.1308Q} the velocity shift signal has been forecasted for
150 globular clusters catalogue in our Milky way.

\begin{figure}[t]
  \centering
    \includegraphics[width=8.5truecm]{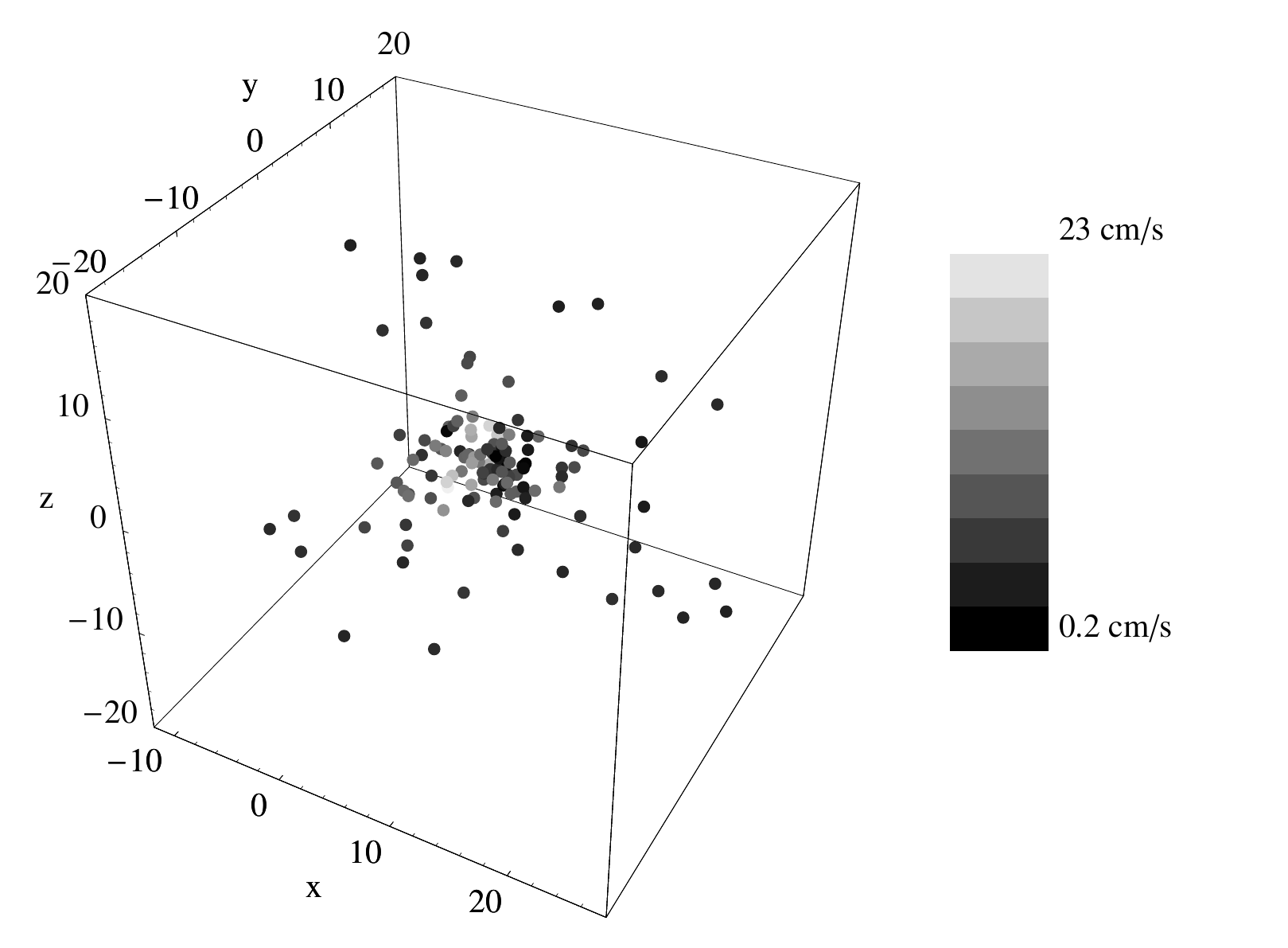}
 \includegraphics[width=8.5truecm]{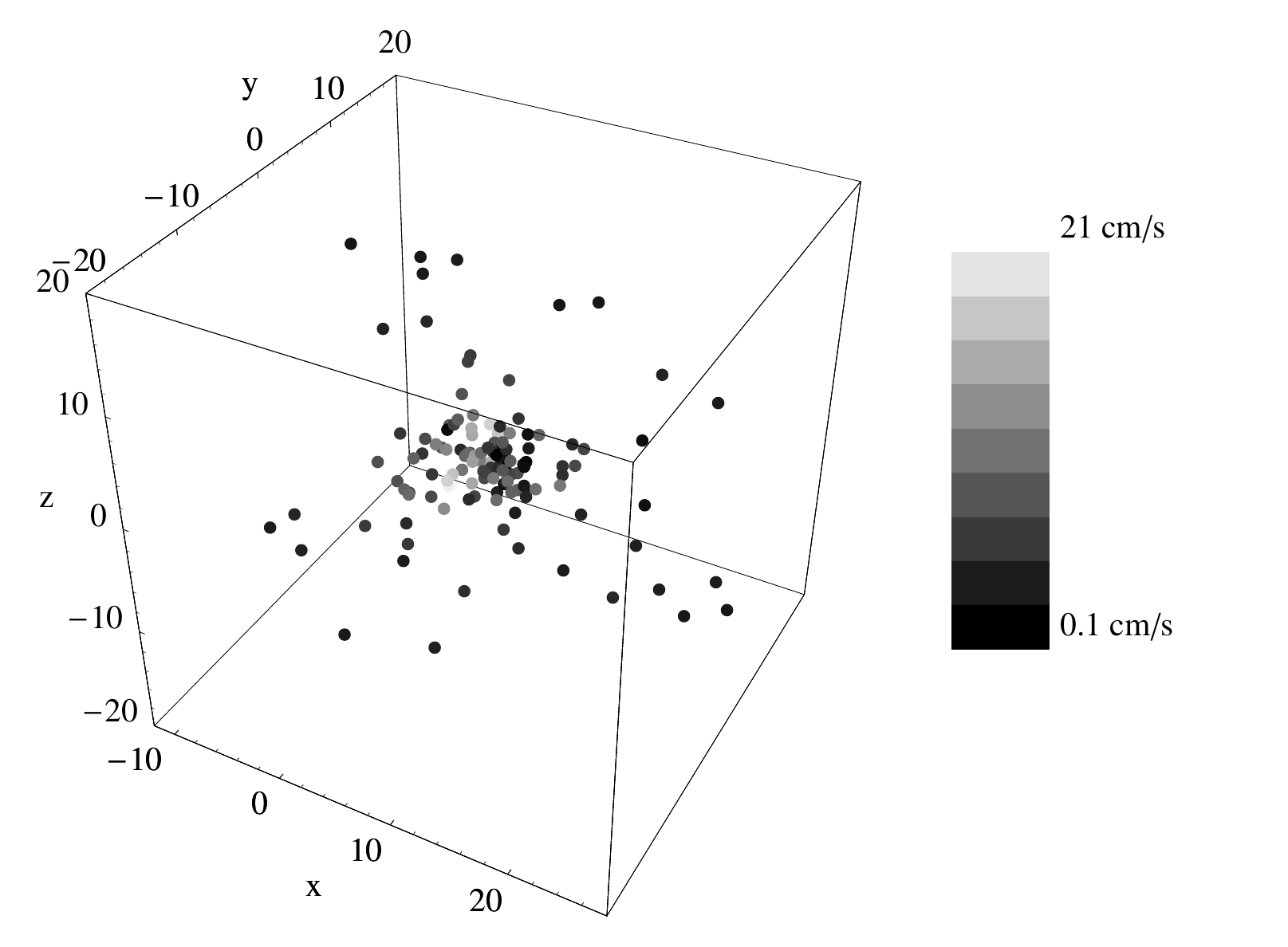}
  \includegraphics[width=8.5truecm]{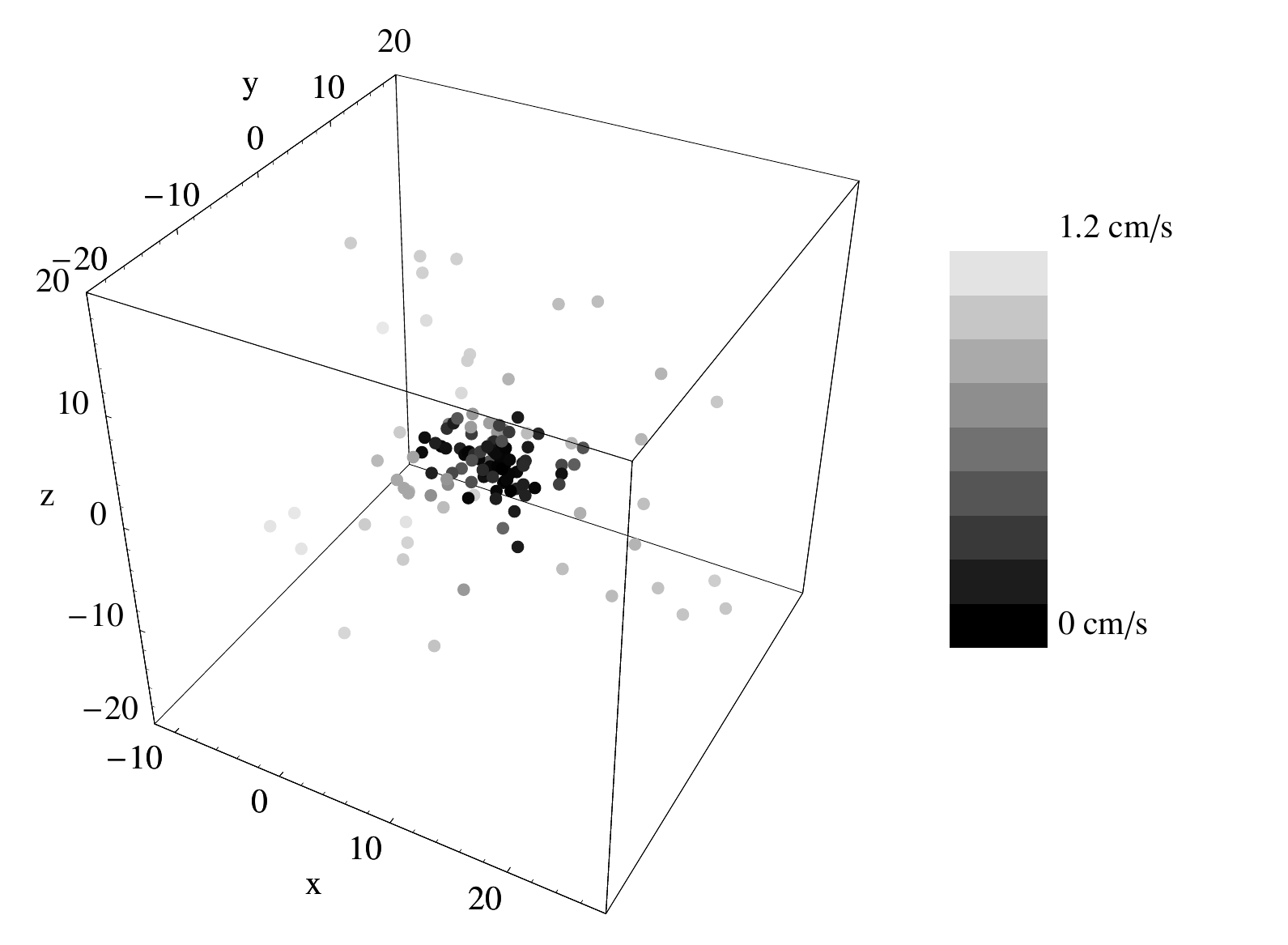}

    \caption{ \small The velocity shift signal (for $\Delta t=15$ years)
expected in the MOND configuration (top left), in the newtonian configuration
(top right) and for their difference (bottom) for the Milky Way globular
clusters taken from Harris (1996) \cite{1996AJ....112.1487H}, plotted in
cartesian coordinates (kpc units). From Ref.~\cite{2008MNRAS.391.1308Q}.}
    \label{fig:mapmond}
\end{figure}

 The effect of being off-centred observers  combined with the non-radial
direction of the acceleration in Kuzmin disc makes the pattern of the effect across the Galaxy
non trivial. However, while again the signal reaches the maximum value close to
the Galactic centre for both scenarios, the difference signal between the
two Galaxy models  is stronger at
high Galactic longitude (close to $\pi$) where the spherical CDM halo is more
important. This can be better seen by eye in Fig.~\ref{fig:mapmond}, where  the
distribution of the peculiar acceleration in Sun-centred cartesian coordinates
has been mapped out.
Unfortunately, the maximum difference is only of order 1 cm/s, making it quite
problematic to observe. However, supposing that such an accuracy will be
achieved in the future, it is interesting to note that the two different
scenarios result in a distinct morphology of the signal distribution, that could
be used as a signature to identify the actual potential (compare Ref.
\cite{2008MNRAS.391.1308Q} for a Mollweide projection of the signals in the sky).
Here the authors  also provided a top-ten list of the globular clusters exhibiting the
highest velocity shift signal and the first ten with  the highest difference
signal between the CDM halo configuration and MOND. The highest signal among the globular clusters
predicted for MOND configuration, CDM halo and their difference is $21.5$cm/s, $20.8$cm/s and $1.18$cm/s,
respectively (in 15 yrs).

The advantage of adopting globular cluster as test particles is related  to the
fact that they not only are bright, but more importantly composed of thousands
of stars and their position is fairly well known; nevertheless, in principle one
might consider  using other probes, like high velocity clouds that consist of neutral hydrogen high velocity interstellar regions or HI regions.
Moreover, one could also use the acceleration in the pulsar timing as another
probe of the acceleration field in the Galaxy (e.g. by measuring the Doppler shift in the pulsar timing).
The practical feasibility of such probes is however still to be demonstrated.

\section{Proper Acceleration }

\label{sec:prop-accel} 

As we mentioned in the Introduction, the fourth application of real-time
cosmology is to the velocity variations in the direction perpendicular to the line of sight (i.e.\ transverse). When this effect is observed in test particles in bound systems, for example within our own Galaxy, we call it \emph{proper acceleration} (since it represents a variation in the proper motion of the source) to distinguish it from the cosmic parallax.  Let us stress that the material presented in this section is set forth here for the first time and it is somewhat preliminary.

Let us assume that a body at distance $r_{s}$ from the Sun has transverse
velocity $v_{T1}$. Then during the time $\Delta t_{M}$, in which
an astrometric mission like GAIA is flying, its angular position will
change by $s_{1}=v_{T1}\Delta t_{M}/r_{s}$. Then another GAIA-like
mission $\Delta t_{S}$ years later will see the change $s_{2}=v_{T2}\Delta t_{M}/r_{s}$
(assuming the same duration for simplicity). If $v_{T2}\not=v_{T1}$
then we will see a proper acceleration. If $r_{s}$ changes also during
the interval $\Delta t_{S}$, we see also an apparent proper acceleration
(see below).

\begin{figure*}[t]
\centering
   \includegraphics[width=12cm]{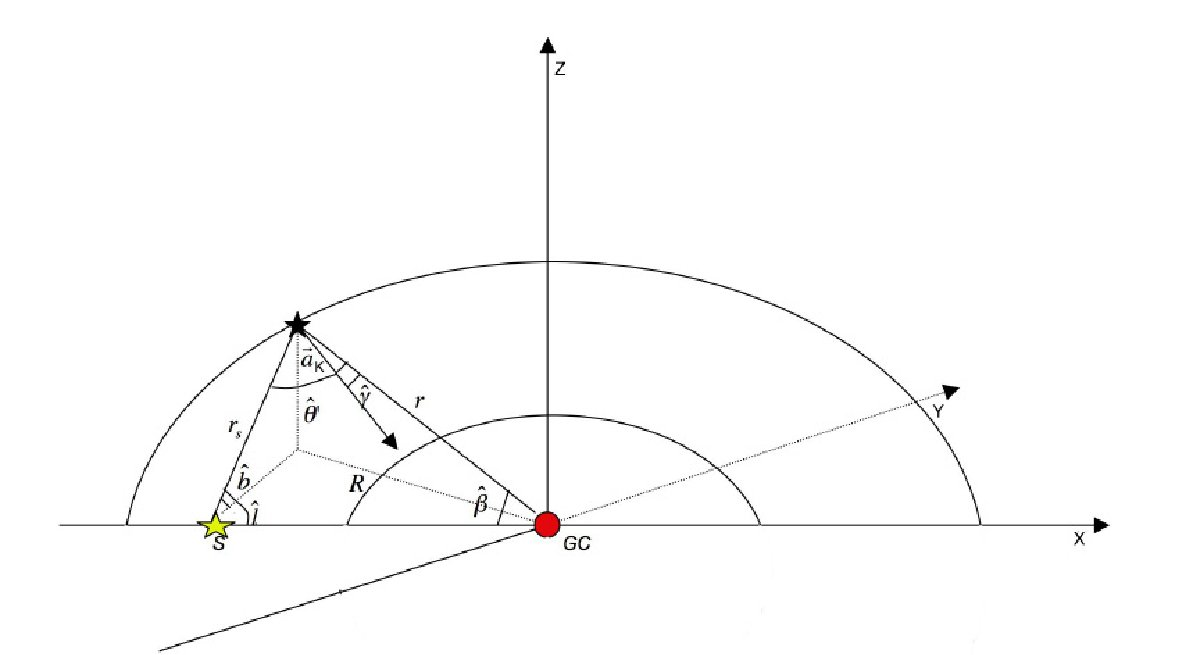}
       \caption{\small Kuzmin disc potential in cartesian coordinates in the Milky Way. The disc lies in the plane (x,y). The star (S) represents the Sun and angles are labelled with a hat. Curved contours show two equipotential levels; $\vec{a}_{K}$ represent the acceleration of a test particle at galactic latitude $\hat\beta$.}
\label{fig:diagram}
\end{figure*}

If the body has a (slowly varying) transverse acceleration $a_{T}$
we have $v_{T2}\approx v_{T1}+a_{T}\Delta t_{S}$ and therefore the
variation in proper motion will be \begin{equation}
\Delta_{t}s=s_{2}-s_{1}=a_{T}\Delta t_{M}\Delta t_{S}/r_{s}.\end{equation}
 We can produce a quick estimate of the effect assuming radial symmetry
around the galactic center. If the body lies in the direction $\vec{r}_{s}$
, is at distance $r$ from the galactic center, and if the galactic
center is in direction $\vec{r}_{c}$, then $r^{2}=r_{c}^{2}+r_{s}^{2}-2r_{c}r_{s}\cos\alpha$,
$\alpha$ being the angle between $\vec{r}_{c}$ and $\vec{r}_{s}$
(see Fig.~\ref{fig:diagram}). Then we may put
\begin{equation}
a_{T}=\frac{v^{2}}{r}\sin\theta'=\frac{v^{2}r_{c}}{r^{2}}\sin\alpha
\end{equation}
 $\theta'$ being the angle between the line of sight and $\vec{r}$
. Then we have
\begin{equation}
\Delta_{t}s=\frac{v^{2}\sin\alpha\Delta t_{M}\Delta t_{S}}{r_{c}^{2}+r_{s}^{2}-2r_{c}r_{s}\cos\alpha}\frac{r_{c}}{r_{s}}.
\end{equation}
 For a star at $r_{s}=100$pc, with $v=300$Km/sec and $\alpha=\pi/2$,
and adopting $\Delta t_{M}=5$yrs and $\Delta t_{S}=10$yrs, we obtain
$\Delta_{t}s\approx1\mu$as. Now suppose GAIA observes $N$ objects
with the same acceleration (e.g. within a cluster or in a small region
of space) with precision $P/\sqrt{N}$$\mu$as each. Therefore it
will be able to observe the proper acceleration if $\Delta_{t}s\ge P\mu{\rm as}/\sqrt{N}$
or equivalently
\begin{eqnarray}
\sqrt{N} & = & 0.86P\Big(\frac{r_{s}}{100{\rm pc}}\Big)\Big(\frac{300{\rm km/sec}}{v}\Big)^{2}\Big(\frac{10{\rm yr}}{\Delta t_{s}}\Big)\Big(\frac{5{\rm yr}}{\Delta t_{M}}\Big)\\
 & \times & \frac{1+(r_{s}/r_{c})^{2}-2(r_{s}/r_{c})\cos\alpha}{\sin\alpha},
 \end{eqnarray}
 where we assumed $r_{c}=8.33$kpc. It appears therefore that for
bodies in the disk near the Sun (e.g. an open cluster at 100pc) and
$\alpha\approx\pi/2$, the effect could be visible if $P\sim10\mu$as
and one has $\sim10^{2}$ stars to average, which seems not impossible.
To reach a distance of $d$kpc with astrometry precision $P\mu$as
one would need to average over roughly $100d^{2}P^{2}$ stars. In
order to realize the measurement with a single GAIA mission one should
put $\Delta t_{M}\approx2$yr and $\Delta t_{S}\approx5$yr. Then
one gets $N\approx2500d^{2}P^{2}$ which appears quite feasible.

The measurement will be subject to several source of errors. We consider
here in particular the apparent acceleration $\Delta_{t}s_{a}$ induced
by a change in $r_{s}$. We have \begin{equation}
\Delta_{t}s_{a}\equiv v_{T1}\Delta t_{M}(\frac{1}{r_{s2}}-\frac{1}{r_{s1}})\approx v_{T1}\Delta t_{M}\frac{\Delta r_{s}}{r_{s}^{2}}=v_{T1}v_{R}\frac{\Delta t_{M}\Delta t_{S}}{r_{s}^{2}}\end{equation}
 which is $r/r_{s}(v_{R}v_{T1}/v_{R}^{2}+v_{T1}^{2})$ times bigger
than the true acceleration. If $v_{R}\approx v_{T1}$ and the object
is $\approx100$pc from the Sun then $\Delta_{t}s_{a}/\Delta_{t}s\approx(r_{c}/r_{s})\approx100$).
A signal-to-noise of order unity is reached for $r_{s}\approx r_{c}$,
but then we would need roughly $10^{4}$ stars to average in order
to detect the effect. However the apparent acceleration can be measured
with good precision by estimating $v_{R}$ with Doppler measurements
and $r_{s}$ via parallax distance. The relative error on the estimate
of $\Delta_{t}s_{a}$ is given by summing the variance on the redshift
and the variance on the distance: \begin{equation}
\sigma_{a}^{2}=\sigma_{z}^{2}+\sigma_{r_{s}}^{2}\end{equation}
 GAIA will measure radial velocities with an error of 15 km/sec, which
amounts to $\sigma_{z}\approx10^{-1}$, and parallax distance with
an error smaller than 10\%, so again $\sigma_{r_{s}}\approx10^{-1}$.
Therefore we can estimate $\sigma_{a}\approx0.1$. Then we need to
have a ratio $\Delta_{t}s_{a}/\Delta_{t}s<10$ to ensure detection
of the true proper acceleration. This gives an lower limit of $r_{s}\ge1$kpc
roughly. Therefore we need roughly at least $10^{4}$ stars to average
over if $P=10\mu$as.

As in Section~\ref{sec:pec-accel} we can now try to see whether proper acceleration
can help distinguishing between the acceleration field induced by
Newtonian and MOND gravity. For a general gravitational potential
the signal is\begin{equation}
\Delta_{t}s=\frac{\Phi_{,r}\sin\alpha\Delta t_{M}\Delta t_{S}}{r_{s}[1+(r_{s}/r_{c})^{2}-2(r_{s}/r_{c})\cos\alpha]^{1/2}}.\end{equation}
 Adopting the configuration introduced in Sec.~\ref{sec:galaxies}
for a disk plus a CDM halo depicted in Fig.~\ref{fig:diagram} the transverse
signal we are after is then:
\begin{eqnarray}
\Delta_{t}s=(a_{t,K}+a_{t,L})\Delta t_{M}\Delta t_{S}/r_{s}.\label{prop
}\end{eqnarray}
 The Kuzmin and the halo accelerations must be projected on the transverse
plane and then added together. While for spherical symmetric logarithmic
potential the acceleration is radial and the angle between the line
of sight and $r$ is simply $\theta'$, the projection angle for the
Kuzmin acceleration, again, does not point towards the origin. The
two transverse accelerations read \begin{eqnarray}
a_{t,K} & = & \frac{MG}{[R^{2}+(|z|+h)^{2}]}\sin{(\theta'\pm\gamma)}\\
a_{t,L} & = & v_{0}^{2}\frac{\sqrt{R^{2}+\frac{z^{2}}{q^{2}}}}{R_{c}^{2}+R^{2}+\frac{z^{2}}{q^{2}}}\sin{\theta'}.\end{eqnarray}
 All the parameters of the gravitational potential will be fixed to
the values of \cite{Read:2005}: $M=1.2\cdot10^{11}M_{\odot}$, $h=4.5$
kpc, $v_{0}=175$ km/s, $R_{c}=13$ kpc and $q=1$.

Let us then write the contribution from proper acceleration to velocity
shift in terms of galactic coordinates, in particular $r_{s}$, the
galactic latitude $b$ and longitude $l$. First, we have to move
from cylindrical coordinates $(z,R,\varphi)$ to galactic coordinates
$(r_{s},b,l)$, and neglecting $\varphi$ due to the axisymmetry of
the potential, the transformation reads \begin{eqnarray}
z=r_{s}\sin{b};\qquad R=\sqrt{r_{c}^{2}+r_{s}^{2}\cos^{2}{b}-2r_{c}r_{s}\cos{b}\cos{l}},\label{transf}\end{eqnarray}

The line of sight corresponds to the direction of the distance from
the object to us, i.e. $r_{s}$. The projection angle $\theta'$ is
related to the new coordinate system as follows: \begin{equation}
\cos{\theta'}=\frac{r_{s}^{2}+r^{2}-r_{c}^{2}}{2r_{s}r},\label{transf}\end{equation}
 where $r^{2}=z^{2}+R^{2}$. In addition, as already mentioned, for
the Kuzmin potential the correction to the non-radial direction of
the acceleration must again be taken into account, so that the projection
angle changes to $\cos\left({\theta'\pm\gamma}\right)$, and \begin{equation}
\gamma=\arccos{\frac{r^{2}+x^{2}-h^{2}}{2rx}};\qquad x^{2}=r^{2}+h^{2}-2rh\sin{\beta}.\label{gam}\end{equation}
 with \begin{eqnarray}
\beta=\arccos{\frac{r^{2}+r_{c}^{2}-r_{s}^{2}}{2rr_{c}}}.\label{beta}\end{eqnarray}
Using these relations we can find the locations where the signal is
strongest. We find that the maximal newtonian signal is $0.003[\Delta t_{M}\Delta t_{S}/(5$yrs$^{2})]\mu$as
at $r_{s}=3$kpc, and decreases to $0.0002[\Delta t_{M}\Delta t_{S}/(5$yrs$^{2})]\mu$as
at $r_{s}=8$kpc. For the difference |MOND-newton| at the same distances we find $0.0018[\Delta t_{M}\Delta t_{S}/(5$yrs$^{2})]\mu$as and $0.00006[\Delta t_{M}\Delta t_{S}/(5$yrs$^{2})]\mu$as, respectively.


\section{Cosmic Microwave Background Radiation}
\label{sec:cmb}
Recently, several authors also entertained the possibility of detecting real-time variations of the CMB angular power spectrum  \cite{2007ApJ...671.1075L,2007PhRvD..76l3010Z,2008PhRvD..77d3505M}.  The future evolution of the temperature anisotropy power spectrum
 should be affected by three  effects: i) the amplitude should reduce due to the $1/a$ scaling of the mean  temperature; ii) the peaks and the minimum should shift to smaller scales, because the radius of the last scattering surface will grow and the same comoving size will be seen under a smaller viewing angle; iii) in presence of either curvature or dark energy the late integrated Sachs-Wolfe (ISW) signal would increase with respect of the first peak.
In addition our own velocity with respect to the CMB frame evolves, hence generating  a temporal change in the dipole.

As it will be clear later on, the prospect for the detection of this signal are at the moment rather weak.

\subsection{Monopole}
\label{sec:monopole}
The first mechanism responsible for the temporal change of the CMB anisotropies is the evolution in time of the mean temperature $T$. Assuming that at present time the CMB radiation energy density  is with a fairly good approximation  decoupled from the rest of the Universe, its conservation equation holds, i.e. $\dot{\rho}_{\gamma}=-4H\rho_{\gamma}$. Due to the Stefan-Boltzmann law the  energy flux density is directly proportional to the fourth power of the black body's thermodynamic temperature, $\rho_{\gamma}\propto T^{4}$, which leads to the evolution equation for the mean temperature:
\begin{equation}
\dot{T}=-HT.
\end{equation}
In \cite{2007PhRvD..76l3010Z} it has been shown how, assuming $H_{0}=73$km s$^{-1}$ Mpc$^{-1}$, this temperature will drop by $1\mu$K  in 5000 yr.

Indeed $T$ will become during the expansion  harder and harder to detect, but there is also a model dependent theoretical limit $T_{lim}=\sqrt{\Lambda/12\pi^{2}}$ below which the CMB will become dominated by the thermal noise of the de Sitter background.
Considering the scaling law of the radiation energy density $\rho_{\gamma}\propto (a/a_{0})^{-4}$ it turns out that we must wait until $a/a_{0}\sim 10^{30}$ before this fundamental limit has been reached; in a $\Lambda$CDM model this corresponds roughly to $t=1$Tyr \cite{2007PhRvD..76l3010Z}.
This monopole effect is clearly out of reach.

\subsection{Dipole}
\label{sec:dipole}

The observed dipole  in the CMB anisotropy power spectrum  is believed to be caused by the
Doppler effect arising from our peculiar velocity, $\boldsymbol{v}$ with respect to a frame of reference external to the Milky Way, for example, the International Celestial Reference Frame. As we have mentioned in the previous sections, due to the local acceleration, this peculiar velocity, and hence the dipole, is expected to evolve with time.

Today we can detect the modulation of the Earth's motion around
the Sun; in the future, with increasing satellite sensitivity, we may be
able to observe the Sun's motion around the Galaxy.  In \cite{2007PhRvD..76l3010Z} the authors assumed that  the motion of the
Sun around the Milky Way is simply a tangential speed
of $220$ km$\,{\rm s}^{-1}$ at a distance of $8.5$ kpc.  Using the current
observed value of $\boldsymbol{v}$ to infer the velocity of the Galactic
centre with respect to the CMB rest frame, the time dependent Sun-CMB
velocity vector then reads
\begin{eqnarray}
\boldsymbol{v}(t)=\bigg[222\sin(\frac{2\pi t}{T})
 -26.3,222\cos(\frac{2\pi t}{T}) - 466.6,\:275.6\bigg]
 {\rm km}\,{\rm s}^{-1}\,,
\label{dop2}
\end{eqnarray}
where the Galactic orbital period is $T=2.35 \times 10^{8}$ yr.

 The most straightforward method to work out whether  a temporal change in the dipole is detectable or not, is
to compute a sky map of the dipole at two epochs. Clearly then, if the temperature
variance of the difference map is greater than the experimental noise
, a detection is likely   to be expected. The variance of the difference map, denoted in \cite{2007PhRvD..76l3010Z} by $C_S$,
is $C_S=[(\delta v_{x})^{2}+ (\delta v_{y})^{2}+
(\delta v_{z})^{2}]/(4 \pi)$, where $\delta \boldsymbol{v}$ is the
difference of the Sun-CMB dipole vector between the two observations.
Using Eq.~(\ref{dop2}), and converting to fractional temperature
variations, the signal variance of the changing dipole is then
\begin{equation}
C_S=8.7 \times 10^{-8} \left[ 1- \cos \left( \frac{2 \pi t}{T}
\right) \right]\,.
\end{equation}
In Section~\ref{sec:cmbobs} whether or not this might be achieved by future observations will be discussed in more details.

It should stressed out that the above discussion relies on the stated usual assumption that the dipole is mostly (or completely) due to a peculiar velocity. This, however, is yet to be proven observationally and there still remains a possibility that the dipole is due (in large or small part) to more exotic origins (see~\cite{Kosowsky:2010jm,Notari:2011sb} for a brief review of the possibilities). A cross-check that the dipole is really due to a peculiar velocity can nevertheless be made by looking for correlations between $a_{\ell m}$ [defined below in~\eqref{eq:alm}] and $a_{\ell' m}$ (see~\cite{Kosowsky:2010jm,Amendola:2010ty,Pereira:2010dn,Notari:2011sb}).

\subsection{Primary anisotropies}
\label{sec:primary}

Besides the time evolution of the dipole, the core of the information about CMB is expanded along all the anisotropy scales through the whole power spectrum. Therefore in principle  it is much less trivial to derive the temporal progress of the other multiples and their correlations. However the authors in \cite{2007PhRvD..76l3010Z} have shown how under simple assumptions it is possible to determine a scaling relation between power spectra at different times. The main approximation is that all of the CMB radiation was emitted from the last scattering surface (LSS)  at some instant
when electrons and photons decoupled, and then propagated freely, supposing that LSS has vanishing thickness.

The temperature fluctuations observed in a direction $\hat{n}$ of the sky can be expanded in spherical harmonics as
\begin{equation}
\frac{\delta T(\hat{n},\tau)}{T(\tau)} = \sum_{\ell m} a_{\ell m}(\tau)Y_{\ell m}(\hat{n}),
\label{eq:spher}
\end{equation}
where $\tau$ is the conformal time. All the details about the map of the CMB at a time $\tau$ are encoded in the expansion coefficients $a_{\ell m}$, that are related to the primordial perturbations $\mathcal{R}$ and the linear transfer function $\mathcal{T}$ as follows:
\begin{equation}
a_{\ell m}(\tau) = \sqrt{\frac{2}{\pi}} \int kdk \mathcal{R}_{\ell m}(k) \mathcal{T}(k,\ell,\tau).
\label{eq:alm}
\end{equation}
The statistical properties of the primary anisotropy $a_{\ell m}s$ are then directly connected to
\begin{equation}
\big<\mathcal{R}_{\ell m}(k) \mathcal{R}^*_{\ell' m'}(k')\big> =
 2\pi^2\delta(k - k')\frac{P_{\mathcal{R}(k)}}{k^3}\delta_{\ell \ell'}\delta_{mm'},
\end{equation}
where $P_{\mathcal{R}}(k)$ is the Gaussian curvature primordial power spectrum. Their variance $C_{\ell}(\tau)$ as a function of time, defined such that $<a_{\ell m}(\tau)a^{*}_{\ell m}(\tau)>=C_{\ell}(\tau)\delta_{\ell \ell'}\delta_{mm'}$, results to be
\begin{equation}
C_{\ell}(\tau) \equiv 4\pi\int\frac{dk}{k}P_{\mathcal{R}}(k)\mathcal{T}^2(k,\ell,\tau).
\end{equation}
From the previous equation it is evident that all the temporal information of the CMB primary anisotropy power spectrum is embedded in the linear transfer function; as the comoving radius of the LSS increases with time, every wavelength  will span a smaller angle.

Since for large $\ell$ one has $\mathcal{T}^{2}(k,\ell',\tau')\simeq (\tau^{2}/\tau^{'2})\mathcal{T}^{2}(k,\ell,\tau)$ (\cite{2007PhRvD..76l3010Z}  ), where the scale $\ell$ has shifted in time to $\ell'=\ell \tau'/\tau$, the final geometrical scaling relation for the power spectrum is
\begin{equation}
\ell'^2C_{\ell'}(\tau') \simeq \ell^2C_{\ell}(\tau).
\label{eq:clscal}
\end{equation}
In particular, the Sachs-Wolfe plateau for a scale invariant power spectrum remains constant in time, as $\ell (\ell+1)C_{\ell}$ is independent of $\ell$ for a Harrison-Zel'dovich initial spectrum. The polarization spectra will scale as well as Eq. (\ref{eq:clscal}), since the main contribution is sourced at LSS. However, minor contributions come from reionization, whose comoving radius will then affect the polarization scaling expression. In addition, in a $\Lambda$CDM universe observers endure a future event horizon, that is a finit $\tau_{f}$ and consequently a maximum $\ell_{f}$.

The measurable real time quantity is the power spectrum difference $\Delta_{\tau} C_{\ell}\equiv C_{\ell}(\tau')-C_{\ell}(\tau)$ at two different times. For small time spans $\Delta \tau$ it reads \cite{2007PhRvD..76l3010Z}
\begin{equation}
\Delta_{\tau} C_{\ell}\simeq \frac{\partial}{\partial \tau} C_{\ell}(\tau)\Delta \tau=-\frac{\Delta \tau}{\tau_{LSS}}\Big(\ell \frac{\partial C_{\ell}(\tau)}{\partial \ell}+2C_{\ell}(\tau)\Big)
\label{eq:delcl1}
\end{equation}
and as all the real time observables it is proportional to the  time interval.
This temporal power spectrum variation can have either negative or positive sign and vanishes for the scale-invariant Sachs-Wolf plateau and  at an acoustic peak. This happens because assuming statistical homogeneity of space the quantity $\ell (\ell+1)C_{\ell}$   does not depend on $\ell$ in the Sachs-Wolfe plateau for a scale-invariant power spectrum, which is valid for the entire acoustic peak structure. The fluctuations of the same physical size in the LSS visible at a time $\tau$ (corresponding to a comoving distance $r$) on angular scale $\theta$ will be visible at a time $\tau'$ (corresponding to $r'$) at a smaller angular scale $\theta'=\theta r/r'$ \cite{2007PhRvD..76l3010Z}.

\subsection{ISW}
\label{sec:ISW}
\begin{figure}[t]
    \includegraphics[width=8.5cm]{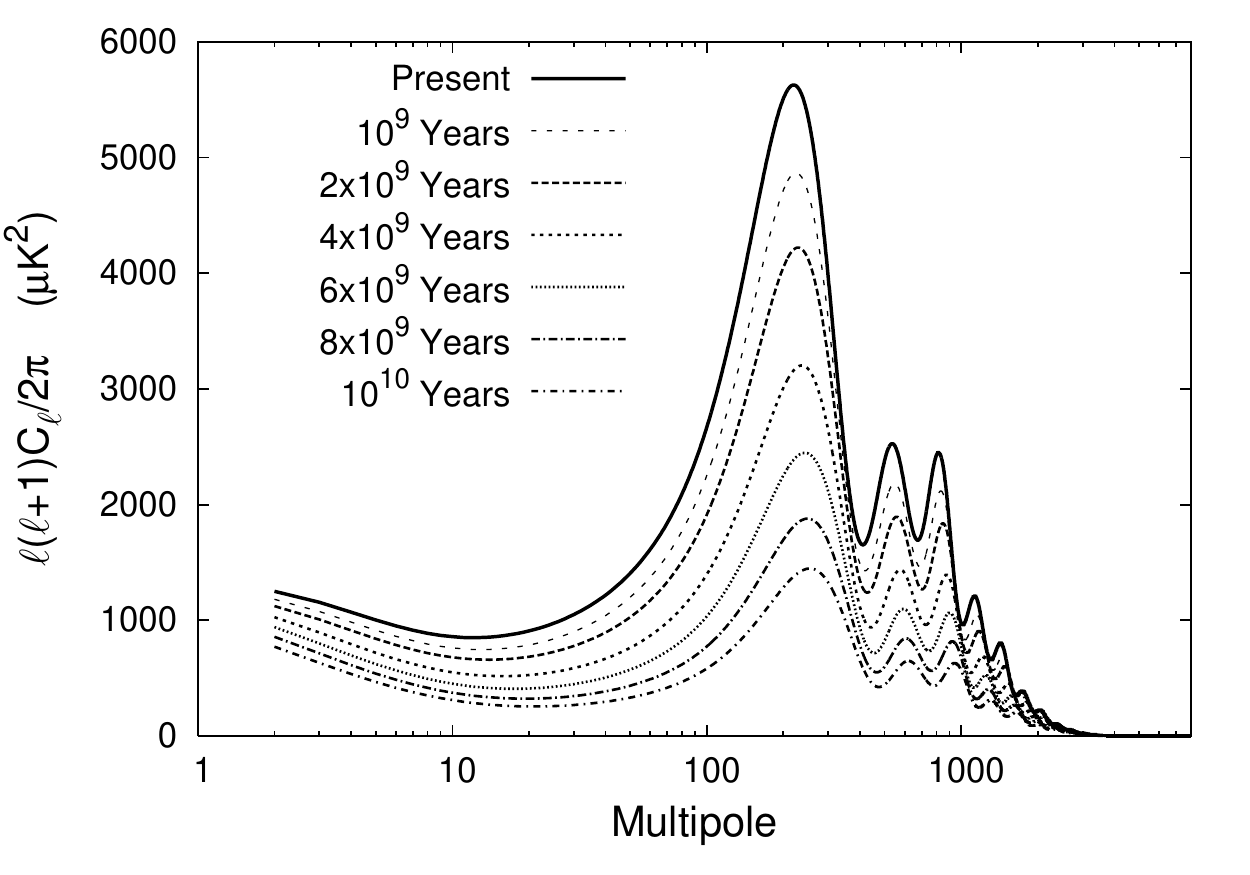}
       \includegraphics[width=8.5cm]{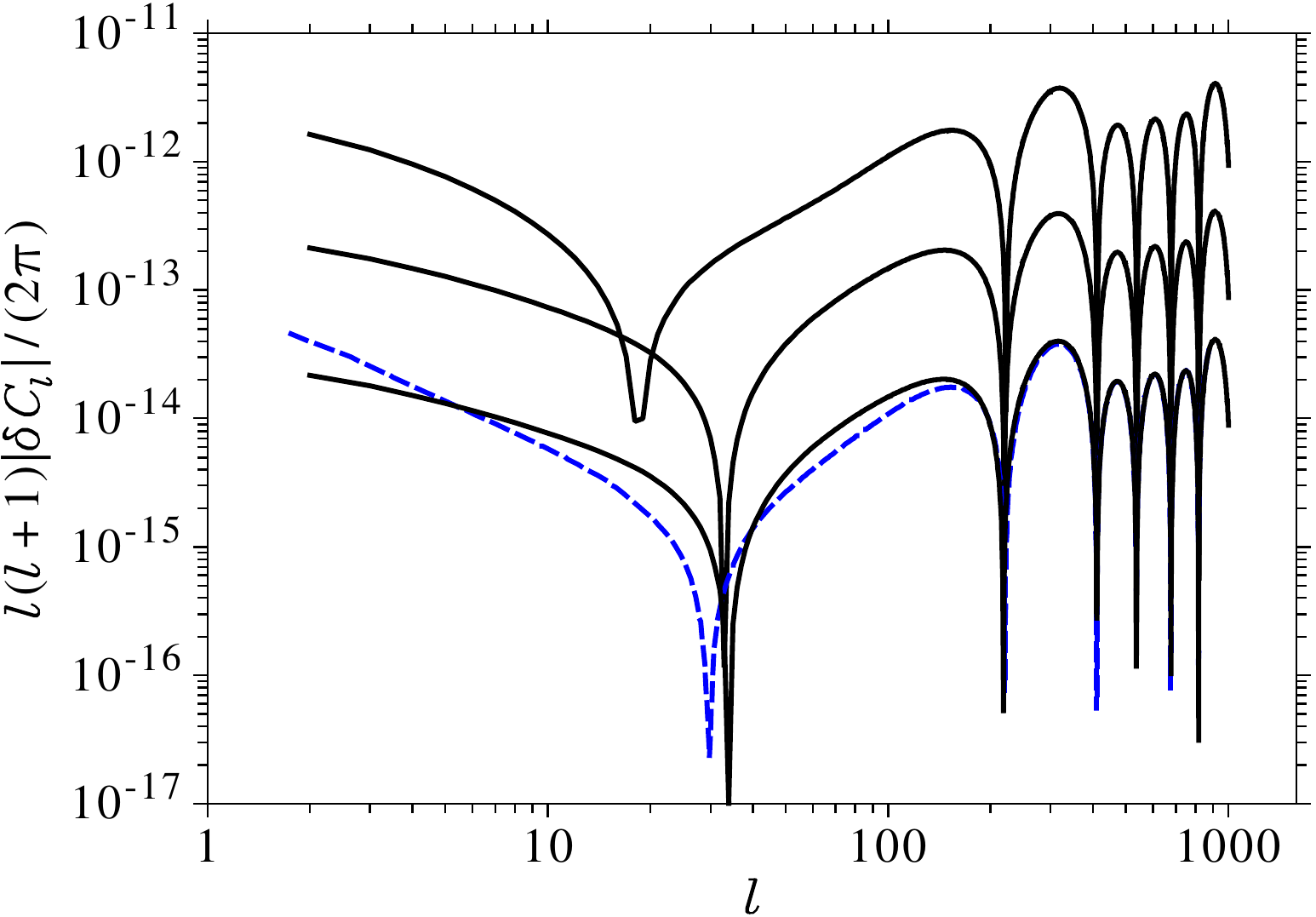}
    \caption{\small {\it Left}: the temperature angular power
spectrum of the CMB at several
representative time steps into the future taken from \cite{2007ApJ...671.1075L}. The scale factors and conformal
distances go from $(1,0)$ to $(602,1\cdot 10^11)$ counting from present time (confront \cite{2007ApJ...671.1075L}). {\it Right}: Absolute value of the difference in the CMB power spectra
$l(l + 1)|\delta C_l/(2\pi)$ between $a_{\rm obs} = 1$ and $a'_{\rm obs}
 = 1 + \delta a$.  The solid curves are from numerical integration in \cite{2007PhRvD..76l3010Z}, and from top to bottom denote $\delta a = 0.01$, $0.001$, and
$10^{-4}$.  The dashed curve was calculated using the analytical
expression for $\delta a = 10^{-4}$. }
    \label{fig:cmb_ev1}
\end{figure}

The integrated Sachs-Wolfe (ISW) contributions to total anisotropy power spectrum comprise an early-ISW generated by the exit from the radiation era and a late-ISW caused by the onset of the dark energy component domination at late times. The former takes place during a time that is relatively close to the LSS, hence superposing coherently to the primary anisotropies; therefore it is supposed to scale as Eq.~(\ref{eq:clscal}).
The large scale temperature fluctuations arising from the late-ISW are due to the small redshift varying gravitational potential, hence it can be described by an extra transfer function $\mathcal{T}(k,\ell,\tau)$.  The ISW power spectrum can then be written as \cite{2007PhRvD..76l3010Z}
\begin{equation}
C_{\ell}^{ISW}(\tau)=4\pi\int \frac{dk}{k}P_{\mathcal{R}}(k)T^{2}_{ISW}(k,\ell,\tau)=\frac{72}{25}\frac{\pi^{2}P_{\mathcal{R}}}{\ell^{3}}\int_{0}^{\tau}d\tau' \Big(\frac{dg(\tau)}{d\tau}\Big)^{2}(\tau-\tau'),
\end{equation}
where $g(\tau)$ is the perturbation growth function. The time derivative of the previous expression gives the temporal variation of the late-ISW  effect, to be added to Eq.~(\ref{eq:delcl1}):
\begin{equation}
\Delta_{\tau} C_{\ell}^{ISW}=\frac{\Delta \tau}{\tau_{LSS}}\Big(\frac{72}{25}\frac{\pi^{2}P_{\mathcal{R}}}{\ell^{3}}\tau_{LSS}\int_{0}^{\tau}d\tau'\Big(\frac{dg(\tau)}{d\tau}\Big)^{2}\Big),
\label{eq:delcl2}
\end{equation}
so that the total CMB shift is the sum of Eq.~(\ref{eq:delcl1}) and (\ref{eq:delcl2}).
Left panel of Fig.~\ref{fig:cmb_ev1} shows the power spectrum at different future times \cite{2007ApJ...671.1075L}. Note the progressive enhancement of the ISW with respect to the acoustic peaks. Of course this is completely model dependent, in that in a Universe with no cosmological constant we do not expect the large scale plateau to change.
On the right panel a numerical derivation of the CMB temporal shift calculated with the Boltzmann code CAMB is compared to the analytical expression, which matches the former reasonably well \cite{2007PhRvD..76l3010Z}.

\subsection{CMB maps}
\label{sec:maps}
\begin{figure*}[t]
\centering
   \includegraphics[width=8cm]{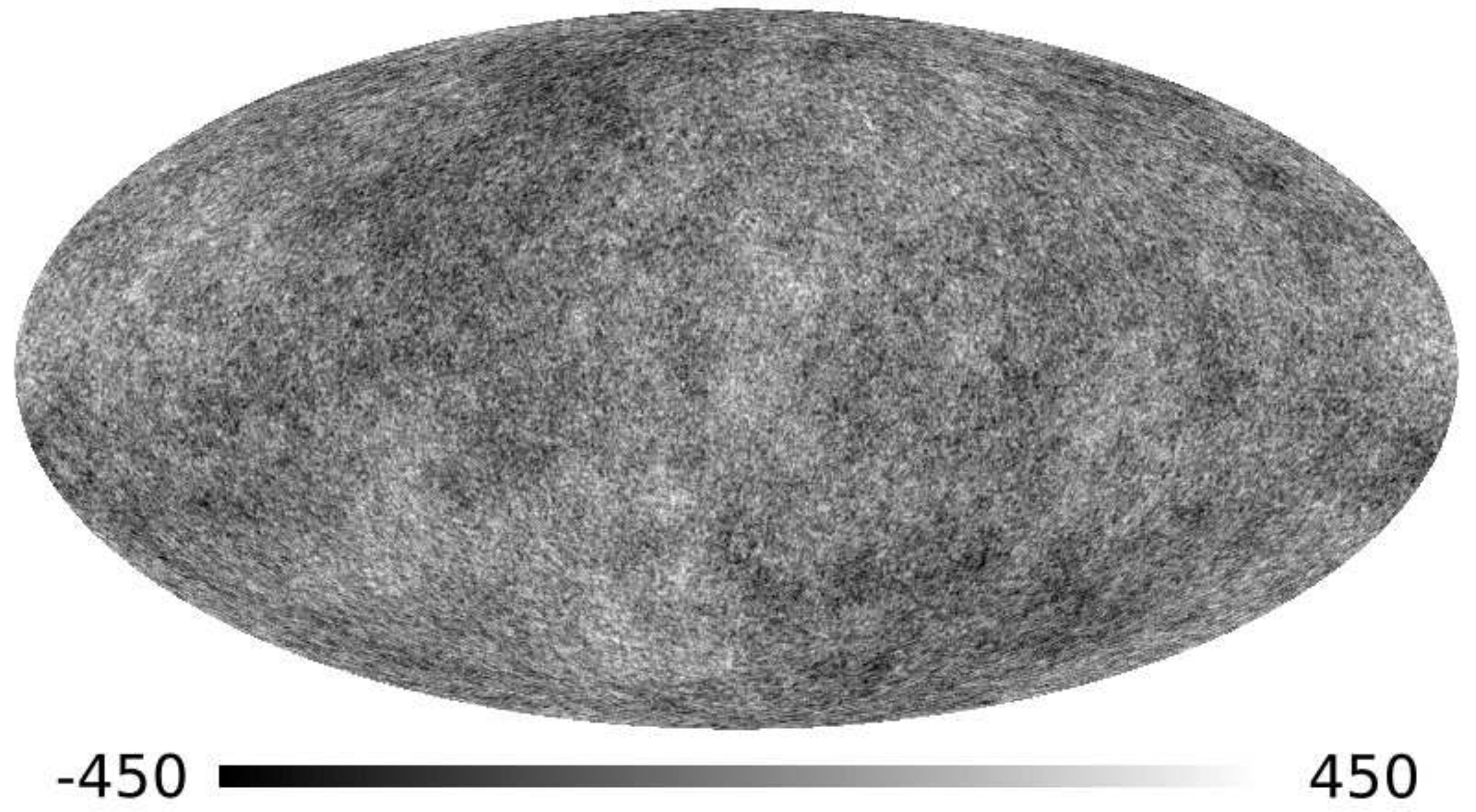}%
   \includegraphics[width=8cm]{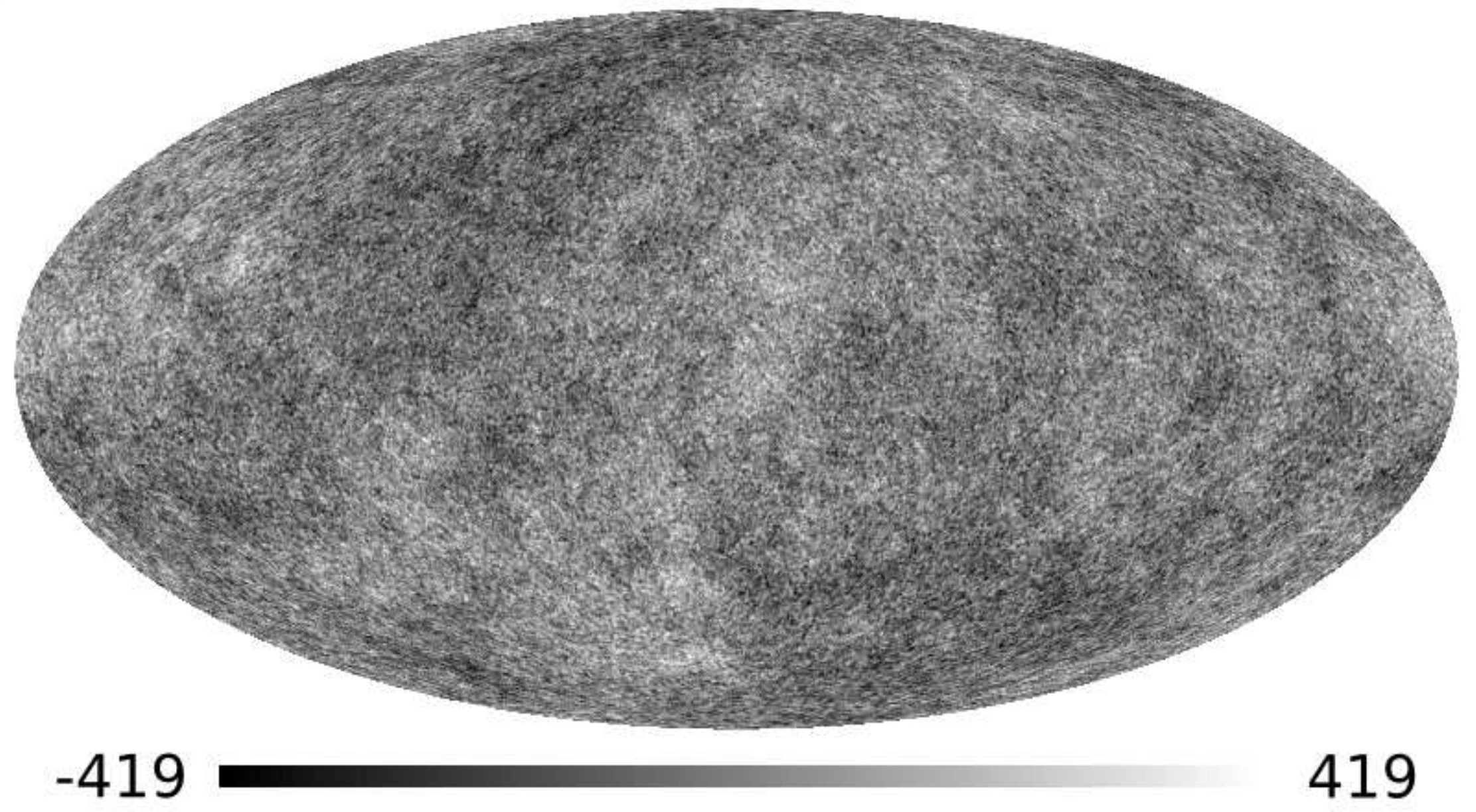}\\
   \includegraphics[width=8cm]{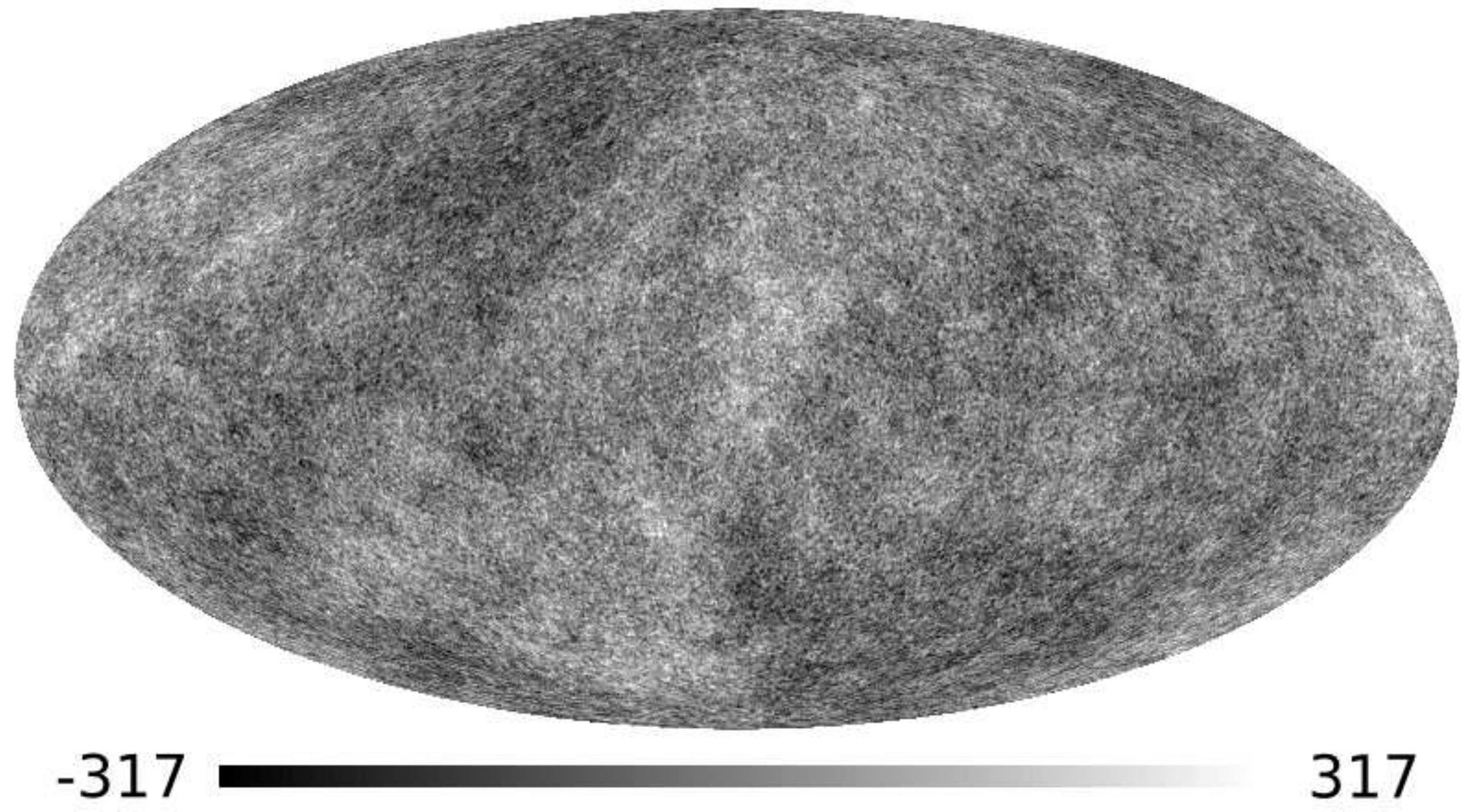}%
   \includegraphics[width=8cm]{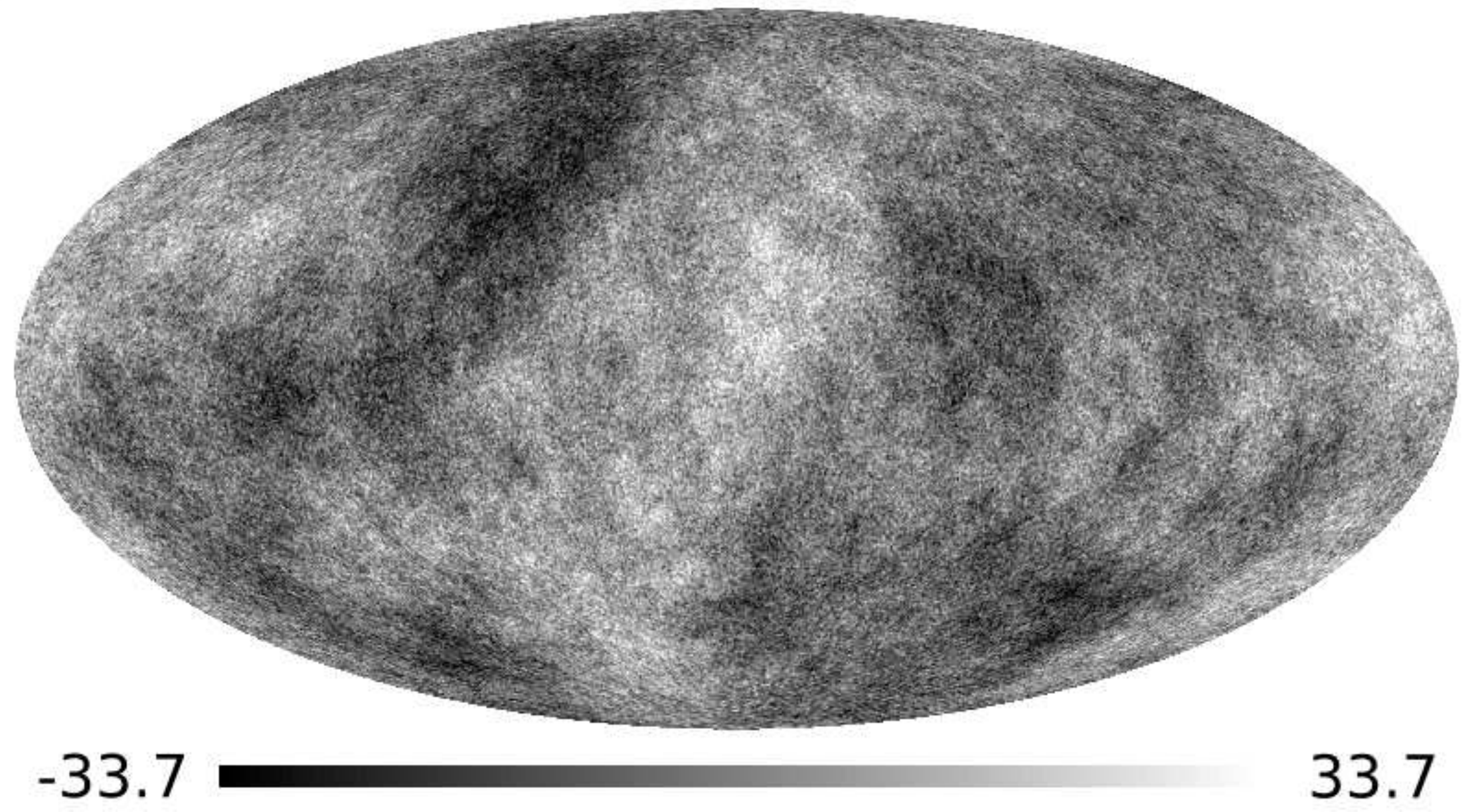}\\
    \caption{\small CMB maps (units on the bar are $\mu$K) for a $\Lambda$CDM model at 4 different future times; from top left clockwise: present, $1\cdot 10^{9}$yr, $5\cdot 10^{9}$yr, $40\cdot 10^{9}$yr in the future. From Ref.~\cite{2007ApJ...671.1075L}.}
\label{fig:allmaps}
\end{figure*}

Knowing the $C_{\ell}(\tau)$ it is possible to draw a series of maps at different times; assuming a gaussian random field all the information is encoded in the $C_{\ell}$ as they are the variance of the $a_{\ell m}s$. In \cite{2007ApJ...671.1075L} the authors have generated random complex values for the multipole coefficients with variance $C_{\ell}$. Since the different maps are supposed to be correlated in times, the $a_{\ell m}s$ need to be derived according to the covariance matrix $\mathbf{C}_{\ell}^{ij}=<a_{\ell m}(\tau_{i})a_{\ell m}^{*}(\tau_{j})>$ and can be expressed as $a_{\ell m}=\mathbf{M}_{\ell}\sqrt{\mathbf{D}_{\ell}}\mathbf{x}$, where $\mathbf{x}$ is a vector of complex random deviates and $\mathbf{D}_{\ell}$ is the diagonalized covariance matrix, such that $\mathbf{C}_{\ell}=\mathbf{M}_{\ell}\mathbf{D}_{\ell}\mathbf{M}^{*}_{\ell}$. In Fig.~\ref{fig:allmaps} four CMB maps at four future  time slices are shown   \cite{2007ApJ...671.1075L} for a WMAP best fit cosmological parameter set. The evolution first affects the small angular scales, shifting them to smaller sizes due to the expansion of the CMB photosphere, then, for larger time span, the larger scales through an evident increase of the ISW.

For small time intervals the maps seems to remain highly correlated, while as the time span increases the correlation vanishes first on smaller scales and then also on larger scales. According to   \cite{2007ApJ...671.1075L} for roughly $10^{10}$yr the future sky decorrelates with present time sky. A detailed analysis of these time correlation can also be found in \cite{2007PhRvD..76l3010Z}. In Section~\ref{sec:cmbobs} the detectability of this real time observable will be examined.

\section{Observations}
\label{sec:observations}

In this Section we review some of the experimental requirements needed to perform the previously described real-time observations. Actually, this issue has barely started to be properly investigated and is far from being settled. All the knowledge to date clearly arises from numerical simulations and forecasts.

In particular, we would like to point out that in general one might be misled by the meaning of the time span $\Delta t$ adopted by all the works on real-time cosmology.
For example, the time interval that enters Eqs.~(\ref{deltaz}), (\ref{eq:deltagamma-full}), (\ref{eq:shift}) and~(\ref{eq:delcl1}) is the time elapsing between two independent measurements. This should not be confused with the effective time needed to perform the observations. In the next subsection the question will be addressed in more details.

\subsection{The redshift drift and ultra-stable spectrographs}
\label{subsec:driftobs}

Recently, two high-precision spectrographs were proposed which could
in principle be used for measuring the redshift drift: the Cosmic
Dynamics Experiment (CODEX)~\cite{2005Msngr.122...10P,2008MNRAS.386.1192L,2007NCimB.122.1165C} at
the European Extremely Large Telescope~(E-ELT)~\cite{2007Msngr.127...11G} and the Echelle Spectrograph for PREcision Super Stable Observations
(ESPRESSO)~\cite{2007MmSAI..78..712D,2008MNRAS.386.1192L,2007NCimB.122.1165C} at the Very Large
Telescope array (VLT). Although proposed later, ESPRESSO would serve
as a prototype implementation on the technology behind CODEX as part
of its feasibility studies and could be operational several years
before CODEX~\cite{2008MNRAS.386.1192L,2007NCimB.122.1165C}.

The possibility of detecting the redshift drift with CODEX was analyzed
in a number of papers~\cite{2007PhRvD..75f2001C,2007MNRAS.382.1623B,2008MNRAS.386.1192L,2007NCimB.122.1165C}.
The achievable accuracy on $\sigma_{\Delta v}$ by the CODEX experiment
was estimated (through Monte Carlo simulations)~\cite{2005Msngr.122...10P} to
be
\begin{equation}
  \sigma_{\Delta v}=1.35\left(\frac{\mbox{S/N}}{2370}\right)^{-1}\!\!\left(\frac{N_{QSO}}{30}\right)^{-\frac{1}{2}} \!\!\left(\frac{1+z_{QSO}}{5}\right)^{q}\mbox{cm/s},
  \label{eq:sigma-v}
\end{equation}
with
\begin{equation}
  q\equiv-1.7\;\mbox{ for }\; z\le4\,,\quad\quad q\equiv-0.9\;\mbox{ for }\; z>4\,,
\end{equation}
where $S/N$ is the signal-to-noise ratio per pixel, $N_{QSO}$ is
the total number of quasar spectra observed and $z_{QSO}$ their redshift.
Note also that the error pre-factor 1.35 corresponds to using all
available absorption lines, including metal lines; using only Lyman-$\alpha$ lines enlarges this pre-factor to 2~\cite{2008MNRAS.386.1192L}.

Achieving a high S/N level (we refer to Appendix~\ref{app:zdot-errorbars} for a more precise formulation of the  S/N level) requires long integration times $t_{{\rm int}}$.
In~\cite{2007MNRAS.382.1623B}, an estimate was performed assuming a  CODEX-like experiment, coupled to a 42 m telescope with approximately 20$\%$ total efficiency, could measure 40 spectra with a  $S/N$ of 3000 in about 15 years (assuming a 20$\%$ use of the telescope, and a $90\%$ of actually usable data). In fact, again according to \cite{2005Msngr.122...10P}, a CODEX-like experiment, coupled to a 60 m telescope with approximately 20$\%$ total efficiency, would give a cumulative $S/N$ of 12000 for a single QSO, requiring roughly 125 hours of observation to get a $S/N$ of 3000 on that spectrum. Then, starting with 10 hours of observation per night, and taking into account a 20$\%$ use of the telescope, and a $90\%$ of actually usable data, one finds that 40 spectra can be measured with that $S/N$ in roughly 7.6 years (this time would actually increase to about 15 years if the telescope aperture is  42 m instead of 60 m). This integration time is not negligible with respect to the interval $\Delta t_{obs}$ over which we expect to measure the redshift drift, as assumed in~\eqref{eq:sigma-v}. Furthermore, it has been claimed in~\cite{2008MNRAS.386.1192L} that in principle it would be preferable to spread the observations more evenly over $\Delta t_{obs}$, inserting an unused temporal  window between the two measurements; in addition  the same authors conclude that the best strategy to minimize the errors would be to concentrate as much as possible the telescope time in both the beginning and ending of $\Delta t_{obs}$. Either way, the error estimate~\eqref{eq:sigma-v} is changed somewhat, but never by more than a factor 2. However, estimating such a correction depends on the details of the observational strategy.

In Ref.~\cite{2010PhRvD..81d3522Q}  a compromise strategy was assumed: a three-period
observation, each of $\Delta t_{obs}/3$ duration, and with observations
contained in the first and third periods. Doing so means that the
effective $\Delta t_{obs}$ for the redshift drift is $2\Delta t_{obs}/3$.
 It is important to note that a larger observational time-frame
allows not only for a larger redshift drift (which is linear in time)
but also for smaller error bars, as more photons are collected and,
therefore, a higher S/N  can
be achieved. In other words, the {}``effective signal'' increases
with $\Delta t_{obs}^{3/2}$ if one assumes a proportional telescope time
is maintained.

\subsection{Cosmic Parallax and high accuracy astrometric missions}
\label{subsec:parobs}

A realistic possibility of observing the cosmic parallax is offered by the forthcoming astrometric Gaia mission. Gaia will produce in five years a full-sky map of roughly $500,000$ quasars with positional error $p$ between $10$ to $200~\mu$as (for quasars with magnitude $V=15$ to $20$). To compare  observations to Gaia we need to evaluate the average $\Delta_{t}\gamma$ with $\Delta t = 5$~years and $N$ sources. The final Gaia error $p$ is obtained by fitting $2N$ independent coordinates from $N^{2}/2$ angular separation measures; the average positional error on the entire sky will scale therefore as $(2N)^{-1/2}$. Over one hemisphere we can therefore estimate that the error scales as $p/\sqrt{N}$. Since the average angular separation of random points on a sphere is $\pi/2$, the average of $\Delta_{t}\gamma$ can be estimated simply as $\Delta_{t}\gamma(\theta=\pi/2)$. Numerically it was found $\Delta_{t}\gamma(\pi/2)=10\, s\,\mu$as, with little dependence on $\Delta z$ \cite{2009PhRvL.102o1302Q}. Therefore Gaia can see the parallax if $p/\sqrt{N}<10\, s\,\mu$as. For $s=4.4\cdot10^{-3}$ (i.e. the current CMB limit) and $p=30\,\mu$as we need $N\gtrsim450,000$ sources: this shows that Gaia can constrain the cosmic anisotropy to CMB levels. An enhanced Gaia mission with $\Delta t=10$~years (or two missions 5~years apart), $p=10\,\mu$as and $N=10^{6}$ would give $s<5\cdot10^{-4}$, i.e. $r_{0}<1$~Mpc if we assume the sources are at 2~Gpc.

Two local effects induce spurious parallaxes: one (of the order of 0.1$\mu$as$\, \mbox{year}^{-1}$) is induced by our own peculiar velocity and the other (of the order of 4-5$\mu$as$\, \mbox{year}^{-1}$\cite{2003A&A...404..743K} by a changing aberration. The latter is the secular aberration drift, that is an alteration of the velocity of distant objects (like quasars) driven by the acceleration of the Solar system barycenter in the Milky Way. Both produce a dipolar signal, just like a LTB: however, the peculiar velocity parallax decreases monotonically with the angular diameter distance, while the aberration is independent of distance and is directed toward the Galactic center \cite{2003A&A...404..743K}. Recently, authors in \cite{2010arXiv1009.3698T} analysed geodetic and astrometric VLBI data from a sample of quasars of 1979-2010 and estimated the acceleration vector, together with a non significant global rotation, to have an amplitude of $5.8\pm1.4$ $\mu$as/yr and an almost constant dipolar function of redshift. In contrast, the LTB signal has a characteristic non-trivial dependence on redshift: for the models investigated here it is vanishingly small inside the void, large near the edge, decreasing at large distances. It is therefore possible in principle to subtract the cosmic signal from the local one, for instance estimating the local effects from sources inside the void, including Milky Way stars. A detailed calculation needs a careful simulation of experimental settings (including possibly effects like source photocenter jitter and relativistic light deflection by solar system bodies) which is outside the scope of this review. Moreover, more general anisotropic models will not produce a simple dipole. One example is offered by the cosmic parallax induced in Bianchi I models \cite{2009PhRvD..80f3527Q}, as analysed in Section~\ref{sec:paralaxBianchi}.

\subsection{Cosmic Microwave Background accuracy}
\label{sec:cmbobs}

The issue of realistically measuring the temporal evolution of the CMB has been addressed in details in two papers \cite{2007ApJ...671.1075L,2008PhRvD..77d3505M}. The comparison between the accuracy reached by an experiment and the variation of the CMB can involve either the power spectrum or the map. First of all it is important to notice that since we are looking at the evolution of a single sky realisation the intrinsic uncertainty of the signal due to the cosmic variance should not be taken into account. All that matters is the instrumental accuracy and the possible foreground contamination. In addition a portion of information is embedded in the time-time correlation that render the simple power spectrum less statistically significant than usual. One can then  define the difference between $a_{\ell m}s$ at two different times $D_{\ell}=<|a_{\ell m}(\tau_{i})-a_{\ell m}(\tau_{j})|^{2}>=C_{\ell}(\tau_{i})+C_{\ell}(\tau_{j})-2C_{\ell}^{ij}$.
In \cite{2008PhRvD..77d3505M} the authors simulated a full sky dataset with no noise at present time. The predicted evolution of the dipole due to galactic motion is $(a_{1-1},a_{10},a_{11})=740.5\cdot 10^{-6}(\cos{x},0,\sin{x})+(-1556.5,919.3,-87.8)\cdot 10^{-6}$, with $x=2\pi t/T$, $T$ being the galactic rotation period. They then use this prediction to build a second noiseless map at some time in the future. The estimator is $\chi^{2}=\sum_{i=1}^{N_{pix}}[T_{i}/T(t)-T_{i}/T_{0}]^{2}/(2C_{N})$, where $C_{N}=<(\delta T_{i}^{noise}/T)^{2}>=\sigma^{2}_{pix}$ is the noise uniform variance. In the noise dominated regime  this statistic reduces to the signal-to-noise  ratio related to the difference signal map  via $\alpha=D_{S}/\Delta D_{s}^{N}$. Considering all the angular scales in a CMB experiment like Planck\footnote{http://www.esa.int/planck} they find $\alpha = 2$ after $t=2\cdot 10^{4}$yr. Performing a harmonic transform of the map they then isolate the $\ell =1$ mode and find for the same time span $\alpha =40$ with $\Delta D_{S}^{N}=3\Delta D_{1}^{N}/(4\pi)$.

\begin{figure}[t]
 \includegraphics[width=12cm]{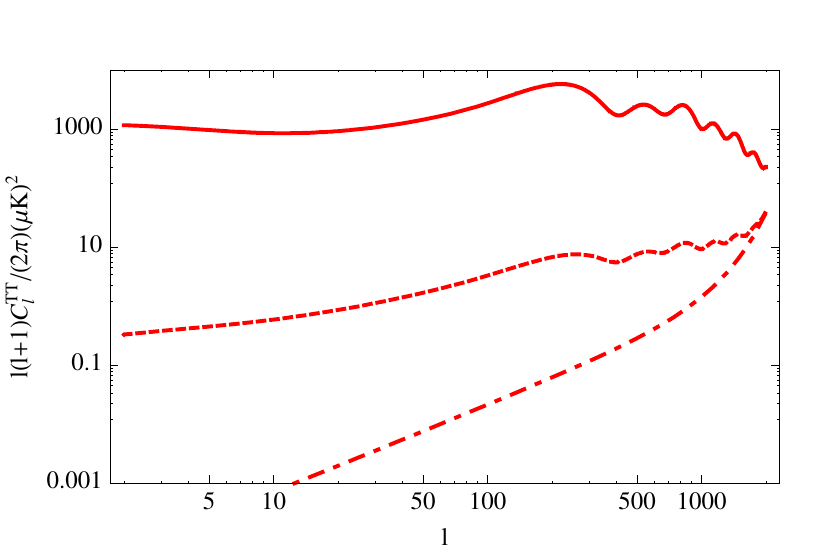}
\caption{\small CMB temperature anisotropy power spectrum (solid curve) confronted with the error $\Delta C_{\ell}^{N}$ (dashed curve) and the pure noise $\sqrt{2/f_{sky}/(2\ell+1)}N_{\ell}$ (dot-dashed curve).}
\label{fig:noisecl}
\end{figure}

Using just the information on the variation of the power spectrum $\delta C_{1}$, where signal-to-noise ratio is $\alpha=|\delta C_{1}|/2\Delta C_{1}^{N}$, gives a lower statistical significance ($\alpha=4$) due to the removal of the time covariance information. As already mentioned previously, the variance of the $C_{\ell}s$
\begin{equation}
\Delta C_{\ell}^{N}=\sqrt{\frac{2N_{\ell}}{2\ell+1}(2<C_{\ell}^{est}>_{N}+N_{\ell})}
\end{equation}
does not contain the cosmic variance term, but includes the uncertainty on the estimated power spectrum in a certain noise realisation ensemble.
In Fig.~\ref{fig:noisecl} a $\Lambda$CDM temperature anisotropy power spectrum is compared to its variance (neglecting the intrinsic cosmic variance) and the simple noise $\sqrt{2/f_{sky}/(2\ell+1)}N_{\ell}$ for a Planck-like experiment. The difference between the two noises arises because the power spectrum is estimated from a map: after a certain number of noise realisation the expected value of the spectrum is  $<C_{\ell}^{est}>$ with a variance that takes into account both the noise and the scatter in the estimate \cite{2008PhRvD..77d3505M}.

\begin{figure}[t]
 \includegraphics[width=8cm]{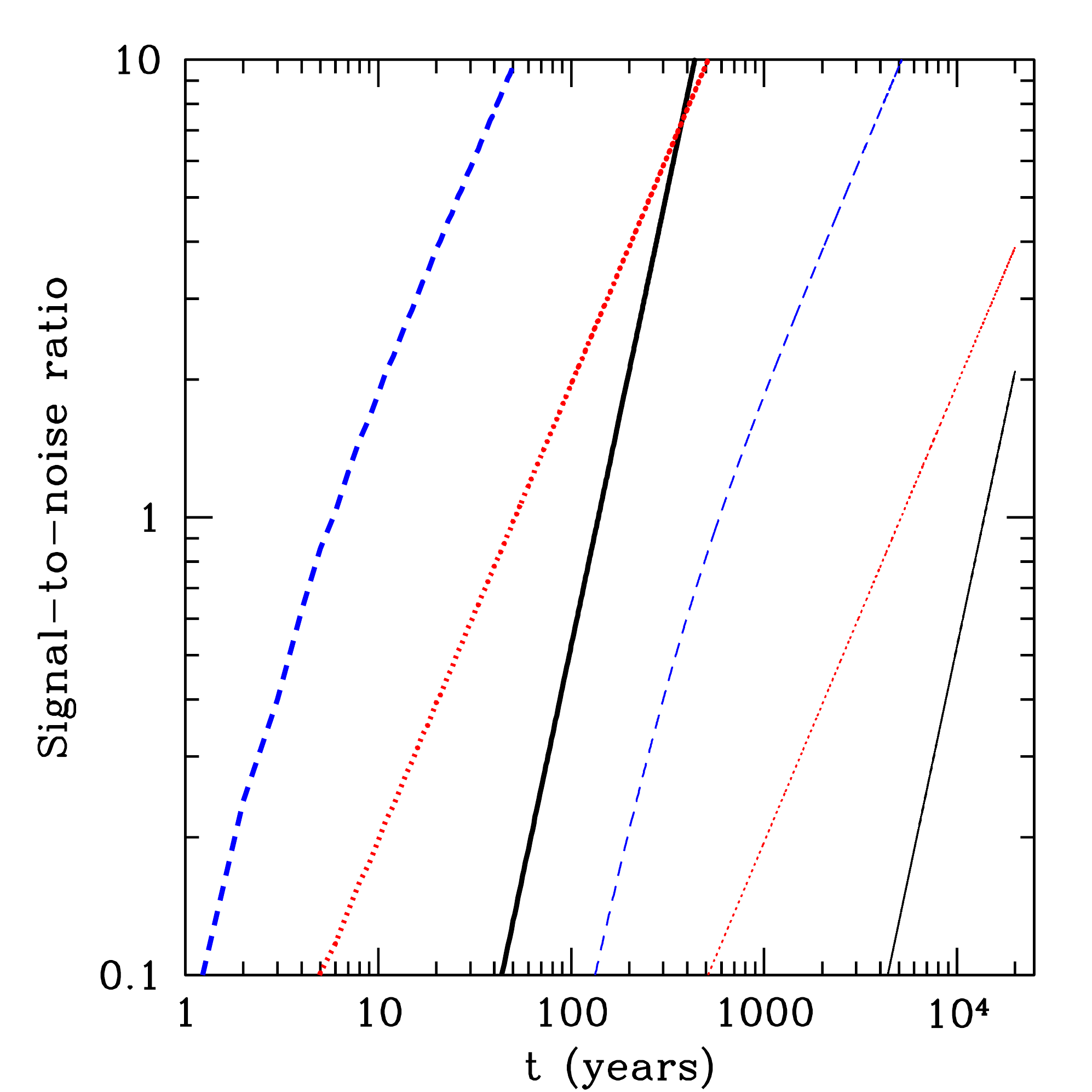}
 \includegraphics[width=9cm]{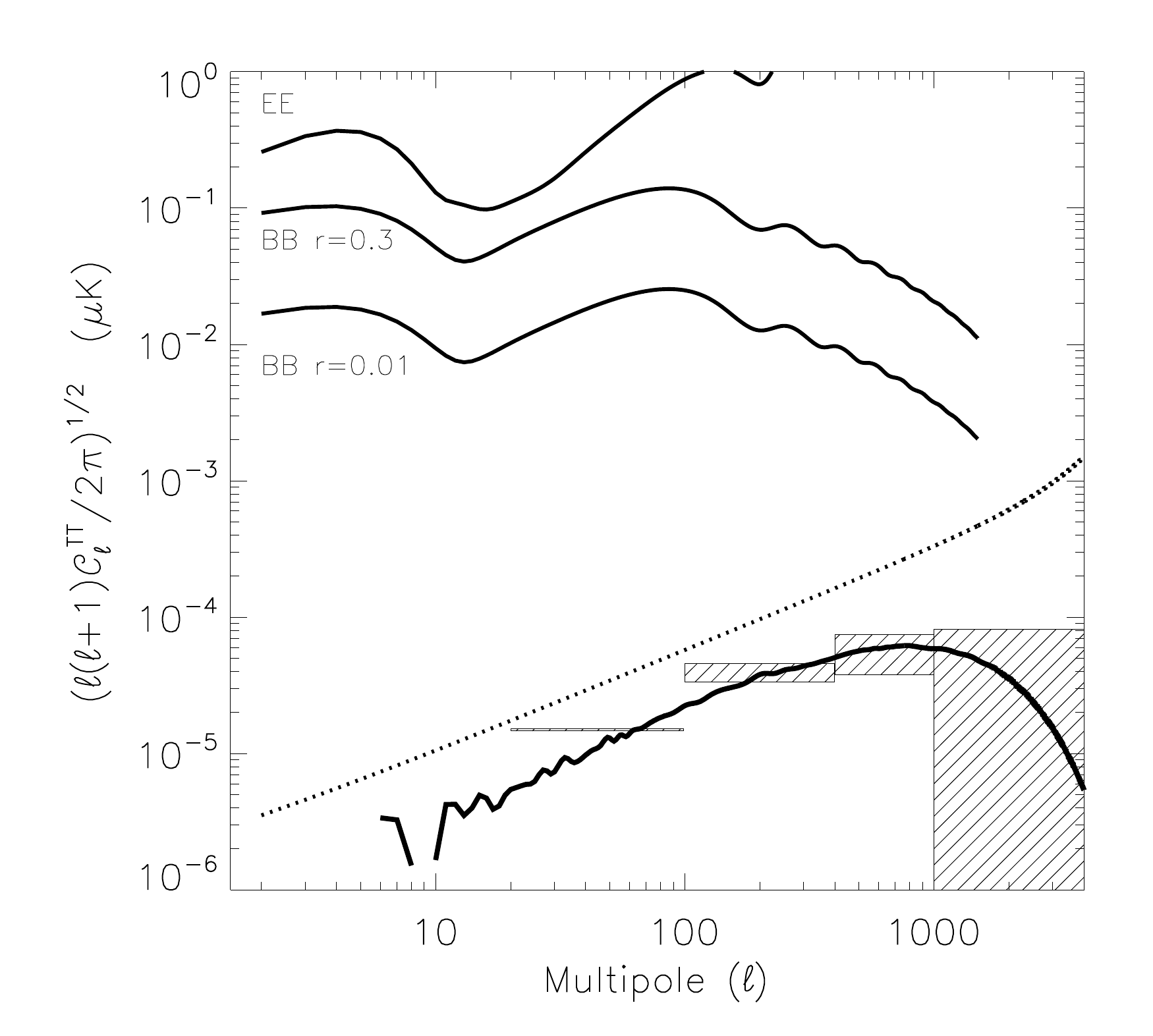}
\caption{\small {\it Left panel}: Signal-to-noise ratio  as a function of time  for observing a change in the CMB dipole. Thin lines are for a Planck-like instrument and thick lines are for an instrument 100 times as sensitive. Solid curves are drawn from differences in maps, while dashed curved after removing small scale noise and dotted curves contain no time-time correlation information. From Ref.~\cite{2008PhRvD..77d3505M}. {\it Right panel}: The bottom curve is the power spectrum of the difference between two maps taken a century apart. The dotted line corresponds to the noise (Eq.~(\ref{eq:bias})) and the hashed boxes are the errors after binning in $\ell$. For comparison the authors superposed on the same plot the EE power spectrum for $\tau=0.1$,  the BB power spectrum for two different values of the tensor-to-scalar ratio. From Ref.~\cite{2007ApJ...671.1075L} }
\label{fig:ston}
\end{figure}

Regarding the dipole, the left panel of Fig.~\ref{fig:ston} shows the signal-to-noise ratio relative to it for a Planck-like mission and a 100-fold enhanced mission. Using the difference map and after removing the small scale contamination it should be possible to detect a time variation of the dipole in 10yr.

Following the formalism in \cite{1995PhRvD..52.4307K} the authors in \cite{2007ApJ...671.1075L} calculated the difference in high precision maps taken a century apart and compared it to the experimental uncertainty, that is the noise with
\begin{equation}
N_{\ell}=\sigma_{pix}^{2}\theta_{fwhm}^{2}\exp{(\ell^{2}\theta_{fwhm}^{2}/8\ln{2})},
\label{eq:bias}
\end{equation}
where $\theta_{fwhm}$ is the FWHM of the gaussian beam profile. The predicted experiment has a sensitivity of $s=40$mK s$^{1/2}$ (where $\sigma_{pix}=s/\sqrt{t_{pix}}$), an angular resolution of $0.86$', a 4 year time mission covering 75$\%$ of the sky. The right panel of  Fig.~\ref{fig:ston} shows the power spectrum of the difference of two maps with a time interval of 100 yr, together with the errors after a binning in $\ell$ and the noise curve of such an experiment.

Among all the contamination that might smudge the detection, the uncertainty on the calibration seems to be the less alarming due to the fact that we are dealing with a difference map. The less promising systematic is probably caused by the time-varying foreground emission. The observer foreground might change with the same time scale as the dipole. Hopefully, multi-wavelength observations will help constraining them. However, it would take thousands of years to detect a change in the spectrum at higher multipoles.

\section{Conclusions}
\label{sec:conclusions}
Time domain information provides critical insight in many branches of astrophysics, even cosmology (distant supernovae and active galactic nuclei being obvious examples). However, the universe as a whole is evolving slowly and so one expects this evolution to be difficult to observe.
Real-time cosmology tests directly cosmic kinematics by observing changes in source positions and velocities.

Measurements of the redshift drift (or velocity shift) with future extremely large telescopes and high-resolution spectrographs could provide interesting information on the source of cosmic acceleration, which would complement other, more traditional cosmological tools.
The observation of redshift drift alone can be affected by strong parameter degeneracies, limiting its ability to constrain cosmological models. Uncertainties on parameter reconstruction (particularly for non-standard dark energy models with many parameters) can be rather large unless strong external priors are assumed. When combined with external inputs, however, the time evolution of redshift could discriminate among otherwise indistinguishable models.

Despite its inherent difficulties, the method has many interesting advantages. One is that it is a direct probe of the dynamics of the expansion, while other tools (e.g.\ those based on the angular diameter distance) are essentially geometrical in nature. This could shed some light on the physical mechanism driving the acceleration. For example, even if the accuracy of future measurements turns out to be insufficient to discriminate among specific models, this test would be still valuable as a tool to support the accelerated expansion in an independent way, or to check the dynamical behaviour of the expansion expected in general relativity compared to alternative scenarios.
In LTB void models for example, its constraining power is particularly evident since the LTB expansion is always decelerated and the effect turns out to be mostly sensitive to the scale of the void, but not to other particular void properties like steepness of the transition.

Furthermore, despite being observationally challenging, the method is conceptually extremely simple. For example, it does not rely on the calibration of standard candles (as it is the case of type Ia SNIa) or on a standard ruler which originates from the growth of perturbations (such as the acoustic scale for the CMB or for transverse BAO) and on effects that depend on the clustering of matter (except on scales where peculiar accelerations start to play a significant role).
Finally, it is at least conceivable that suitable sources at lower redshifts  could be used to monitor the redshift drift in the future. This would be extremely valuable, since some non-standard models have a stronger parameter dependence at low and intermediate redshifts, that could be exploited as a discriminating tool.  Exploring the feasibility of such proposals will certainly be an interesting topic for further studies from observers.

Peculiar and proper accelerations would also be an interesting target
for such future measurements, since they could give
an independent measurement of the mass profile of clusters of galaxies,
and possibly even of individual galaxies. Measurements of the velocity shift in real time for different mass scales could
test the relation between the concentration parameter of clusters
and the virial mass.

The expected signal for  some objects (like Coma cluster) might be  comparable
to the cosmological change in redshift
expected for distant quasars due to the expansion dynamics.  It then  seems plausible
that it could be observable as well, although the objects used as test particles in the two cases
are different in nature. This would  also be useful to the extent of
assessing the effect of peculiar acceleration as a source of errors in the measurement
of the cosmological velocity shift.

Experimental and observational details
of future instruments are still too blurry to make any clear statement
on the actual feasibility of using peculiar acceleration as an astrophysical
and cosmological tool (even less for proper acceleration). Many issues are still to be discussed before
forecasting the results of actual observations, from the stability
of the galactic spectra to the effect of local deviations from spherical
symmetry. However, it is exciting to entertain the possibility
that peculiar and also proper accelerations might actually be observable in the not
too distant future, and hope that more detailed investigations of
their applicability in astrophysics might take place in the next few years.


In particular, the possibility of reconstructing the gravitational potential of a galaxy by means of the redshift drift  signal is also appealing. Assuming that either the  cosmological redshift of the galaxy is almost vanishing or it can be averaged out, the velocity shift produced by peculiar acceleration of test particles orbiting outside the galactic disc might be a probe of a modified gravity configuration, in which the CDM halo is absent and the Poisson equation is modified following MOND: although the two scenarios are undistinguishable using rotation velocity curve,  the question of whether peculiar acceleration might be employed to discriminate between them seems quite interesting.

As expected, the closer is the test particle to the galactic center the higher is the signal, reaching values comparable to the cosmological ones predicted by simulations and possibly detectable from high-resolution and ultra-stable spectrograph coupled to new generation telescope.

In our Galaxy, the 150  known globular clusters orbiting outside the disc  could play the role of test particles. The advantage of adopting globular cluster as test particles is related  to the fact that they not only are bright, but more importantly that they are made of thousands of stars and that their position is fairly well known.  As in this case we clearly are internal observers, the projected acceleration of the Sun on the disc, which has its maximum value on the order of a few cm/s, is assumed to be subtracted, and the cluster-centered acceleration of single stars in globular clusters is assumed to be averaged out.  The effect of being off-centered observers  combined with the non-radial direction of the acceleration in Kuzmin disc makes the pattern of the contours non trivial. However, while  the signal reaches the maximum value close to the Galactic centre for both scenarios, the difference signal is stronger at high Galactic longitude  where the spherical CDM halo is more influent.

 Although the feasibility of such observations is still under debate, this new technique would open a new window to the ability of testing the gravitational potential and of constraining the parameters involved in its expression, such as the mass of the galaxy.


Any anisotropy will leave an imprint on the angular
distribution of objects that are able to trace cosmic expansion. Planned space-based astrometric missions aim at accuracies of the order of a few micro-arcseconds. In the framework of real-time cosmology, the cosmic parallax of distant sources in an anisotropic expansion  might be observable employing the same missions. A positive detection of large-scale cosmic parallax would disprove therefore one of the basic tenets of modern cosmology, isotropy.  This test may probe a different range of scales depending on the quasar redshift distribution. Contrary to CMB limits, the cosmic parallax method cannot be completely undermined by the observer's peculiar motion, and is limited only by source statistics instead of cosmic variance. The major source of systematics would be the subtraction of the aberration change parallax.

 If an anisotropy is present before the last scattering surface,
CMB maps will also be affected. The temperature field will carry
extra anisotropies mainly caused by the angular dependence of the
redshift at decoupling. By resolving geodesic equations and expanding
temperature anisotropies in spherical harmonics, it is straightforward
to relate the low multipole components to the eccentricities of the
model. In Bianchi I models the first notable
multipoles related to the CMB are the monopole and the quadrupole.
The observed value of the latter puts constraints on the shear \textit{at
last scattering} of order $10^{-5}$, taking into account the cosmic
variance.  In this still exploratory stage
of analysis it is worth stressing that CMB puts indeed a very powerful
constraint on the shear at the time of decoupling, but with almost
no direct impact on late time expansion history. Complementary to
that, cosmic parallax is a direct and potentially powerful test of anisotropy
at small redshifts.

The anisotropic stress of dark energy is expected to have a leading
role in the generation of anisotropy at late times. It can be parameterized
by skewness parameters in the stress-energy tensor formulations, which
may be constant or time dependent functions. The only way to test these models
is to use either the angular dependence of the magnitude or the angular
distribution of objects in the sky at recent time, i.e.\ either distant
source angular distribution or the real-time cosmic parallax, the
two relying on different techniques and having independent systematics
that complement each other.

Gaia will be able to constrain the skewness parameters up to $10^{-3}\div10^{-4}$
at 2$\sigma$, comparable to CMB tests at decoupling time
(a Gaia+ experiment would improve them by one order of magnitude).

Unlike the LTB models with
off-centre observers, the cosmic parallax
signal in Bianchi I models is a combination of two quadrupole functions
of the two angular coordinates. Since the most important systematic
noises, caused by peculiar velocities and aberration changes, have
a dipolar functional form, Bianchi I models seem to be ideally testable,
though even in LTB models specific observational strategy aiming at
distinguishing the signal from the noise are possible.

CMB and cosmic parallax detect anisotropy at two different times and,
from an observational point of view, are completely independent on
each other: by combining them together one will have the opportunity
to reconstruct the evolution of the anisotropy and test with high
accuracy the Copernican Principle.

The two effects, the redshift drift together with cosmic parallax, can be detected with the E-ELT and with Gaia or an enhanced version of Gaia. These tests can be employed to distinguish a LTB void from an accelerating FRW universe, possibly eliminating an exotic alternative explanation to dark energy.

A 4$\sigma$ separation can be achieved with E-ELT in less than 10 years, much before the same experiment will be able to distinguish between competing models of dark energy. A Gaia-like mission, on the other hand, can only achieve a reasonable detection of a void-induced cosmic parallax in the course of 30 years.

Nevertheless, cosmic parallax remains an important tool and in fact one of the most promising way to probe general late-time cosmological anisotropy. In particular, even if it only lasts 6 years, Gaia should constrain late-time anisotropies similarly to current supernovae catalogs, but in an independent way. Also,
in a FRW model it can be used to measure our own peculiar velocity with respect to the quasar reference frame and consequently to the CMB, therefore providing a new and promising way to break the degeneracy between the intrinsic CMB dipole and our own peculiar velocity.  Combined, they will form an important direct test of the FRW metric.

As any proper motion signal increases linearly with time, any future mission with a global astrometric accuracy as good as Gaia can be used to detect the cosmic parallax signal.

It's really compelling that two great tools like Gaia and E-ELT are
being planned now when we begin to realize the importance
of ultra-precise astrometric and spectroscopic measurements for
cosmology.

The question of how long one would have to wait to see new information from the CMB has also been addressed.  We expect to observe a change on time scales much shorter than cosmological.  By considering optimal estimates for differences in observed skies, using currently available detector technology like Planck, it turns out that the dipole might  be observed to change due to our Galactic motion in about a decade, while unfortunately it would take thousands of years to detect even a statistical change in the higher order multipoles.


\subsection*{Acknowledgements}
We would like to thank Bruce Bassett for fruitful discussions about the topics and suggestions for the manuscript. We also would like to thank Giancarlo de Gasperis, for his technical and scientific support, and Christian Marinoni and David Mota for the opportunity to talk over real-time cosmology. L.A. is supported by the DFG through TRR33 ``The Dark Universe''. M.Q. is supported by the Brazilian research agency CNPq.
\appendix

\section{Geodesics in LTB Models}
\label{app:ltb-geodesics}

Due to the axial symmetry and the fact that photons follow a path which preserves the 4-velocity identity $u^{\alpha}u_{\alpha}=0$, the four second-order geodesic equations for $(t,r,\theta,\phi)$ can be written as five first-order ones. We will choose as variables the center-based coordinates $t$, $r$, $\theta$, $p\equiv\dd r/\dd\lambda$ and the redshift $z$, where $\lambda$ is the affine parameter of the geodesics. We shall refer also to the conserved angular momentum
\begin{equation}\label{eq:J}
    J\equiv R^{2} \frac{\dd\theta}{\dd\lambda} = \textrm{\emph{const}}=J_{0}\,.
\end{equation}
For a particular source, the angle $\xi$ is the coordinate equivalent to $\theta$ for the observer, and in particular $\xi_{0}$ is the coordinate $\xi$ of a photon that arrives at the observer at the time of observation $t_{0}$. Obviously this coincides with the measured position in the sky of such a source at $t_{0}$.

As per Figure~\ref{fig:CP-overview} and Section~\ref{sec:driftLTB} we will refer to ($t$, $r$, $\theta$, $\phi$) as the comoving coordinates with origin on the center of a spherically symmetric model. Peculiar velocities aside, the symmetry of such a model forces objects to expand radially outwards, keeping $r$, $\theta$ and $\phi$ constant. In terms of these variables, and defining $\lambda$ such that $u(\lambda)<0$, the autonomous system governing the geodesics is written as
\begin{equation}
\begin{aligned}
    \frac{\dd t}{\dd\lambda}\;=\; & -\sqrt{\frac{(R')^{2}}{1+\beta}\, p^{2} + \frac{J^{2}}{R^{2}}}\,,\\
    \frac{\dd r}{\dd \lambda} \;=\; & p\,,\\
    \frac{\dd\theta}{\dd\lambda} \;=\; & \frac{J}{R^{2}}\,, \\
    \frac{\dd z}{\dd\lambda} \;=\; & \frac{(1+z)}{\sqrt{\frac{(R')^{2}}{1+\beta}\, p^{2} + \frac{J^{2}}{R^{2}}}} \left[\frac{R'\dot{R}'}{1+\beta}\, p^{2} + \frac{\dot{R}}{R^{3}}J^{2} \right],\\
    \frac{\dd p}{\dd\lambda} \;=\; & 2\dot{R}'\, p\,\sqrt{\frac{p^{2}}{1+\beta}\, + \frac{J^{2}}{R^{2}\, R'^{2}}}\,+\,\frac{1+\beta}{R^{3}R'}J^{2} \, + \left[\frac{\beta'}{2+2\beta} - \frac{R''}{R'}\right] p^{2}\,.
\end{aligned}
\end{equation}

The angle $\xi$ along a geodesic is given by~\cite{Alnes:2006a}:
\begin{equation}
    \cos\xi = - \frac{R'(t,r)\, p}{u\,\sqrt{1+\beta(r)}} \,,
\end{equation}
from which we obtain, exploiting the remaining freedom in the definition of $\lambda$, the relations~\cite{Alnes:2006a} $J_{0}=J=R(t_{0},r_{0})\sin(\xi_{0})$
and $p_{0}\,=\,-\sqrt{1+\beta(r_{0})}\cos(\xi_{0})\,/\, R'(t_{0},r_{0})$. Therefore, our autonomous system is completely defined by the initial conditions $t_{0}$, $r_{0}$, $\theta_{0}=0$, $z_{0}=0$ and $\xi_{0}$. The first two define the instant of measurement and the offset between observer and center, while $\xi_{0}$ stands for the direction of incidence of the photons.

Following~\cite{2009PhRvL.102o1302Q}, an algorithm for predicting the variation of an arbitrary angular separation and redshift with time can be written as follows:

\begin{enumerate}
\item Denote with $(z_{a1},\xi_{a1})$ the observed coordinate
of a source at a given time $t_{0}$ and observer position
$r_{0}$; \vspace{-0.1cm}

\item Solve numerically the autonomous system with initial conditions $(t_{0},r_{0},\theta_{0}=0,z_{0}=0,\xi_{0}=\xi_{a1})$
    and find out the values of $\lambda_{a}^{\ast}$ such that $z(\lambda_{a}^{\ast})=z_{a1}\,$;
\vspace{-0.1cm}

\item Take note of the values $r_{a1}(\lambda_{a}^{\ast})\,$ and $\theta_{a1}(\lambda_{a}^{\ast})\,$
(since the sources are assumed comoving with no peculiar velocities,
these values are constant in time); \vspace{-0.1cm}

\item Define $\lambda_{a}^{\dagger}$ as the parameter value for which $r_{a2}(\lambda_{a}^{\dagger})=r_{a1}(\lambda_{a}^{\ast})$,
    where $r_{a2}$ is the geodesic solution for a photon arriving a time
    $\Delta t$ later with an incident angle $\xi_{a2}$, and vary $\xi_{a2}$
    until $\theta_{a2}(\lambda_{a}^{\dagger}) = \theta_{a1}(\lambda_{a}^{\ast})\,$;
\vspace{-0.1cm}

\item Compute the difference $\Delta z=z_{a2}-z_{a1}$ (to obtain the redshift drift);

\item Repeat the above steps for source $b$, and compute the difference
    $\Delta_{t}\gamma\equiv\gamma_{2}-\gamma_{1} = (\xi_{a2}-\xi_{b2})-(\xi_{a1}-\xi_{b1})$ (to obtain the cosmic parallax).
\end{enumerate}
\vspace{-0.1cm}

The above algorithm provide two (in principle) coupled observables: the Cosmic Parallax (CP) (see Sec.~\ref{sec:parallax}), and the redshift drift (see Sec.~\ref{sec:driftLTB}).

A remark on the above procedure is in order. Due to the intrinsically smallness of both the cosmic parallax and the redshift drift (in the course of a decade), a carefully constructed numerical code is needed to correctly compute either. To give an idea of the amount of precision required, consider the following: if one naively calculates $\Delta_{t}\gamma$ for a $\Delta t$ of 10~years, one needs to evaluate $\xi_{a1}$ and $\xi_{a2}$ with at least 13 digits of precision (as the CP is of the order of $0.2\,\mu\mbox{as}\sim10^{-12}$~rad) \cite{2010PhRvD..81d3522Q}.

\section{Error Bars Estimates with SDSS Quasars}
\label{app:zdot-errorbars}

The signal-to-noise ratio per pixel (of size $0.0125$~{\AA}) in~\eqref{eq:sigma-v} was estimated in~\cite{2008MNRAS.386.1192L}
to be
\begin{equation}
    \frac{\mbox{S}}{\mbox{N}}=700\left[\frac{Z_{X}}{Z_{r}}\;10^{0.4(16-m_{X})}\; \left(\frac{D}{42\,\mbox{m}}\right)^{2}\;\frac{t_{{\rm int}}}{10\,\mbox{h}}\;\frac{\epsilon}{0.25}\right]^{\frac{1}{2}},\label{eq:StoN}
\end{equation}
where $Z_{X}$ and $m_{X}$ are the source zero point and apparent
magnitude in the ``X'' band and $D$, $t_{{\rm int}}$ and $\epsilon$
are the telescope diameter, total integration time and total efficiency
respectively. We assumed a central obscuration of the telescope's primary collecting area of $10\%$~\cite{2008MNRAS.386.1192L}. Note that $D=42$~m corresponds to the reference design for the E-ELT~\cite{gilmozzi07}.

The reason magnitudes are quoted in terms of an arbitrary ``X''
band is because one should use the magnitude of the bluest filter
that still lies entirely redwards of the quasar's Lyman-$\alpha$ emission
line~\cite{2008MNRAS.386.1192L}. This means that for $z_{QSO}<2.2$ one should
use the magnitude in the $g$-band; for $2.2<z_{QSO}<3.47$ the one
in the $r$-band; for $3.47<z_{QSO}<4.61$ the $i$-band; for $z_{QSO}>4.61$
the $z$-band. A good estimate for $m_{X}$ can be achieved with the
SDSS DR7, selecting the brightest quasars in each redshift bin using
the appropriate band for such bin.

Following~\cite{2007MNRAS.382.1623B}, Ref.~\cite{2010PhRvD..81d3522Q} selected 40 quasars in 5 redshift bins, centered at $z=\{2,\,2.75,\,3.5,\,4.25,\,5\}$, all of the same redshift width of $0.75$. The corresponding bands are, in order, $\{g,r,r,i,z\}$ (where the $i$-band could equally be chosen for the middle bin). Doing so, one gets for the average (amongst the 8 brightest quasars) apparent magnitude $m_{X}$ for each bin the following: $m_{X}=\{15.45,\,16.54,\,16.40,\,17.51,\,18.33\}$. Finally, the zero point magnitude ratio in each bin was estimated in~\cite{2010PhRvD..81d3522Q} to be: $Z_{X}/Z_{r}=\{1.01,\,1.00,\,1.00,\,0.98,\,0.93\}$.
The accuracy of this last estimate is however quite unimportant in the results shown.

\bibliography{Brevfin}
\end{document}